\documentclass[journal]{IEEEtran}

\usepackage{float}
\usepackage{amssymb}
\usepackage{bm}
\usepackage{amsmath,amsfonts}
\usepackage[ruled]{algorithm2e}
\usepackage{array}
\usepackage{textcomp}
\usepackage{stfloats}
\usepackage{url}
\usepackage{verbatim}
\usepackage{graphicx}
\usepackage{subfigure}
\usepackage{cite}
\usepackage{booktabs}    
\usepackage{mathrsfs} 
\usepackage{setspace}
\usepackage{geometry}
\usepackage{multirow}
\usepackage{amsmath} 
\allowdisplaybreaks[4] 
\usepackage{autobreak} 
\usepackage{color}
\usepackage{hyperref} 
\usepackage{geometry}  
\makeatletter
\renewcommand{\maketag@@@}[1]{\hbox{\m@th\normalsize\normalfont#1}}%
\makeatother

\begin{document}
\begin{spacing}{0.95}

\title{Sensing User's Channel and Location with Terahertz Extra-Large Reconfigurable Intelligent Surface under Hybrid-Field Beam Squint Effect}

\author{Zhuoran Li, Zhen Gao, Tuan Li
	\thanks{Z. Li, Z. Gao, and T. Li are with the School of Information and Electronics, Beijing Institute of Technology, Beijing 100081, China, and also with the Advanced Research Institue of Multidisciplinary Sciences, Beijing Institute of Technology, Beijing 100081, China
		(e-mails: \mbox{\{lizhuoran, gaozhen16, 6120210230\}@bit.edu.cn}).\textit{(Corresponding author: Zhen Gao.)}}
}


\maketitle
\begin{abstract}
	This paper investigates the sensing of user's uplink channel and location in terahertz extra-large reconfigurable intelligent surface (XL-RIS) systems, where the unique hybrid far-near field effect and the beam squint effect caused by the XL array aperture as well as the XL bandwidth are overcome.
	Specifically, we first propose a joint channel and location sensing scheme, which consists of a location-assisted generalized multiple measurement vector orthogonal matching pursuit (LA-GMMV-OMP) algorithm for channel estimation (CE) and a complete dictionary based localization (CDL) scheme, where a frequency selective polar-domain redundant dictionary is proposed to overcome the hybrid field beam squint effect. 
	The CE module outputs coarse on-grid angle estimation (respectively observed from the BS and RIS) to the localization module, which returns the fine off-grid angle estimation to improve CE.
	Particularly, with RIS, CDL can obtain user's location via line intersection, and a polar-domain gradient descent (PGD) algorithm at the base station is proposed to achieve the off-grid angle estimation with super-resolution accuracy.
	Additionally, to further reduce the sensing overhead, we propose a partial dictionary-based localization scheme, which is decoupled from CE, where RIS is served as an anchor to lock the user on the hyperbola according to time difference of arrival and the user's off-grid location can be obtained by using the proposed PGD algorithm.
	Simulation results demonstrate the superiority of the two proposed localization schemes and the proposed CE scheme over state-of-the-art baseline approaches.
\end{abstract}

\vspace{-5mm}
\section{Introduction}\label{sec:Introduction}
\vspace{-1mm}
\subsection{Prior Works}
\IEEEPARstart{T}{he} cellular network localization is a prerequisite for various critical applications in 6G. Compared to satellite localization, cellular network localization is more suitable for indoor and urban environment.
Conventional cellular network localization methods can be divided into three categories, depending on the receive signal strength (RSS), time of arrival (ToA)/time difference of arrival (TDoA), and angle of arrival (AoA)/angle of departure (AoD), respectively\cite{2018_CST_localization}.
In contrast to RSS-based localization methods, which are mainly used indoors due to their poor accuracy for outdoor localization\cite{2018_TCOM_RSSindoor,2018_CST_localization}, ToA/TDoA and AoA/AoD-based localization methods can be used both indoors and outdoors.
ToA-based methods estimate the delays between multiple anchors and the user equipment (UE) to obtain the UE's location according to the intersection of several circles\cite{2017_TSP_DirectLoc_ToA}, while the TDoA-based methods resort to the time differences between anchors and the UE to conduct hyperbolic localization\cite{2015_CC_TDoA}.
The advantage of TDoA-based methods over ToA-based methods is that the former only requires accurate synchronization of time between anchors\cite{2015_CC_TDoA,2ToA1TDoA}, which avoids the affection of clock offset between the UE and the BS.
Additionally, in the OFDM frequency-domain model, the cyclic prefix can only be used to obtain the delay of the multipath with respect to the first path, and the absolute delay of each path is not available.
Therefore, the communication protocol for localization is based on the TDoA and round trip time\cite{ref_PositioningIn5GNetworks,ref_3GPP_38215}.
Benefiting from high angle resolution of massive multiple-input multiple-output (mMIMO) and even extra-large MIMO (XL-MIMO) systems, the AoA-based UE localization has been studied in \cite{2017_TSP_DirectLoc_ToA,2018_TWC_AoA}.
In \cite{2017_TSP_DirectLoc_ToA}, multiple base stations (BSs) were utilized to sense the UE's location, where the matching filtering method and compressed sensing (CS) method were used to estimate the ToA and AoA, respectively.
However, this work required the collaboration of multiple BSs and therefore it cannot be applied to the case of a single BS.
In \cite{2018_TWC_AoA}, distributed compressed sensing-simultaneous orthogonal matching pursuit (OMP) algorithm was utilized to estimate the AoA, AoD and ToA, which were then refined using the expectation maximization algorithm to sense the UE's location indoors.
However, the UE needed to locate scatterers in order to calculate its own location.
Moreover, this scheme cannot work well if the UE is equipped with only one antenna.

In addition, some literature has conducted the studies of reconfigurable intelligent surface (RIS)-assisted UE localization \cite{2022_JSAC_RISaidedSensing,ref_TWC_wangwei,ref:RIS_assisted_localization_bound}.
A RIS-self-sensing system was proposed in \cite{2022_JSAC_RISaidedSensing}, which designed the phases of the RIS and adopted multiple signal classification (MUSIC) algorithm to estimate the AoA.
However, this work only estimated the AoA, rather than the specific location of the UE.
Using the degree of freedom of observation brought by multiple RISs, random beamforming and maximum likelihood estimation method were utilized in \cite{ref_TWC_wangwei} to estimate the AoD and sense the UE's location.
However, this work requires the UE to have perfect knowledge of the locations of the RISs, which is difficult to achieve in practice.
Moreover, the downlink localization imposes a certain computational burden on the UE compared to the uplink localization.
In \cite{ref:RIS_assisted_localization_bound}, the RIS-assisted localization error bound was analyzed, which brought theoretical guidance to the deployment of the RIS.
However, the process of designing the phase of the RIS and analyzing the error bound requires the UE's location in advance, which is difficult to achieve in practice.

As for channel sensing or channel estimation (CE), the path loss is severe for terahertz (THz) signals, so the angle-domain representation of mMIMO and XL-MIMO channels presents sparse features.
To exploit the angle-domain sparsity, various CS methods (e.g., OMP algorithms and its derivatives) have been proposed to sense the channels \cite{ref_ShicongLiu, ref_ZiweiWan, ref:OMP,ref:ZhenGao_Letter,2022_TCOM_LinglongDai_PolarDomain,gOMP}.
For wideband mMIMO systems, a distributed grid matching pursuit (DGMP) algorithm was proposed in \cite{ref:ZhenGao_Letter}.
For the CE problem in the near-field region, the polar-domain simultaneous OMP (PSOMP) algorithm has been proposed in \cite{2022_TCOM_LinglongDai_PolarDomain}, which only works in the case that the beam squint effect (BSE) is not obvious.
However, the aforementioned algorithms rely on the on-grid processing, which suffer from the limited estimate resolution due to the continuously distributed AoA/AoD.
Therefore, off-grid super-resolution channel sensing algorithms were proposed to improve the channel sensing accuracy \cite{offgrid_2018_TVT,offgrid_2021_TWC}.
\begin{small}
\begin{table*}[]
	\centering
	\begin{tiny}
	\centering
	\caption{A brief comparison of the related literature with our work}
	\label{tab:intro}
	\begin{tabular}{|c|ccc|cccc|ccc|c|c|c|}
		\hline
		\multirow{3}{*}{\textbf{Reference}}                 & \multicolumn{3}{c|}{\textbf{\begin{tabular}[c]{@{}c@{}}Categories of\\ Localization Methods\end{tabular}}}                                               & \multicolumn{4}{c|}{\textbf{MIMO Channel}}                                                                                                                                                                                                                                                     & \multicolumn{3}{c|}{\textbf{\begin{tabular}[c]{@{}c@{}}Precoding/Beamforming\\ architecture\end{tabular}}}                                                                        & \multirow{3}{*}{\textbf{\begin{tabular}[c]{@{}c@{}}Combine\\ with CE\end{tabular}}} & \multirow{3}{*}{\textbf{\begin{tabular}[c]{@{}c@{}}Assisted\\ by RIS\end{tabular}}} & \multirow{3}{*}{\textbf{Algorithm}}                                                                 \\ \cline{2-11}
		& \multicolumn{1}{c|}{\begin{tabular}[c]{@{}c@{}}ToA/\\ TDoA\end{tabular}} & \multicolumn{1}{c|}{\begin{tabular}[c]{@{}c@{}}AoA/\\ AoD\end{tabular}} & RSS & \multicolumn{1}{c|}{\begin{tabular}[c]{@{}c@{}}Beam \\ Squint\end{tabular}} & \multicolumn{1}{l|}{\begin{tabular}[c]{@{}l@{}}Far-\\ Field\end{tabular}} & \multicolumn{1}{c|}{\begin{tabular}[c]{@{}c@{}}Near-\\ Field\end{tabular}} & \begin{tabular}[c]{@{}c@{}}Hybrid-\\ Field\end{tabular} & \multicolumn{1}{c|}{\begin{tabular}[c]{@{}c@{}}Single\\ antenna\end{tabular}} & \multicolumn{1}{c|}{\begin{tabular}[c]{@{}c@{}}Full-\\ Digital\end{tabular}} & Hybrid &                                                                                     &                                                                                     &                                                                                                     \\ \hline
		\cite{2018_TCOM_RSSindoor}       & \multicolumn{1}{c|}{}                                                    & \multicolumn{1}{c|}{{\checkmark}}                                                & {\checkmark} & \multicolumn{1}{c|}{}                                                       & \multicolumn{1}{c|}{{\checkmark}}                                                  & \multicolumn{1}{c|}{}                                                      &                                                         & \multicolumn{1}{c|}{}                                                         & \multicolumn{1}{c|}{{\checkmark}}                                                     &        &                                                                                     &                                                                                     & \begin{tabular}[c]{@{}c@{}}Hybrid RSS-AoA positioning scheme\end{tabular}                         \\ \hline
		\cite{2017_TSP_DirectLoc_ToA}   & \multicolumn{1}{c|}{{\checkmark}}                                                 & \multicolumn{1}{c|}{{\checkmark}}                                                &     & \multicolumn{1}{c|}{}                                                       & \multicolumn{1}{c|}{{\checkmark}}                                                  & \multicolumn{1}{c|}{}                                                      &                                                         & \multicolumn{1}{c|}{}                                                         & \multicolumn{1}{c|}{{\checkmark}}                                                     &        &                                                                                     &                                                                                     & \begin{tabular}[c]{@{}c@{}}Direct Source Localization\end{tabular}                                \\ \hline
		\cite{2015_CC_TDoA}              & \multicolumn{1}{c|}{{\checkmark}}                                                 & \multicolumn{1}{c|}{}                                                   &     & \multicolumn{1}{c|}{}                                                       & \multicolumn{1}{c|}{{\checkmark}}                                                  & \multicolumn{1}{c|}{}                                                      &                                                         & \multicolumn{1}{c|}{{\checkmark}}                                                      & \multicolumn{1}{c|}{}                                                        &        &                                                                                     &                                                                                     & \begin{tabular}[c]{@{}c@{}}Maximum likelihood estimation\end{tabular}                             \\ \hline
		\cite{2018_TWC_AoA}              & \multicolumn{1}{c|}{{\checkmark}}                                                 & \multicolumn{1}{c|}{{\checkmark}}                                                &     & \multicolumn{1}{c|}{}                                                       & \multicolumn{1}{c|}{{\checkmark}}                                                  & \multicolumn{1}{c|}{}                                                      &                                                         & \multicolumn{1}{c|}{}                                                         & \multicolumn{1}{c|}{{\checkmark}}                                                     &        &                                                                                     &                                                                                     & \begin{tabular}[c]{@{}c@{}}Modified OMP, Expectation maximization\end{tabular}                  \\ \hline
		\cite{2022_JSAC_RISaidedSensing} & \multicolumn{1}{c|}{}                                                    & \multicolumn{1}{c|}{{\checkmark}}                                                &     & \multicolumn{1}{c|}{}                                                       & \multicolumn{1}{c|}{{\checkmark}}                                                  & \multicolumn{1}{c|}{}                                                      &                                                         & \multicolumn{1}{c|}{}                                                         & \multicolumn{1}{c|}{{\checkmark}}                                                     &        &                                                                                     & {\checkmark}                                                                                 & \begin{tabular}[c]{@{}c@{}}Customized MUSIC\end{tabular}                                          \\ \hline
		\cite{ref_TWC_wangwei}           & \multicolumn{1}{c|}{}                                                    & \multicolumn{1}{c|}{{\checkmark}}                                                &     & \multicolumn{1}{c|}{}                                                       & \multicolumn{1}{c|}{{\checkmark}}                                                  & \multicolumn{1}{c|}{}                                                      &                                                         & \multicolumn{1}{c|}{}                                                         & \multicolumn{1}{c|}{}                                                        & {\checkmark}    &                                                                                     & {\checkmark}                                                                                 & \begin{tabular}[c]{@{}c@{}}Maximum likelihood estimation\end{tabular}                             \\ \hline
		\cite{ref_ICC_loc_RIS}          & \multicolumn{1}{c|}{}                                                    & \multicolumn{1}{c|}{}                                                   & {\checkmark} & \multicolumn{1}{c|}{}                                                       & \multicolumn{1}{l|}{}                                                     & \multicolumn{1}{c|}{{\checkmark}}                                                   &                                                         & \multicolumn{1}{c|}{{\checkmark}}                                                      & \multicolumn{1}{c|}{}                                                        &        &                                                                                     & {\checkmark}                                                                                 & \begin{tabular}[c]{@{}c@{}}Maximum likelihood estimation\end{tabular}                             \\ \hline
		\cite{ref_GFF_loc}               & \multicolumn{1}{c|}{}                                                    & \multicolumn{1}{c|}{{\checkmark}}                                                &     & \multicolumn{1}{c|}{{\checkmark}}                                                    & \multicolumn{1}{l|}{}                                                     & \multicolumn{1}{c|}{{\checkmark}}                                                   &                                                         & \multicolumn{1}{c|}{}                                                         & \multicolumn{1}{c|}{}                                                        & {\checkmark}    &                                                                                     &                                                                                     & \begin{tabular}[c]{@{}c@{}}Time-delay lines-assisted localization\end{tabular}                   \\ \hline
		\cite{2019_TSP_AoA}              & \multicolumn{1}{c|}{{\checkmark}}                                                 & \multicolumn{1}{c|}{{\checkmark}}                                                &     & \multicolumn{1}{c|}{}                                                       & \multicolumn{1}{c|}{{\checkmark}}                                                  & \multicolumn{1}{c|}{}                                                      &                                                         & \multicolumn{1}{c|}{}                                                         & \multicolumn{1}{c|}{{\checkmark}}                                                     &        & {\checkmark}                                                                                 &                                                                                     & \begin{tabular}[c]{@{}c@{}}Successive localization and beamforming\end{tabular}                   \\ \hline
		Our Work                                            & \multicolumn{1}{c|}{{\checkmark}}                                                 & \multicolumn{1}{c|}{{\checkmark}}                                                &     & \multicolumn{1}{c|}{{\checkmark}}                                                    & \multicolumn{1}{l|}{}                                                     & \multicolumn{1}{c|}{}                                                      & {\checkmark}                                                     & \multicolumn{1}{c|}{}                                                         & \multicolumn{1}{c|}{}                                                        & {\checkmark}    & {\checkmark}                                                                                 & {\checkmark}                                                                                 & \begin{tabular}[c]{@{}c@{}}LA-GMMV-OMP algorithm along with CDL Scheme, PDL Scheme\end{tabular} \\ \hline
	\end{tabular}
	\end{tiny}
\end{table*}
\end{small}

On the other hand, due to the high carrier frequency of millimeter-wave (mmWave)/THz and the large aperture of XL-MIMO or XL-RIS, the Rayleigh distance becomes significantly large in cellular networks, therefore the conventional far-field assumption is not always valid.
In contrast, the near-field communications have attracted much attention recently.
In \cite{2022_TCOM_LinglongDai_PolarDomain}, the phenomenon that the coexistence of the near-field and the far-field is called the hybrid far-near field (HFNF) effect.
Meanwhile, the BSE \cite{2021_JSAC_AnwenLiao} induced by ever-increasing bandwidth severely limits the performance of communications and network sensing of XL-MIMO systems.
In \cite{2022_TCOM_LinglongDai_PolarDomain}, to acquire better performance in the near-field, polar-domain transform matrix (PTM) was proposed to replace the Fourier transform matrix (FTM) in the OMP-based channel sensing schemes.
In contrast to the FTM, only having the angle-domain resolution, the PTM has both the angle-domain resolution and distance-domain resolution, which can overcome the energy spread effect\cite{2022_TCOM_LinglongDai_PolarDomain}.
A solution to sense the UE's location under the near-field BSE was proposed in\cite{ref_GFF_loc}, but the localization accuracy was poor.
In \cite{ref_ICC_loc_RIS}, the RIS was used as lens and its phase was specifically designed, where a maximum likelihood estimation method was utilized to estimate the UE's location in the near-field region without BSE.
Although sophisticated methods have been proposed in \cite{2021_JSAC_AnwenLiao},\cite{ref_hybrid_GaoFeifei} to overcome the BSE in communication systems, the influence of such effect on cellular network sensing has not been well studied at the time of writing.

By far, the aforementioned localization methods seldom considered the HFNF BSE, which is common in mmWave/THz XL-MIMO systems with very large bandwidth\cite{2017_TSP_DirectLoc_ToA,2018_TWC_AoA,2019_TSP_AoA,2022_JSAC_RISaidedSensing,ref_TWC_wangwei,ref_ICC_loc_RIS}.
Even if the HFNF BSE was taken into consideration, the accuracy of localization did not meet the requirements of 6G communications\cite{ref_GFF_loc}.
In addition, although UE localization and channel sensing have been jointly investigated in \cite{2018_TWC_AoA,2019_TSP_AoA} in the far-field region without the BSE, no current research has investigated the joint localization and channel sensing in the HFNF channel with BSE. 
Therefore, the study of joint UE localization and channel sensing in RIS-assisted \textcolor{black}{XL-array} systems under the HFNF BSE is still in its early stage.

\vspace{-4mm}
\subsection{Our Contributions}
\begin{figure}[!t]
	\centering
	\color{black}
	\includegraphics[width=3.4in]{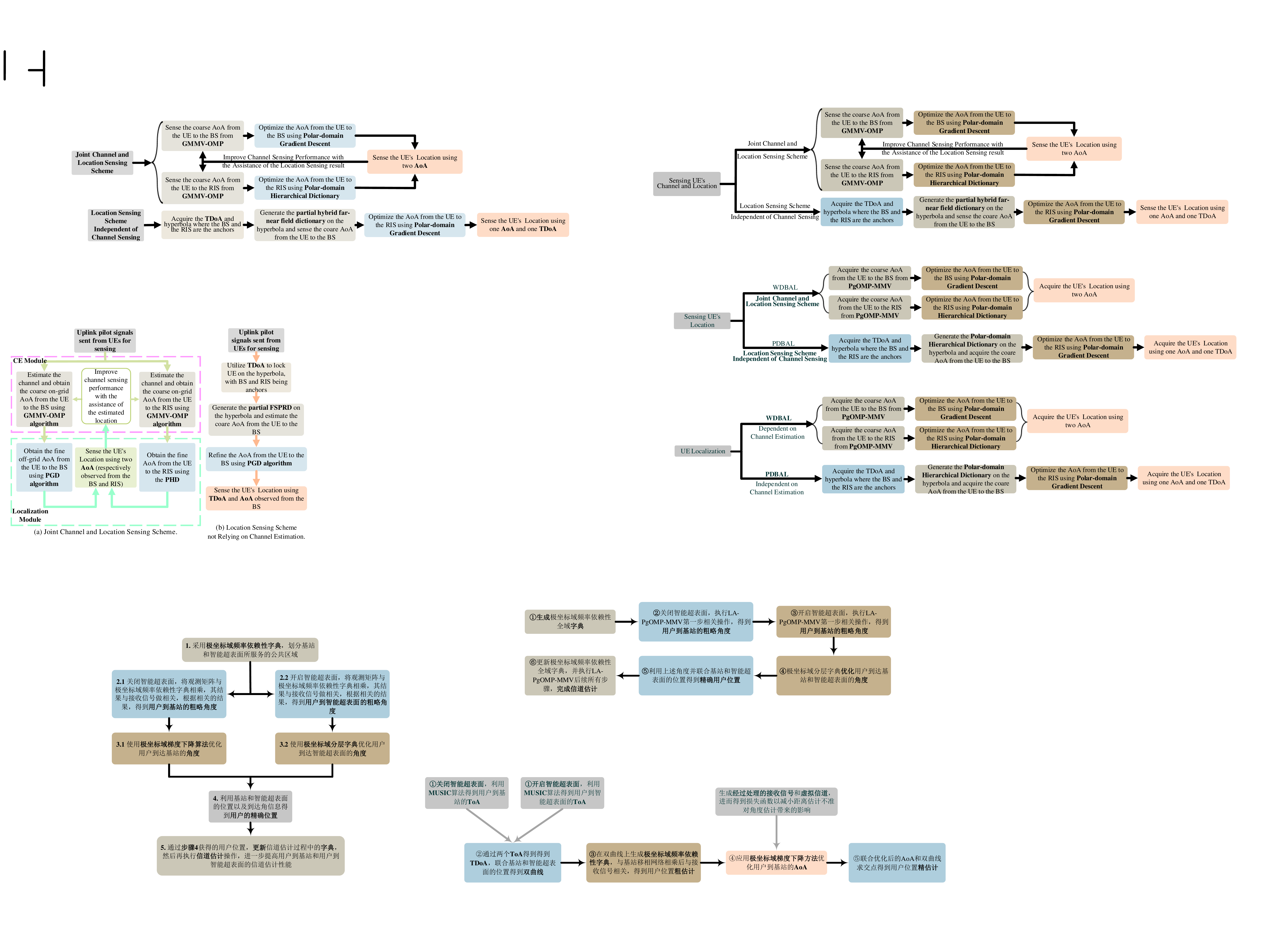}
	\vspace{-3mm}
	\caption{The sensing procedure of the joint channel and location sensing scheme and the location sensing scheme not relying on channel estimation.}
	\label{fig_loc_flow}
\end{figure}

This paper proposes two RIS-assisted localization paradigms for network sensing, where HFNF channel with BSE is considered.
Specifically, we propose a joint channel and location sensing scheme in Section \ref{sec:WDBAL} and a pure location sensing scheme not relying on channel estimation in Section \ref{sec:PDL}, respectively.
The sensing procedure for the UE's channel and location is summarized in Fig. \ref{fig_loc_flow}.

Our contributions\footnote{Simulation codes are provided to reproduce the results in this paper: \href{https://github.com/LiZhuoRan0}{https://github.com/LiZhuoRan0}} are summarized as follows:
\begin{itemize}
	\item \textbf{We design a frequency-selective polar-domain redundant dictionary (FSPRD) for sensing the UE's channel and location under the HFNF BSE.}
	The conceived FSPRD is developed from the PTM\cite{2022_TCOM_LinglongDai_PolarDomain}, so that the angle-distance parameters can be reliably estimated under HFNF channels. Moreover, BSE indicates that the virtual angle-distance representation under HFNF channels shifts as the subcarrier deviates from the central carrier\footnote{BSE originally refers to that the virtual angle representation in far-field shifts as the subcarrier deviates from the central carrier. Here we extend this concept to HFNF channels.}. 
	The proposed FSPRD can compensate the offsets for different subcarriers so that the identical physical angle-distance parameters among different subcarriers can be ensured and exploited for enhanced sensing.				
	
	\item \textbf{We propose a joint channel and location sensing scheme.}
	This solution consists of a location-assisted generalized multiple measurement vector orthogonal matching pursuit (LA-GMMV-OMP) algorithm for CE and a complete dictionary based localization (CDL) scheme.
	The CE module outputs coarse on-grid angle estimation (respectively observed from the BS and RIS) to the localization module, which returns the fine off-grid angle estimation to improve CE.
	Specifically, the correlation operation of the LA-GMMV-OMP algorithm outputs a coarse AoA estimation to the CDL scheme for localization.
	In CDL scheme, a polar-domain gradient descent (PGD) algorithm is proposed to obtain the fine off-grid estimation of AoA seen from the BS, and the polar-domain hierarchical dictionary (PHD) is utilized to obtain the fine estimation of AoA observed from the RIS. 
	On this basis, we can obtain the accurate UE’s location, which can further facilitate the line-of-sight (LoS) channel reconstruction and therefore improve the channel sensing performance of LA-GMMV-OMP algorithm.
	Note that, by adding multiple atoms to the support set in each iteration, the LA-GMMV-OMP algorithm can better estimate NLoS paths in channels with cluster structure. At the same time, we apply a novel adaptive iterative stopping criterion, which has more stable performance than the conventional residual-based criterion.

	\item \textbf{We propose a PGD algorithm to acquire the fine off-grid estimation of AoAs at the BS.}
	Since the on-grid estimation of AoA based on the quantized FSPRD has limited resolution, the sensing performance of UE's channel and location has a limited precision. 
	Therefore, an off-grid PGD algorithm is dedicatedly designed.
	By carefully designing the combiner of the BS, the AoA estimation can be decoupled from distance and we can obtain an equivalent LoS path channel, which can be used to obtain a loss function with good local convexity, since it only consists of the channel gain and the HFNF steering vector.
	On this basis, the PGD algorithm is proposed to obtain the off-grid AoA estimation at the BS without knowing the exact distance.
	
	\item \textbf{We propose a pure location sensing scheme that does not rely on CE.}
	In the case that only the UE's location is required, the sensing signal overhead can be further reduced. 
	Therefore, to directly sense the UE's location, we further propose a partial dictionary based localization (PDL) scheme, where the TDoA is utilized to lock the UE on the hyperbola and both the BS and RIS are served as anchors.
	Particularly, partial FSPRD are generated on the hyperbola to obtain the coarse AoA and then the PGD algorithm is utilized to improve the accuracy of the AoA estimation.
	Since the UE is locked on the hyperbola, the size of FSPRD and the involved computational complexity can be considerably reduced.        
\end{itemize}

\vspace{-3mm}
\subsection{Notation}

Throughout this paper, scalar variables are denoted by normal-face letters, while boldface lower and uppercase letters denote column vectors and matrices, respectively;
the transpose and conjugate transpose operators are denoted by $(\cdot)^T$ and $(\cdot)^H$, respectively;
$j=\sqrt{-1}$ is the imaginary unit;
$\mathbb{C}$ is the sets of complex-valued numbers;
$|\mathcal{A}|_c$ is the cardinal number of set $\mathcal{A}$;
$\emptyset$ is the empty set;
$\mathbf{X}_{:,m_1:m_2}$ is the matrix composed of column vectors from $m_1$-th column to $m_2$-th column of matrix $\mathbf{X} \in \mathbb{C}^{N\times M}$;
$[m]$ in ${\bf{X}}[m](\theta)$ means extracting some elements of ${\bf{X}}$ indexed by $[m]$, where $\theta$ is the argument of ${\bf{X}}[m](\theta)$;
$\text{diag}(\mathbf{a})$ is a diagonal matrix with elements of $\mathbf{a}$ on its diagonal;
$\text{tr}(\cdot)$ is the trace operator;
$|s|$ is the magnitude of $s$, whether $s$ is a real number or a complex-valued number;
$\left\|\mathbf{s}\right\|_{\text{F}}$ and $\left\|\mathbf{S}\right\|_{\text{F}}$ is the Frobenius norm of vector $\mathbf{s}$ and matrix $\mathbf{S}$, respectively;
$\mathcal{CN}(\mu, \Sigma)$ is the Gaussian distribution with mean $\mu$ and covariance $\Sigma$;
$\mathcal{U}(a, b)$ is the uniform distribution between $a$ and $b$;
$\mathscr{R}(s)$ is the real part of the complex-valued number $s$;
$\left\lceil s \right\rceil$ represents finding the smallest integer greater than or equal to $s$;
$\partial(\cdot)$ is the first-order partial derivative operation;
$\odot$ is the Hadamard product;
$\mathbf{0}_n$, $\mathbf{1}_n$ and $\mathbf{I}_n$ are the vector of size $n$ with all the elements being 0, 1 and the $n\times n$ identity matrix, respectively;
$c$ is the speed of light.

\section{System Model}\label{sec:SystemModel}
\subsection{Channel Model}
We consider that each UE is equipped with one omni-directional antenna\footnote{\textcolor{black}{This paper can be directly extended to the multi-user scenario by assigning orthogonal pilots, i.e., orthogonal time-frequency resources, to different users. Since different users are completely orthogonal, without loss of generality, we can analyze the performance of the proposed scheme by taking only one user as an example\cite{ref_Tse,2022_TCOM_LinglongDai_PolarDomain}.}}. 
In order to reduce the prohibitive cost and power consumption in \textcolor{black}{XL-array} systems, hybrid beamforming is adopted.
Specifically, the BS is equipped with $N$-element uniform linear array (ULA) while only $N_{\text{RF}}$ radio frequency (RF)-chains are adopted ($N_{\text{RF}}<N$), the RIS has $N_{\text{RIS}}$ elements, and $M$ subcarriers are assigned to each UE.

\begin{figure}[!t]
	\vspace{-5mm}
	\centering
	\includegraphics[width=3.4in]{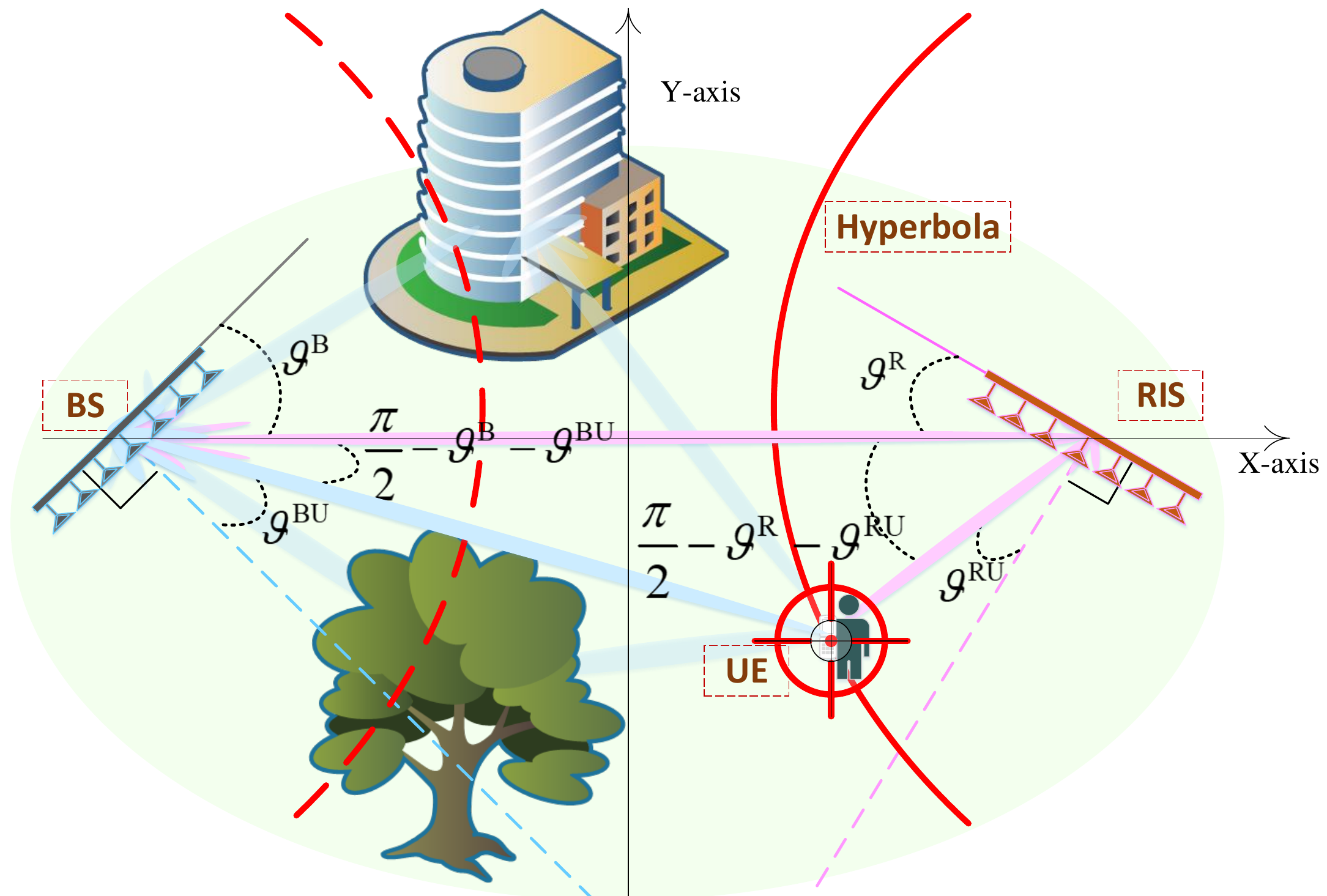}
	\vspace{-3mm}
	\caption{The model of a RIS-assisted localization system, and this is also a schematic diagram of the proposed location sensing scheme not relying on channel estimation in Section \ref{sec:PDL}.}
	\label{fig_hyperbola}
\end{figure}

In Fig. \ref{fig_hyperbola}, $\vartheta^{\text{B}}$ ($\vartheta^{\text{R}}$) is the angle between the BS (RIS) array and the $x$-axis.
$\vartheta^{\text{BU}}$ ($\vartheta^{\text{RU}}$) is the AoA from the UE to the BS (RIS) relative to the normal direction of the array.
For convenience, we define $\theta$ as the sine of the true AoA $\vartheta$ in radians, i.e., $\theta=\text{sin}(\vartheta)$.

In Fig. \ref{fig_hyperbola}, if the distance between the BS and the UE is less than the Rayleigh distance or more accurately the effective Rayleigh distance as will be described in the section \ref{subsec:boundary}, far-field planar wavefront assumption is no longer valid.
In this case, the near-field channel between each antenna element and the UE is not only determined by the AoA, but also by the distance.
Therefore, in \textcolor{black}{XL-array} systems, in order to model the near-field and the far-field channel simultaneously, the HFNF channel is adopted, and the channel between the BS (or RIS) and UE on the $m$-th subcarrier \textcolor{black}{${\mathbf{h}}[m] \in \mathbb{C}^{N}$ (or $\mathbb{C}^{N_{\text{RIS}}}$)} can be modeled as follows
\textcolor{black}{
\begin{equation}
	\label{near_field_channel}
	{\mathbf{h}}[m] = \sum\limits_{l = 0}^L {\sum\limits_{g = 1}^{{G_l}} {e^{ - j{k_m}\bar{r}_{l,g}}} {\boldsymbol{\alpha} _{l,g}[m] \odot {\mathbf{b}}_{l,g}[m]({f_m},\theta _{l,g},r_{l,g})} },
\end{equation}}
where $L$ denotes the number of clusters from the UE to the BS (or RIS),
$G_{l}$ denotes the number of paths in the $l$-th cluster,
$\boldsymbol{\alpha} _{l,g}[m] \in \mathbb{C}^{N}$  (or $\mathbb{C}^{N_{\text{RIS}}}$) denotes the channel gain of the $g$-th path in the $l$-th cluster on the $m$-th subcarrier,
$f_m={{f_c} - B/2 + (m - 1)B/M}$ denotes the frequency of the $m$-th subcarrier,
$B$ denotes the bandwidth,
$\lambda_m$ denotes the wavelength of the $m$-th subcarrier,
${k_m} = \frac{{2\pi }}{{{\lambda _m}}}$ denotes the wavenumber of the $m$-th subcarrier,
$\bar{r}_{l,g}$ denotes the total distance between the UE and the reference point of the BS (or RIS) array associated with the $g$-th path in the $l$-th cluster while $r_{l,g}$ denotes the distance of the last hop between the scatterer and the reference point of the BS (or RIS) array\footnote{$\bar{r}_{l,g}$ degrades to $r_{l,g}$ if the path is the LoS path.},
and $\theta _{l,g}$ denotes the AoA of the $g$-th path in the $l$-th cluster between the reference point of the BS (or RIS) array and the UE (or scatterers).
The reference point of the BS (or RIS) array is set to the center of the array.
\textcolor{black}{
$\boldsymbol{\alpha} _{l,g}[m] = {\alpha}^{\text{S}}_{l,g}[m] \boldsymbol{\alpha}^{\text{F}}_{l,g}[m] \odot
\boldsymbol{\alpha}^{\text{A}}_{l,g}[m]$, where $\boldsymbol{\alpha}^{\text{F}}_{l,g}[m]$ denotes the large-scale fading and can be depicted by Friis formula, ${\alpha}^{\text{S}}_{l,g}[m]$ denotes the small-scale fading, and $\boldsymbol{\alpha}^{\text{A}}_{l,g}[m]$ denotes the attenuation due to atmospheric gases (mainly water vapour and oxygen) and can be modeled based on the data in ITU-R P.676-12\cite{ref_2022_TCOM_NanYang_ChongHan, ref_2011_TWC_663,ref_2022_WC_ChongHan,2023_IEEENetwork_NanYang_ChongHan}.
It is worth noting that it is difficult to communicate efficiently in some frequency bands due to severe molecular absorption.
Therefore, the THz band is divided into a number of less heavily absorbed subbands, also known as transmission windows\cite{ref_2022_WC_ChongHan}.
The frequency bands considered in this paper fall within these transmission windows.}
In addition, we define the path with $l=0$ as the LoS and $l>0$ as the non-line-of-sight (NLoS) paths.
Therefore, $G_0 = 1$ and we only use the subscript $l=0$ to represent the channel gain $\boldsymbol{\alpha}_{0}[m]$, distance $r_0$, and AoA $\theta_{0}$ of the LoS channel.
The HFNF steering vector on the $m$-th subcarrier \textcolor{black}{${\mathbf{b}}_{l,g}[m] \in \mathbb{C}^{N}$ (or  $\mathbb{C}^{N_{\text{RIS}}}$)} can be acquired as
\begin{equation}
	\label{near_field_steervector}
	\begin{gathered}
		{\mathbf{b}}_{l,g}[m]({f_m},\theta _{l,g},r_{l,g}) \hfill \\
		= {[{e^{ - j{k_m}(r_{{l,g},0} - r_{l,g})}},\; \cdots ,\;{e^{ - j{k_m}(r_{{l,g},N - 1} - r_{l,g})}}]^T}/{\sqrt N } \hfill \\ 
	\end{gathered},
\end{equation}
where $N$ in (\ref{near_field_steervector}) can be replaced by $N_{\text{RIS}}$ if the HFNF steering vector is from the UE to the RIS (the same holds in the following, where the specific value can be determined according to the context), $r_{l,g,n} = \sqrt {({r_{l,g}})^2 - 2{{r_{l,g}}}{{\delta}}_n d\theta _{l,g} + {\delta} _n^2d^2}$ \cite{2022_TCOM_LinglongDai_PolarDomain} denotes the distance between the UE and the $n$-th element of the BS (or RIS) array associated with the $g$-th path in the $l$-th cluster,
$d=\frac{\lambda_c}{2}$ denotes the elements spacing in the BS (or RIS) array,
$\lambda_c$ is the carrier wavelength, and ${{\delta} _n} = (2 \times n - N + 1)/2, {\text{ }}n = 0, \cdots ,N - 1$.
When $r_{l,g}$ is large enough, in other words, the UE is in the far-field region of the BS (or RIS), the HFNF steering vector in (\ref{near_field_channel}) can be degenerated to the far-field steering vector \textcolor{black}{${\mathbf{a}}_{l,g}[m] \in \mathbb{C}^{N}$ (or $\mathbb{C}^{N_{\text{RIS}}}$)} as follows
	\vspace{-1mm}

\textcolor{black}{
\begin{align}
			\vspace{-2mm}
	\label{far_field_steervector}
		& {{\mathbf{a}}_{l,g}}[m]({f_m},{\theta _{l,g}}) \notag\\
		& = {[{e^{j{k_m}{\theta _{l,g}}d{\delta _0}}}, \cdots ,{e^{j{k_m}{\theta _{l,g}}d{\delta _{N - 1}}}}]^T}/\sqrt N  \notag\\ 
		& = {[{e^{j\frac{{2\pi }}{c}{f_m}{\theta _{l,g}}d{\delta _0}}}, \cdots ,{e^{j\frac{{2\pi }}{c}{f_m}{\theta _{l,g}}d{\delta _{N - 1}}}}]^T}/\sqrt N  \notag\\
		& \triangleq {{\mathbf{a}}_{l,g}}[m](\Xi_{m,l,g}) \notag\\
		& = {[{e^{j\pi\Xi_{m,l,g}{\delta _0}}}, \cdots ,{e^{j\pi\Xi_{m,l,g}{\delta _{N-1}}}}]^T}/\sqrt N  ,\notag\\ 
\end{align}
where $\Xi_{m,l,g} = \frac{f_m}{f_c}\theta_{l,g}$ by using $d=\frac{\lambda_c}{2}=\frac{c}{2f_c}$, which is consistent with the beam angle in \cite{ref_hybrid_GaoFeifei, ref_RevisionPrecode1, ref_RevisionPrecode2}.}
The parameters of (\ref{far_field_steervector}) are defined the same as (\ref{near_field_steervector}).

\vspace{-3mm}
\subsection{Boundary Between Near-Field and Far-Field}\label{subsec:boundary}
$A=Nd$ (or $A=N_{\text{RIS}}d$) is the aperture of the ULA at the BS (or RIS).
Conventionally, the boundary between the near-field spherical wave and the far-field plane wave can be defined according to the classical Rayleigh distance \cite{2021_arXiv_LinglongDai_effectiveRD}, i.e.,
\begin{equation}
	\label{Rayleigh_distance}
	Z = {2{A^2}}/{{\lambda_c}},
\end{equation}
which only depends on the aperture $A$ and the central frequency $f_c$.
However, when the number of the elements at the BS (or RIS) and the system bandwidth become large, i.e., for the THz \textcolor{black}{XL-array} systems, the definition (\ref{Rayleigh_distance}) is not always accurate, since the boundary is also determined by the frequency of the subcarrier and the AoA of the incident signal.
Therefore, in \cite{2021_arXiv_LinglongDai_effectiveRD}, an effective Rayleigh distance is further defined as
\begin{equation}
	\label{effective_Rayleigh_distance}
	{Z^{\text{eff}}_m}(\theta) = \epsilon  (1 - {\theta ^2}) {2{A^2}}/\lambda_m,
	_{}\end{equation}
where $\epsilon$ can be defined artificially based on the requirements.
According to the  \cite{2021_arXiv_LinglongDai_effectiveRD}, we can obtain $\epsilon$ by defining the constraint as
\begin{equation}
	\label{effective_Rayleigh_distance_define}
	{|{{\chi (f_m, \theta ,{ Z^{\text{eff}}_{m}}(\theta )) - \rho (f_m, \theta , {Z^{\text{eff}}_{m}}(\theta ))}}|^2} = \hbar,
\end{equation}
where $\chi (f_m,\theta, {Z^{\text{eff}}_{m}}(\theta ))=|{\mathbf{b}}[m]{({f_m},\theta, {Z^{\text{eff}}_{m}}(\theta ))^H}{\mathbf{b}}[m]({f_m}$
$,\theta,{Z^{\text{eff}}_{m}}$
$(\theta ))|$
, $\rho (f_m,\theta,{Z^{\text{eff}}_{m}}(\theta ))=|{\mathbf{a}}[m]({f_m},\theta )^H$
${\mathbf{b}}[m]{({f_m}}$
${,\theta, {Z^{\text{eff}}_{m}}(\theta ))}|$.
$\chi$ represents the autocorrelation of the HFNF steering vector.
$\rho$ represents the correlation between the HFNF steering vector and the far-field steering vector.
Since $\hbar$ is defined as the loss if a spherical wave is approximated as a plane wave, the greater $\hbar$ indicates the more significant difference between $\chi$ and $\rho$.
Given $\hbar$, $\epsilon$ and the corresponding $Z^{\text{eff}}_m(\theta)$ in (\ref{effective_Rayleigh_distance}) can be obtained according to simulations, e.g., if $f_m=f_c=0.1$ THz, $A=0.384$ m (i.e., $N=256$), $\hbar=0.1$, $\theta=0.5$, $\mathbf{a}[m]$ and  $\mathbf{b}[m]$ are taken from (\ref{far_field_steervector}) and (\ref{near_field_steervector}), respectively, $\epsilon$ is 0.4 and ${Z^{\text{eff}}_{c}}(\theta )$ is 29.5 m.
However, if (\ref{Rayleigh_distance}) is utilized, the classical Rayleigh distance is 98.3 m, which is much greater than the effective Rayleigh distance.
Meanwhile, if the classical Rayleigh distance is used as the boundary between the near-field region and the far-field region, the loss $\hbar=0.0096$, which is ten times smaller than the one if the effective Rayleigh distance is used.
Therefore, this paper adopts the effective Rayleigh distance rather than the classical Rayleigh distance to distinguish the boundary between near-field region and far-field region.

\vspace{-3mm}
\subsection{Hybrid Far-Near Field Beam Squint Phenomenon}\label{subsec:HFNF_BS}
We begin with the far-field case.
\textcolor{black}{$\Xi_{m}\approx\theta$ when $B$ is small since $\frac{f_m}{f_c}\approx1$, which indicates that $\Xi_m$ can be used to express $\theta$.
However, when $B$ is large, $\frac{f_m}{f_c}$ cannot be approximated as 1 and $\Xi_{m}$ cannot be approximated as $\theta$.
This is the beam squint phenomenon in the far-field case.
If we omit this phenomenon and think $\Xi_{m}\approx\theta$ is still hold, we will get
\begin{equation}
	\label{far_field_beam_squint}
	{f_{m_1}} {\theta_{m_1}} = {f_{m_2}} {\theta_{m_2}},
\end{equation}
i.e., different frequencies will correspond to different $\theta_m$, while in fact there can only be one ture physical angle $\theta$.
}
Therefore, if the system bandwidth $B$ is large enough compared with the center frequency $f_c$, e.g., $B=10$ GHz, $f_c=0.1$ THz and $B/{{f_c}} = 1/{10}$, which is the typical THz communication system parameters and will be adopted in our latter simulation, the maximum difference between $\theta_{m_1}$ and $\theta_{m_2}$ can be ${{\theta_{m_2}}}/{{\theta_{m_1}}} = {{f_{m_1}}}/{{f_{m_2}}} = {{f_{\text{max}}}}/{f_{\text{min}}} = {105\text{ GHz}}/{95\text{ GHz}} \approx 1.1$. 
If $\theta_1$ takes $0.5$, then $\theta_2$ takes $0.55$.
When we  convert $\theta$ to $\vartheta$, we have ${\vartheta _1} = \arcsin ({\theta _1}) = 30^\circ $ and ${\vartheta _2} = \arcsin ({\theta _2}) = 33.37^\circ$.
The angle-domain resolution can be approximated as $1/N=1/256\approx0.0039$.
Therefore, the BSE results in an angle differences across almost $(0.55-0.5)/0.0039\approx12.8$ angle-domain resolutions.
Besides, as the distance between the UE and the BS (or RIS) increases, there will be a greater localization error when this estimated angle is used for localization.
For example, if the distance between the UE and the BS (or RIS) is 30 m, the location offset between $f_{\text{max}}$ and $f_{\text{min}}$ can be up to 1.76 m, while if the distance between the UE and the BS (or RIS) increases to 60 m, the location offset can be 3.53 m.
Since the BS (or RIS) has the ultra-high angular resolution benefited from the XL-arrays for localization, the angle deviation caused by the BSE will have a severe impact on communications and sensing.
\begin{figure*}[!t]
	\color{black}
	\centering
	\subfigure[Far-field.]{
		\begin{minipage}[t]{0.45\linewidth}
			\centering
			\includegraphics[width=3.4in]{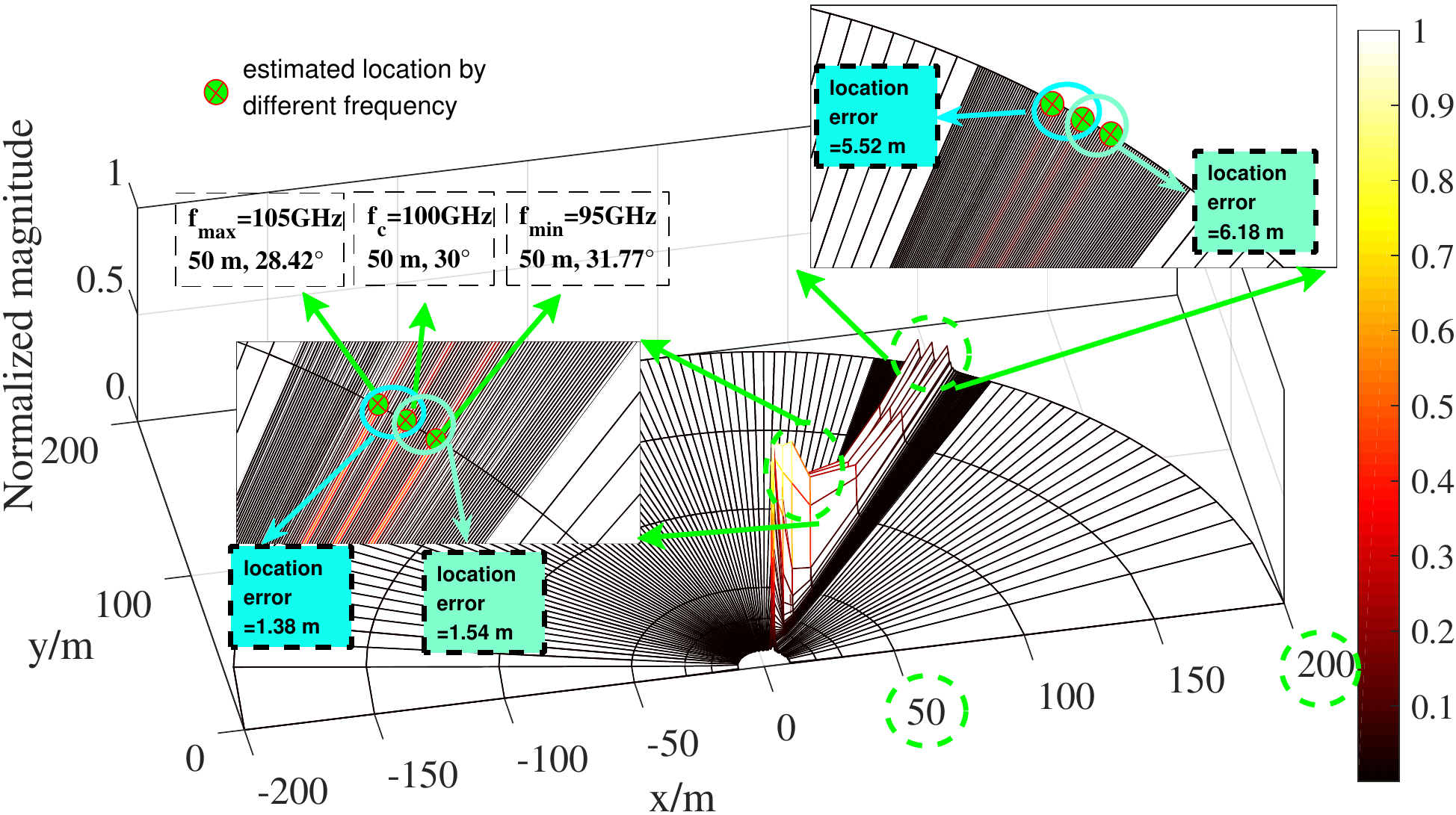}
			\label{fig_BS_far_side}
		\end{minipage}%
	}%
	\hspace{5mm}
	\subfigure[Near-field.]{
		\begin{minipage}[t]{0.45\linewidth}
			\centering
			\includegraphics[width=3.4in]{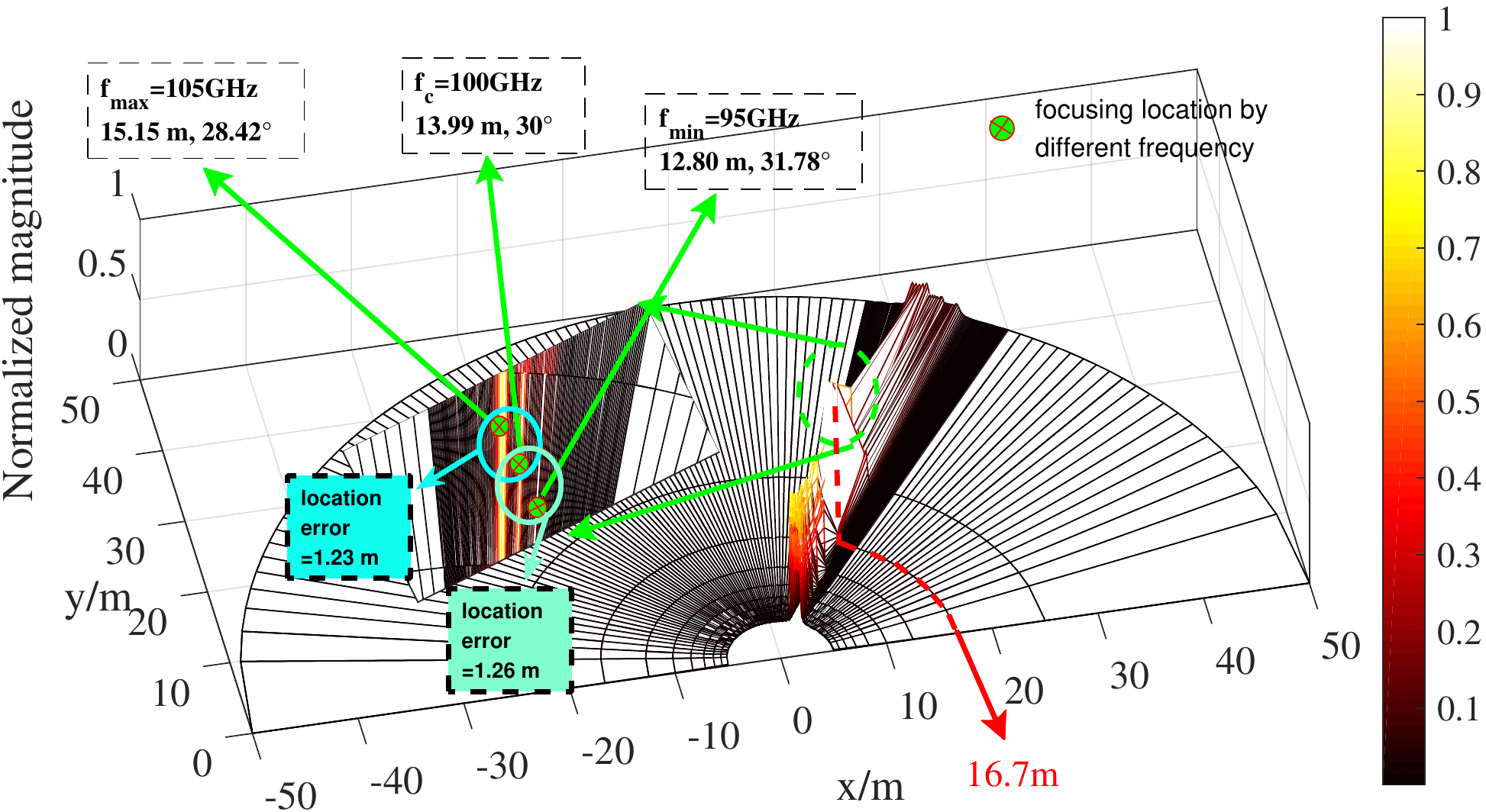}
			\label{fig_BS_hybrid_side}
		\end{minipage}%
	}%
	\vspace{-3mm}	
	\caption{The schematic diagram to illustrate the localization problem under the HFNF BSE through the inner product of the real channel (only LoS path) and HFNF steering vectors.}
	\label{fig_beamsquint}
\end{figure*}

The analysis of the near-field BSE is much more complex than that of the far-field one, since the near-field BSE involves frequency, angle, and distance ($f_m, \theta_m, r_m$), while the far-field one only involves frequency and angle ($f_m, \theta_m$).
\textcolor{black}{
The relation between $f_m$ and $\theta_m$ can be accurately characterized by $\Xi_m=\frac{f_m}{f_c}\theta$ in the far-field case, whereas the relation among $f_m, \theta_m, r_m$ in the near-field case are difficult to characterize.
To maintain the consistence between the far-field steering vector and the near-field steering vector, we will adopt $\mathbf{a}[m](f_m,\theta)$ instead of $\mathbf{a}[m](\Xi_m)$ in future expressions.}
We introduce the cross-correlation of the near-field steering vector $C=|{{\mathbf{b}}}[m_1]({f_{m_1}},{\theta_{m_1}},{r_{m_1}})^H {{\mathbf{b}}[m_2]}({f_{m_2}},{\theta_{m_2}},{r_{m_2}})|$ between ${\bf{b}}[m_1]$ and ${\bf{b}}[m_2]$ taken from (\ref{near_field_steervector}) to analyze the near-field BSE.
It is difficult to obtain the closed-form solution of the near-field beam squint as concise as the far-field one in (\ref{far_field_beam_squint}).
Although some existing literature have worked towards the closed-form solution of the near-field BSE \cite{ref_GFF_loc}, some approximations are made in them for the sake of simple analysis.
Therefore, we focus our attention on qualitatively understanding how the near-field BSE affects localization and how to eliminate this effect.
Specifically, we plot simulated Fig. \ref{fig_beamsquint} to get some enlightening conclusions, which gives an illustration of the localization problem under the HFNF BSE.
In order to illustrate the BSE better, we set
$N=512$,
$f_c=0.1$ THz,
$\hbar=0.5$.
Therefore, according to (\ref{effective_Rayleigh_distance}) and (\ref{effective_Rayleigh_distance_define}), we can obtain $\epsilon=0.16$ and the effective Rayleigh distance is about 46 m.
In the far-field region, the real location of the UE is set as $\theta_{m_1}=0.5$ and $r_{m_1}=50$ m, where $f_{m_1}$ is fixed to $f_c=0.1$ THz.
If $f_{m_2}=f_c$, the direction of the beam will be precisely at $\theta_{m_2}=0.5$.
However, if $f_{m_2}$ takes other values, such as $f_{m_2}=f_{\text{min}}=95$ GHz or $f_{m_2}=f_{\text{max}}=105$ GHz, the beam direction will squint, as shown in Fig. \ref{fig_BS_far_side}.
Since there is no obvious beam-focusing phenomenon when $r_{m_1}=50$ m, it is more reasonable to use the effective Rayleigh distance as the boundary between the far-field region and the near-field region, where the classical Rayleigh distance is 400 m, which is far larger than the effective Rayleigh distance.
Meanwhile, as can be seen in Fig. \ref{fig_BS_far_side}, the farther the distance from the UE to the BS (or RIS), the larger the location error.
In the near-field region, the real location of the UE is set as $\theta_{m_1}=0.5$ and $r_{m_1}=16.7$ m, where $f_{m_1}$ is fixed to $f_c=0.1$ THz.
If $f_{m_2}=f_c$, the beam-focusing region will be located at $\theta_{m_2}=0.5$ and $r_{m_2}=13.99$ m.
The focusing point is $r_{m_2}=13.99$ m rather than $r_{m_2}=16.7$ m since large-scale fading cannot be overlooked in reality.
If $f_{m_2}$ takes other values, such as $f_{m_2}=f_{\text{min}}=95$ GHz or $f_{m_2}=f_{\text{max}}=105$ GHz, the beam-focusing region will also squint, as shown in Fig. \ref{fig_BS_hybrid_side}.
The HFNF BSE makes the localization more difficult, and thus, two schemes are proposed to overcome this problem in the following two sections.

\section{Proposed Joint Channel and Location Sensing Scheme}\label{sec:WDBAL}	
\vspace{-2mm}
\subsection{Problem Formulation}
The proposed scheme aims to jointly estimate the channels between the BS (or RIS) and the UE  ${\mathbf{h}}^{{\text{BU}}}$ (or ${\mathbf{h}}^{{\text{RU}}}$) and sense the UE's location.

Here we consider the UE transmits pilot signals to facilitate the joint channel and location sensing at the BS, and the channel is assumed to remain unchanged during the joint channel and location sensing stage.
Without the assistance of the RIS, the received uplink pilot on the $m$-th subcarrier in the $p$-th time slot, denoted by ${{\mathbf{y}}^{\text{NRIS}}}[p,m] \in {\mathbb{C}^{{N_{\text{RF}}}}}$, can be expressed as
\begin{equation}
	\label{NRIS_rcv_pilot_unit}
	{{\mathbf{y}}^{\text{NRIS}}[p,m]} = {{\mathbf{W}}^{\text{NRIS}}[p]}{{\mathbf{h}}^{\text{BU}}[m]}{x[p,m]} + {{\mathbf{n}}^{\text{NRIS}}[p,m]},
\end{equation}
where ${x[p,m]}$, ${\mathbf{h}}^{\text{BU}}[m] \in {\mathbb{C}^{N}}$, and ${{\mathbf{W}}^{\text{NRIS}}[p]} \in {\mathbb{C}^{{N_{\text{RF}}} \times N}}$ denote the pilot transmitted by the UE, the HFNF channel between the BS and the UE, and the combiner of the BS, respectively.
Each element of ${{\mathbf{W}}^{\text{NRIS}}[p]}$ satisfies the constant modulus constraint $|{{\mathbf{W}}^{\text{NRIS}}_{i,j}}[p]|=\frac{1}{\sqrt{N_{\text{RF}}}}$.
${{\mathbf{n}}^{\text{NRIS}}[p,m]} = {{\mathbf{W}}^{\text{NRIS}}[p]}{\bar {\mathbf{n}}^{\text{NRIS}}[p,m]}$, and ${\bar {\mathbf{n}}^{\text{NRIS}}[p,m]} \in {\mathbb{C}^{N}}$ denotes the Gaussian complex noise, which follows $\mathcal{C}\mathcal{N}(\mathbf{0}_{N}, {\sigma ^2}{{\mathbf{I}}_{N}})$.
In (\ref{NRIS_rcv_pilot_unit}), the RIS is turned off and ${{\mathbf{y}}^{\text{NRIS}}[p,m]}$ is assumed to be not affected by the RIS.
Similarly, the received pilot with the assistance of the RIS on the $m$-th subcarrier in the $p$-th time slot, denoted by ${{\mathbf{y}}^{\text{RIS}}}[p,m] \in {\mathbb{C}^{{N_{\text{RF}}}}}$, can be expressed as
\begin{equation}
	\begin{aligned}
		\label{RIS_rcv_pilot_unit}
		{{\mathbf{y}}^{\text{RIS}}[p,m]} & = {{\mathbf{W}}^{\text{RIS}}[p]}{{\mathbf{H}}^{\text{BR}}[m]}{{\mathbf{\Phi }}^{\text{RIS}}[p]}{{\mathbf{h}}^{\text{RU}}[m]}{x[p,m]}\\
		& + {\mathbf{W}}^{{\text{RIS}}}[p]{\mathbf{h}}^{{\text{BU}}}[m]{x[p,m]} + {{\mathbf{n}}^{\text{RIS}}[p,m]},
	\end{aligned}
\end{equation}
where ${{\mathbf{\Phi }}^{\text{RIS}}[p]} \in {\mathbb{C}^{{N_{\text{RIS}}} \times {N_{\text{RIS}}}}}$, ${{\mathbf{H}}^{\text{BR}}[m]} \in {\mathbb{C}^{N \times {N_{\text{RIS}}}}}$, and ${{\mathbf{h}}^{\text{RU}}[m]} \in {\mathbb{C}^{{N_{\text{RIS}}}}}$ denote the phase of the RIS, the HFNF channel between the BS and the RIS, and the HFNF channel between the RIS and the UE, respectively.
${\mathbf{\Phi }}^{{\text{RIS}}}[p] \triangleq \text{diag}[{e^{j{\iota _1}[p]}}, \cdots , e^{j{\iota _{n}}[p]},\cdots ,{e^{j{\iota _{{N_{{\text{RIS}}}}}}[p]}}]$ is a diagonal matrix accounting for the reflection phases of the RIS, where $e^{j{\iota _{n}}[p]}$ is the phase of the $n$-th RIS element in the $p$-the time slot and it has a precision of 1 bit\footnote{
\textcolor{black}{
RIS reflections in this paper take random values from -1 and 1 such that signals of approximately equal power can be received at all subcarriers without knowing the UE location.
The optimal RIS reflection design in the HFNF BSE scenario still needs further investigation, and this is a promising research direction, whose essence can be formulated as how to jointly design frequency-independent RIS reflections with finite quantization based on unknown angles and distances under the HFNF BSE to achieve better received power at each subcarrier.
}}
In (\ref{RIS_rcv_pilot_unit}), ${{\mathbf{y}}^{\text{RIS}}[p,m]}$, ${{\mathbf{W}}^{\text{RIS}}[p]}$, ${x[p,m]}$, ${\mathbf{h}}^{\text{BU}}[m]$ and ${{\mathbf{n}}^{\text{RIS}}[p,m]}$ are similar to what we defined before in (\ref{NRIS_rcv_pilot_unit}).
Since the RIS is deployed in advance to ensure a LoS link between the RIS and the BS,  ${{\mathbf{H}}^{\text{BR}}[m]}$ can be assumed to be known.
\textcolor{black}{
Moreover, ${{\mathbf{H}}^{\text{BR}}[m]}$ can be assumed to have only one LoS link, since the energy of the multi-hop path is drastically reduced in the THz band.
The phase of each element in ${{\mathbf{H}}^{\text{BR}}[m]}$ can be obtained from the distance between each element of the RIS and each element of the BS antenna array, and the amplitude of each element in ${{\mathbf{H}}^{\text{BR}}[m]}$can be calculated from the Friis formula.
${{\mathbf{W}}^{\text{RIS}}[p]}$ can be designed to minimize the effect of incoming signals from directions other than the one from the RIS to the BS, and to estimate the information from the UE to the RIS since spatial filtering can weaken the signal strength from other directions.} 
Meanwhile, $\mathbf{h}^{\text{BU}}[m]$ can also be weakened since ${{\mathbf{W}}^{\text{RIS}}[p]}$ is not aligned along the directions contained in $\mathbf{h}^{\text{BU}}[m]$.
We assume ${x[p,m]}=1$, and stack the received $P^{\text{NRIS}}$ pilots without the assistance of RIS as follows \textcolor{black}{${\mathbf{Y}}^{{\text{NRIS}}}[m]={[{({\mathbf{y}}^{{\text{NRIS}}}[1,m])^T},\cdots,{({\mathbf{y}}^{{\text{NRIS}}}[{P^{{\text{NRIS}}}},m])^T}]^T}\in \mathbb{C}^{P^{\text{NRIS}}N_{\text{RF}}},\text{ }{{{\mathbf{\bar W}}}^{{\text{NRIS}}}}={[{({\mathbf{W}}^{{\text{NRIS}}}[1])^T},\cdots,{({\mathbf{W}}^{{\text{NRIS}}}[P^{{\text{NRIS}}}])^T}]^T}$
$\in \mathbb{C}^{P^{\text{NRIS}}N_{\text{RF}}\times N}$ and ${\mathbf{N}}^{{\text{NRIS}}}[m]={[({\mathbf{n}}^{{\text{NRIS}}}[1,m])^T,\cdots,}$
${{({\mathbf{n}}^{{\text{NRIS}}}[P^{{\text{NRIS}}},m])^T}]^T}\in \mathbb{C}^{P^{\text{NRIS}}N_{\text{RF}}}$}.
Therefore, we can obtain 
\begin{equation}
	\label{NRIS_rcv_pilot}
	{\mathbf{Y}}^{{\text{NRIS}}}[m] = {{\mathbf{\bar W}}^{{\text{NRIS}}}}  {\mathbf{h}}^{{\text{BU}}}[m] + {\mathbf{N}}^{{\text{NRIS}}}[m].
\end{equation}
Similarly, we stack the received $P^{\text{RIS}}$ pilots with the assistance of RIS and have
\begin{equation}
	\label{RIS_rcv_pilot}
	{\mathbf{Y}}^{{\text{RIS}}}[m] = {\mathbf{\bar W}}^{{\text{RIS}}}[m]  {\mathbf{h}}^{{\text{RU}}}[m] + {\mathbf{\check W}}^{{\text{RIS}}}  {\mathbf{h}}^{{\text{BU}}}[m] + {\mathbf{N}}^{{\text{RIS}}}[m],
\end{equation}
where
\textcolor{black}{${\mathbf{Y}}^{{\text{RIS}}}[m]={[{({\mathbf{y}}^{{\text{RIS}}}[1,m])^T},\cdots,{({\mathbf{y}}^{{\text{RIS}}}[{P^{{\text{RIS}}}},m])^T}]^T}\in \mathbb{C}^{P^{\text{RIS}}N_{\text{RF}}}$, 
${\mathbf{\bar W}}^{{\text{RIS}}}[m]={[{({\mathbf{W}}^{{\text{RIS}}}[1]  {\mathbf{H}}^{{\text{BR}}}[m]  {\mathbf{\Phi }}^{{\text{RIS}}}[1])^T}{\text{ }}  ,\cdots,}$
${{({\mathbf{W}}^{{\text{RIS}}}}}[P^{{\text{RIS}}}]{\mathbf{H}}^{{\text{BR}}}[m]{{{\mathbf{\Phi}}^{{\text{RIS}}}[P^{{\text{RIS}}}])^T}]^T}\in \mathbb{C}^{P^{\text{RIS}}N_{\text{RF}}\times N_{\text{RIS}}}$, ${{{\mathbf{\check W}}}^{{\text{RIS}}}}=$
${[{({\mathbf{W}}^{{\text{RIS}}}[1])^T},}$
${\cdots,{({\mathbf{W}}^{{\text{RIS}}}[P^{{\text{RIS}}}])^T}]^T}\in \mathbb{C}^{P^{\text{RIS}}N_{\text{RF}}\times N}$, and ${\mathbf{N}}^{{\text{RIS}}}[m]={[({\mathbf{n}}^{{\text{RIS}}}[1,m])^T,\cdots,}$
${{({\mathbf{n}}^{{\text{RIS}}}[P^{{\text{RIS}}},m])^T}]^T}\in \mathbb{C}^{P^{\text{RIS}}N_{\text{RF}}}$}.

Since there is no prior information about the UE's location, with the RIS turned off, the ${{\mathbf{W}}_{i,j}^{{\text{NRIS}}}[p]}$ for all $i, j, p$ needs to be set to be omnidirectional to receive signals in all directions, i.e., 
\begin{equation}
	\label{eq_W_NRIS}
	{{\mathbf{W}}_{i,j}^{{\text{NRIS}}}[p]} = e^{j2\pi z_{i,j,p}}/\sqrt{N_{\text{RF}}},
\end{equation}
where $z_{i,j,p}$ $\forall i,j,p$ follows $\mathcal{U}(0, 1)$.
However, in order to better estimate the AoA from the UE to the BS, we will carefully design some values of ${{\mathbf{W}}_{i,j}^{{\text{NRIS}}}[p]}$ later in Section \ref{subsubsec_PGD}.
Similarly, when the RIS is turned on, the phase of the RIS also needs to be set to be omnidirectional to receive signals from all directions, i.e., for all $p$, the probability that the diagonal elements of ${{\mathbf{\Phi }}^{\text{RIS}}[p]}$ take values 1 and -1 is both 0.5.
Meanwhile, ${{\mathbf{W}}^{\text{RIS}}[p]}$ for all $p$ needs to be set to be directed towards the RIS so as to maximize the signal energy from the RIS direction and reduce the signal energy from other directions to the BS.
However, since ${\mathbf{H}}^{{\text{BR}}}[m]$ is frequency selective while ${{\mathbf{W}}^{\text{RIS}}[p]}$ is frequency invariant, if ${{\mathbf{W}}^{\text{RIS}}[p]}$ is designed based on the central frequency using (\ref{near_field_steervector}) as follows,
\begin{equation}
	\label{eq_W_centralFreq}
	{\bf{W}}_{i,:}^{\text{RIS}}[p]=
	(\ {\bf{b}}[\frac{M}{2}+1](f_c, \text{sin}(\frac{\pi}{2}-\vartheta^{\text{B}}), r^{\text{B2R}})\ )^H \frac{\sqrt{N}}{\sqrt{N_{\text{RF}}}},\forall i,p,
\end{equation}
where $\frac{\pi}{2}-\vartheta^{\text{B}}$ denotes the real AoA from the RIS to the BS and it can be seen clearly in Fig. \ref{fig_hyperbola}
, and $r^{\text{B2R}}$ denotes the known distance between the BS and the RIS
, the energy in other subcarriers will be weakened due to the large bandwidth in the THz systems.
To solve this problem, we design ${{\mathbf{W}}^{\text{RIS}}[p]}$ using (\ref{near_field_steervector}) as follows,
\begin{small}
\begin{equation}
	\label{eq_W_allFreq}
	{\bf{W}}_{i,:}^{\text{RIS}}[p]=
	(\ {\bf{b}}[\bar m(i,p)](f_{\bar m(i,p)}, \text{sin}(\frac{\pi}{2}-\vartheta^{\text{B}}), r^{\text{B2R}})\ )^H \frac{\sqrt{N}}{\sqrt{N_{\text{RF}}}},\forall i,p,
\end{equation}
\end{small}
where $f_{\bar m(i,p)} = f_c - B/2 + \frac{B}{N_{\text{RF}}P^{\text{RIS}}}((p-1)N_{\text{RF}}+i)$ and $\bar m(i,p) = \frac{M}{N_{\text{RF}}P^{\text{RIS}}}((p-1)N_{\text{RF}}+i) + 1$.
The meaning of (\ref{eq_W_allFreq}) is that at each RF-chain in each time slot, the BS's analog combiner is varied by using (\ref{near_field_steervector}), where the AoA and the distance use the value from the center of the RIS to the center of the BS, and the frequency is evenly valued throughout the bandwidth.

To better explain our problem and the variables intended to be solved, we formulate the following optimization problem,
\textcolor{black}{
\begin{small}
\begin{align}
	\label{eq_OptPro1}
		& \mathop {\min }\limits_{{\small{{{{\mathbf{\hat h}}}^{{\text{BU}}}},{{{\mathbf{\hat h}}}^{{\text{RU}}}},\hat x^{{\text{UE}}},\hat y^{{\text{UE}}}}}} \Sigma_{m = 1}^M {\Big|\!\Big| {{{\mathbf{Y}}^{{\text{NRIS}}}}[m] - {{{\mathbf{\bar W}}}^{{\text{NRIS}}}}{{{\mathbf{\hat h}}}^{{\text{BU}}}}[m](\hat x^{{\text{UE}}},\hat y^{{\text{UE}}})} \Big|\!\Big|_{\text{F}}^2}   \notag\\
 		&\qquad\qquad\ +\Sigma_{m = 1}^M {\Big|\!\Big| {{{\mathbf{Y}}^{{\text{RIS}}}}[m] - {{{\mathbf{\bar W}}}^{{\text{RIS}}}}[m]{{{\mathbf{\hat h}}}^{{\text{RU}}}}[m](\hat x^{{\text{UE}}},\hat y^{{\text{UE}}}) }} \notag\\
		&\qquad\qquad\qquad\qquad\qquad\qquad\quad{{
		 - {{{\mathbf{\check W}}}^{{\text{RIS}}}}{{{\mathbf{\hat h}}}^{{\text{BU}}}}[m](\hat x^{{\text{UE}}},\hat y^{{\text{UE}}})} \Big|\!\Big|_{\text{F}}^2}   \notag\\
		&\;\;\;\;\;\;{\text{s}}.{\text{t}}.  \;\;{\text{C1:}}\;\left| {{\mathbf{W}}_{i,j}^{{\text{NRIS}}}[p]} \right| = \{ 0,\;1/{\sqrt {N_{\text{RF}}} }\} ,\;\forall i,j,p  \notag\\
		&\qquad\quad\;\,{\text{C2:}}\;\left| {{\mathbf{W}}_{i,j}^{{\text{RIS}}}[p]} \right| = \{ 0,\;1/{\sqrt {{N_{{\text{RF}}}}} }\} ,\;\forall i,j,p  \notag\\
		&\qquad\quad\;\,{\text{C3:}}\;{\mathbf{\Phi }}_{i,i}^{{\text{RIS}}}[p] = \{  - 1,\;1{\text{\} }},\;\forall i,p  \notag\\
		&\qquad\quad\;\,{\text{C}}4{\text{:}}\;{{{\mathbf{\bar W}}}^{{\text{NRIS}}}} = {[{({{\mathbf{W}}^{{\text{NRIS}}}}[1])^T}, \cdots ,{({{\mathbf{W}}^{{\text{NRIS}}}}[{P^{{\text{NRIS}}}}])^T}]^T}  \notag\\
		&\qquad\quad\;\,{\text{C5:}}\;{{{\mathbf{\bar W}}}^{{\text{RIS}}}}[m] = [{({{\mathbf{W}}^{{\text{RIS}}}}[1]{{\mathbf{H}}^{{\text{BR}}}}[m]{{\mathbf{\Phi}}^{{\text{RIS}}}}[1])^T}{\text{ }}, \cdots ,  \notag\\
		&\qquad\qquad\qquad\quad\ {({{\mathbf{W}}^{{\text{RIS}}}}[{P^{{\text{RIS}}}}]{{\mathbf{H}}^{{\text{BR}}}}[m]{{\mathbf{\Phi }}^{{\text{RIS}}}}[{P^{{\text{RIS}}}}])^T}{]^T}, \forall m \notag\\ 
		&\qquad\quad\;\,{\text{C}}6{\text{:}}\;{{{\mathbf{\check W}}}^{{\text{RIS}}}} = {[{({{\mathbf{W}}^{{\text{RIS}}}}[1])^T}, \cdots ,{({{\mathbf{W}}^{{\text{RIS}}}}[{P^{{\text{RIS}}}}])^T}]^T},  \notag\\
\end{align}
\end{small}
}
where $(\hat x^{{\text{UE}}},\hat y^{{\text{UE}}})$ is the estimation of UE's location, ${{{{\mathbf{\hat h}}}^{{\text{BU}}}}}$ and ${{{{\mathbf{\hat h}}}^{{\text{RU}}}}}$, determined by  $(\hat x^{{\text{UE}}},\hat y^{{\text{UE}}})$, are estimations of ${{{{\mathbf{ h}}}^{{\text{BU}}}}}$ and ${{{{\mathbf{ h}}}^{{\text{RU}}}}}$, respectively, constraints C1 and C2 are the constant modulus constraint of the analog combiner at the BS (if some phase shifters are switched off, those will take on a value of zero),
constraint C3 is the limitation on the value of the RIS phase shifter with a precision of 1 bit,
constraints C4, C5 and C6 denote the sensing matrices collected from multiple time slots.
Our goals are to estimate ${{{{\mathbf{\hat h}}}^{{\text{BU}}}}}$ from ${{{\mathbf{Y}}^{{\text{NRIS}}}}}$,
estimate ${{{{\mathbf{\hat h}}}^{{\text{RU}}}}}$ from ${{{\mathbf{Y}}^{{\text{RIS}}}}}$,
and estimate the UE's location (${\hat x^{{\text{UE}}},\hat y^{{\text{UE}}}}$), which can be utilized to improve the CE performance of ${{{{\mathbf{\hat h}}}^{{\text{BU}}}}}$ and ${{{{\mathbf{\hat h}}}^{{\text{RU}}}}}$, jointly from ${{{\mathbf{Y}}^{{\text{NRIS}}}}}$ as well as ${{{\mathbf{Y}}^{{\text{RIS}}}}}$.

\begin{figure}[!t]
	\centering
	\includegraphics[width=3.3in]{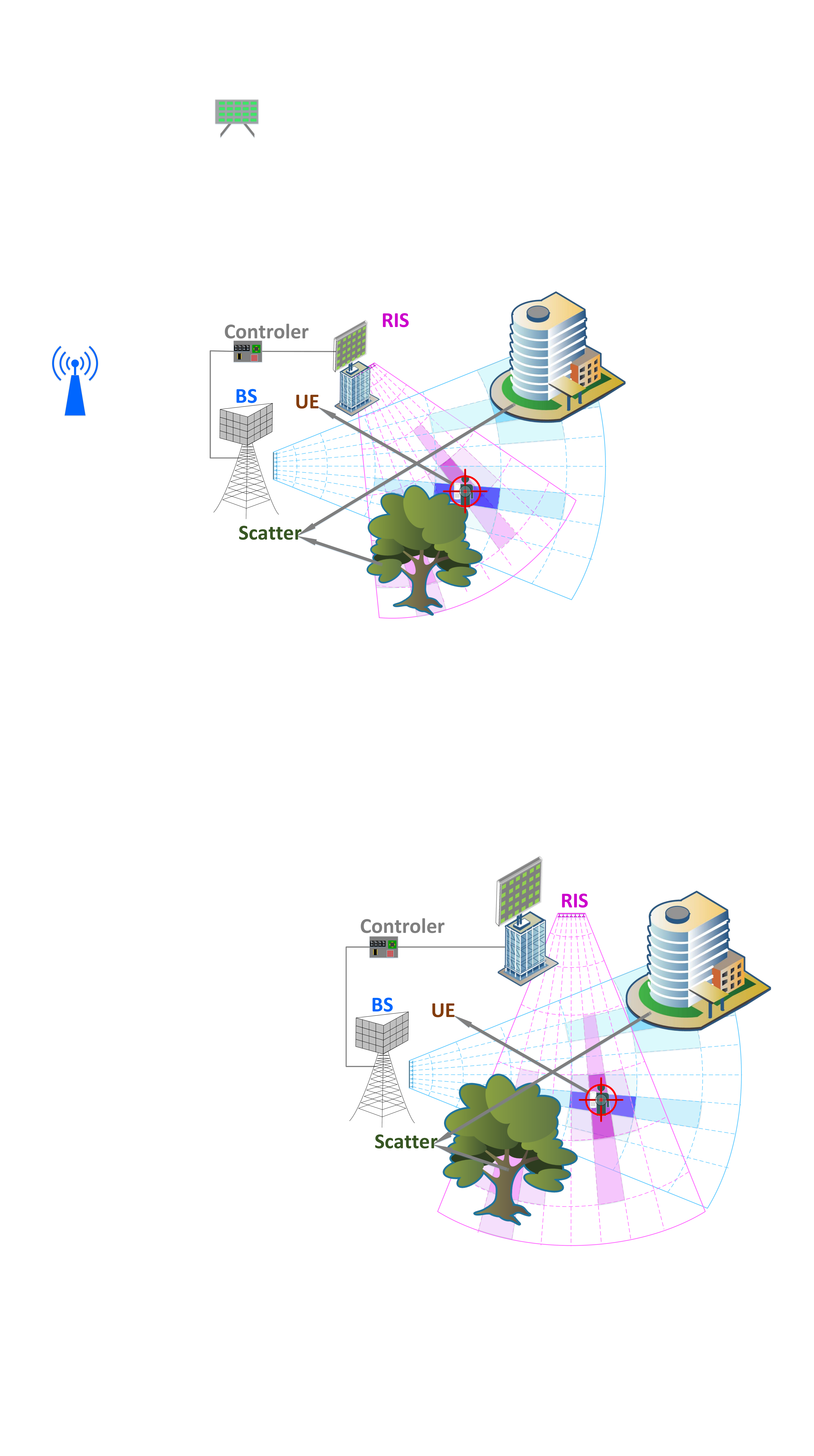}
	\vspace{-3mm}
	\caption{Schematic diagram of the proposed joint channel and location sensing scheme.}
	\label{WDBAL_fig}
\end{figure}
To solve the optimization problem in (\ref{eq_OptPro1}), we propose a dictionary design scheme, a CE algorithm and a localization scheme, which are described in detail in Section \ref{subsec_FSPRD}, \ref{subsec_CE} and \ref{subsec_loc}, respectively.
Among them, the CE module outputs coarse on-grid angle estimation (respectively observed from the BS and RIS) to the localization module, which returns the fine off-grid AoA estimation to improve CE.
In the following equations, if one equation is divided into subequation (a) and subequation (b), (a) stands for sensing without the assistance of the RIS and (b) for sensing with the assistance of the RIS.

\subsection{Frequency Selective Polar-Domain Redundant Dictionary}\label{subsec_FSPRD}
\begin{algorithm}
	\begin{small}
		\label{alg_FreqSelec}
		\caption{Generating Procedure of the Frequency Selective Polar-domain Redundant Dictionary}
		\LinesNumbered 
		\KwIn{the number of array elements of the BS (or RIS) $N$, the number of subcarriers $M$, carrier frequency $f_c$, bandwidth $B$, the number of distance grids $S$, redundancy rate
			$\varsigma$}
		\KwOut{the frequency selective polar-domain redundant dictionary $\mathbf{D} \in \mathbb{C}^{N\times \varsigma N S \times M}$}
		\For{$m = \{ 1, 2,  \cdots, M\}$}{
			$f_m={{f_c} - B/2 + (m - 1)B/M}$\;
			\For{$n = \{ 0, 1,  \cdots, \varsigma N-1\}$}{
				Generate the $n$-th angle grid $\theta_n$ as (\ref{theta_sample})\;
				\For{$s = \{ 0, 1, \cdots, S-1\}$}{
					\eIf{$s=0$}{
						Generate far-field steering vector $\mathbf{a}$ as (\ref{far_field_steervector}) by using $\theta_n$\;
						$\mathbf{D}[m]=[\mathbf{D}[m] \; \mathbf{a}[m]]$\;
					} 
					{
						Generate the $s$-th distance grid as ${r_{s,n}} = 2{Z^{\text{eff}}_{c}}(0)(1 - {\theta_n ^2})/s$\; 
						Generate HFNF steering vector $\mathbf{b}$ as (\ref{near_field_steervector}) by using $f_m, \theta_n$ and $r_{s,n}$\; $\mathbf{D}[m]=[\mathbf{D}[m] \; \mathbf{b}[m]]$\;
					}
				}
			}
		}
	\end{small}
\end{algorithm}
We take ${\bf{h}}^{\text{BU}}$, which is the same as ${\bf{h}}^{\text{RU}}$, as an example.
In far-field region, (\ref{far_field_steervector}) can be used to model the steering vector and the phase of each element in the steering vector is linear to the antenna index. Thus, we can use the Fourier transform to sparse the spatial-domain channel to the angle-domain channel as 
\begin{equation}
	\label{far_field_channel_angle}
	{\mathbf{h}}^{{\text{BU}}}[m] = {\mathbf{F}}  {\mathbf{h}}^{{\text{BU,A}}}[m],
\end{equation}
where $\mathbf{F}$ denotes the FTM and ${\mathbf{h}}^{{\text{BU,A}}}[m]$ denotes the angle-domain channel.
Since the number of paths is limited, ${\mathbf{h}}^{{\text{BU,A}}}[m]$ is sparse and we can recover signals from higher dimensionalities with a small number of pilots through the CS-based methods.

Nevertheless, in the HFNF region, we use (\ref{near_field_steervector}), which is determined not only by the AoA from the UE to the BS, but also by the distance between the UE and the BS, to model the steering vector.
Thus, the FTM $\mathbf{F}$ cannot be utilized to sparse the near-field channel because of the energy spread effect described in \cite{2022_TCOM_LinglongDai_PolarDomain}.
Instead, the new transform matrix ${\mathbf{D}}[m] \in {\mathbb{C}^{N \times \varsigma N S}}$, which is developed from the PTM in \cite{2022_TCOM_LinglongDai_PolarDomain} and called frequency selective polar-domain redundant dictionary (FSPRD), is proposed.
$m$ means the $m$-th subcarrier, $N$ is the number of angle grids as well as the BS (RIS) elements, $S$ is the number of distance grids and $\varsigma \geq 1$ is the redundant rate.
Since ${\bf{D}}[m]$ takes into account the differences in the HFNF steering vectors across different subcarriers, it can overcome the BSE compared to FTM and PTM.
We can obtain relatively good correlation property when $\theta$ is uniformly sampled from $(-1, 1)$ as 
\begin{equation}
	\label{theta_sample}
	{\theta _n} = ({2n - \varsigma N + 1})/(\varsigma N),{\text{ }}n = 0,1,\cdots,\varsigma N - 1,
\end{equation}
and $r$ is sampled as
\begin{equation}
	\label{eq_r_sample}
r_{s,n} = 2{Z^{\text{eff}}_{c}}(0)(1 - {\theta_n ^2})/s,{\text{ }}s = 1,2,\cdots,S - 1,
\end{equation}
where ${Z^{\text{eff}}_{c}}(0)$ is the effective Rayleigh distance when $f_m=f_c$ and $\theta=0$.
$r$ is sampled in the manner of inverse proportional function to make the correlation of elements in the FSPRD smaller.
On account of the effective Rayleigh distance in (\ref{effective_Rayleigh_distance}), $r$ is associated with $\theta$.
Moreover, the coefficient "2" in (\ref{eq_r_sample}) can be utilized to cover UEs within the effective Rayleigh distance better, i.e., to generate more distance grids around the effective Rayleigh distance.
Thus, we can sparse the HFNF channel ${\mathbf{h}}^{{\text{BU}}}[m]$ to the polar-domain channel ${\mathbf{h}}^{{\text{BU,P}}}[m]$ approximately by the FSPRD ${\mathbf{D}[m]}$ as 
\begin{equation}
	\label{near_field_channel_polar}
	{\mathbf{h}}^{{\text{BU}}}[m] = {\mathbf{D}[m]}  {\mathbf{h}^{{\text{BU,P}}}}[m].
\end{equation}
Therefore, considering both BSE and HFNF, we can obtain the generation step of FSPRD as Algorithm \ref{alg_FreqSelec}.
\subsection{Proposed LA-GMMV-OMP Algorithm}\label{subsec_CE}
The FSPRD is generated as $\mathbf{D}^{\text{NRIS}}[m]$ and $\mathbf{D}^{\text{RIS}}[m]$ using Algorithm \ref{alg_FreqSelec}.
Based on the representation of FSPRD in (\ref{near_field_channel_polar}), we can get the equivalent measurement matrix ${{\mathbf{\tilde W}}^{{\text{NRIS}}}}[m] \in {\mathbb{C}^{{N_{{\text{RF}}}}{P^{{\text{NRIS}}}} \times \varsigma NS}}$ and ${{\mathbf{\tilde W}}^{{\text{RIS}}}}[m] \in {\mathbb{C}^{{N_{{\text{RF}}}}{P^{{\text{RIS}}}} \times {\varsigma N_{{\text{RIS}}}}S}}$ as follows
\begin{subequations}
	\label{eq_A_tilde_BS}
	\begin{align}
		{\mathbf{\tilde W}}^{{\text{NRIS}}}[m] =& {{\mathbf{\bar W}}_{}^{{\text{NRIS}}}}{\mathbf{D}}^{{\text{NRIS}}}[m],\label{eq_A_tilde_BS_a}\\
		{\mathbf{\tilde W}}^{{\text{RIS}}}[m] =& {{\mathbf{\bar W}}^{{\text{RIS}}}[m]}{\mathbf{D}}^{{\text{RIS}}}[m].\label{eq_A_tilde_BS_b}
	\end{align}	
\end{subequations}
At the beginning of each iteration, we calculate ${\mathbf{\Gamma }}^{{\text{NRIS}}}[m] \in {\mathbb{C}^{\varsigma NS}}$ (${\mathbf{\Gamma }}^{{\text{RIS}}}[m] \in {\mathbb{C}^{\varsigma N_{\text{RIS}}S}}$), the correlation matrix on the $m$-th subcarrier between ${{\mathbf{\tilde W}}^{{\text{NRIS}}}}[m]$ (${{\mathbf{\tilde W}}^{{\text{RIS}}}}[m]$) and the residual $\mathbf{R}^{\text{NRIS}}[m]$ ($\mathbf{R}^{\text{RIS}}[m]$), as follows
\begin{subequations}
	\label{eq_cor_A_R}
	\begin{align}
		{\mathbf{\Gamma }}^{{\text{NRIS}}}[m] =& |{({\mathbf{\tilde W}}^{{\text{NRIS}}}[m])^H}  {{\mathbf{R}}^{{\text{NRIS}}}[m]}|,\\
		{\mathbf{\Gamma }}^{{\text{RIS}}}[m] =& |{({\mathbf{\tilde W}}^{{\text{RIS}}}}[m])^H  {{\mathbf{R}}^{{\text{RIS}}}[m]}|,
	\end{align}	
\end{subequations}
where $\mathbf{R}^{\text{NRIS}}[m]$ ($\mathbf{R}^{\text{RIS}}[m]$) is ${\mathbf{Y}}^{{\text{NRIS}}}[m]$ (${\mathbf{Y}}^{{\text{RIS}}}[m]$) in the first iteration, and is calculated by (\ref{eq_UpdateResidual}) in the subsequent iterations.
For ${\mathbf{\Gamma }}^{{\text{NRIS}}}[m]$ (${\mathbf{\Gamma }}^{{\text{RIS}}}[m]$),  the positions where peaks appear are determined by $(\theta^{\text{BU}} ,r^{\text{BU}})$ ($(\theta^{\text{RU}} ,r^{\text{RU}})$), where $\theta^{\text{BU}}$, $r^{\text{BU}}$, $\theta^{\text{RU}}$, and $r^{\text{RU}}$ denote the AoAs from the UE to the BS, the distances from the UE to the BS, the AoAs from the UE to the RIS, and the distances from the UE to the RIS, respectively.
Through the correlation matrix in (\ref{eq_cor_A_R}), we can obtain
\begin{subequations}
	\label{eq_FindSupSet}
	\begin{align}	
		{{\mathbf{\Upsilon }}_{i}^{{\text{NRIS}}}} =& \sum\nolimits_{m = 1}^M {|{\mathbf{\Gamma }}_{i}^{{\text{NRIS}}}[m]{|^2}},\label{eq_FindSupSet_NRIS}\\
		{{\mathbf{\Upsilon }}_{i}^{{\text{RIS}}}} =& \sum\nolimits_{m = 1}^M {|{\mathbf{\Gamma }}_{i}^{{\text{RIS}}}[m]{|^2}}, \label{eq_FindSupSet_RIS}
	\end{align}	
\end{subequations}
where ${\mathbf{\Upsilon }}^{{\text{NRIS}}} \in {\mathbb{C}^{\varsigma NS}}$ and ${{\mathbf{\Upsilon }}^{{\text{RIS}}}} \in {\mathbb{C}^{\varsigma N_{\text{RIS}}S}}$.
Since we adopted the FSPRD, the peaks of ${\mathbf{\Gamma }}^{\text{NRIS}}$ and ${\mathbf{\Gamma }}^{\text{RIS}}$ on different subcarriers have the same AoA and distance indices, so we can take advantage of this MMV property to improve the robustness of finding $(\theta^{\text{BU}} ,r^{\text{BU}})$ and $(\theta^{\text{RU}} ,r^{\text{RU}})$.
For the first iteration, since there is a single path of the LoS channel, we only pick out the largest elements from ${\mathbf{\Upsilon }}^{{\text{NRIS}}}$ and ${{\mathbf{\Upsilon }}^{{\text{RIS}}}}$ as $\gamma^{\text{NRIS}}$ and $\gamma^{\text{RIS}}$, which correspond to the rough AoAs and distances of the LoS paths (respectively from the UE to the BS and from the UE to the RIS).
These coarse AoAs and distances can provide good initial values for further localization in Section \ref{subsec_loc}, which returns the fine AoA estimation to improve CE.
For subsequent iterations, we pick out the $N_s^{\text{NRIS}}$ and $N_s^{\text{RIS}}$ largest elements from ${\mathbf{\Upsilon }}^{{\text{NRIS}}}$ and ${{\mathbf{\Upsilon }}^{{\text{RIS}}}}$ as
\begin{subequations}
	\label{eq_UpdateSupSet}
	\begin{align}
		\gamma^{\text{NRIS}}  = &\{ {\gamma _1^{\text{NRIS}}},{\gamma _2^{\text{NRIS}}},\cdots,{\gamma^{\text{NRIS}} _{{N_{\text{s}}^{\text{NRIS}}}}}\}, 	\label{eq_UpdateSupSet_a}	\\
		\gamma^{\text{RIS}}  = &\{ {\gamma _1^{\text{RIS}}},{\gamma _2^{\text{RIS}}},\cdots,{\gamma^{\text{RIS}} _{{N_{\text{s}}^{\text{RIS}}}}}\}, \label{eq_UpdateSupSet_b}
	\end{align}
\end{subequations}
respectively.
Then, the support sets $\Omega^{\text{NRIS}}$ and $\Omega^{\text{RIS}}$, which are $\emptyset$ at the beginning, can be updated as $\Omega^{\text{NRIS}} = \Omega^{\text{NRIS}} \cup \gamma^{\text{NRIS}}$ and $\Omega^{\text{RIS}} = \Omega^{\text{RIS}} \cup \gamma^{\text{RIS}}$.
\textcolor{black}{
We have the following consideration for adding multiple atoms to the support set in estimating the channel of NLoS paths in each iteration.
Since the distance is taken into account in the FSPRD, the dimensionality of the FSPRD is so large that the true path has strong correlations with several FSPRD elements.
Additionally, the size of the scatterers can not be negligible in the HFNF region, so the channel is a cluster-sparse multi-path channel where many paths are contained in one cluster.
Therefore, one scatterer corresponds to multiple dictionary atoms with similar angles and distances.
If multiple atoms with large energies are selected in each iteration, we can relatively correctly select the atoms corresponding to the scatterer with the largest energy in the current iteration.
However, if only one atom with the largest energy is selected each time, the energy of the other atoms with the same angle and distance will be weakened in future iterations, since the energy of these paths has already been weakened from the residual after each iteration of the OMP-based algorithm.
These atoms with reduced energy are difficult to pick out correctly in the future iterations and hence the CE performance degrades, as verified by simulations.}
After updating the support set, for every subcarrier, the orthogonal projection \textcolor{black}{${\mathbf{\Phi}}^{\text{NRIS}}[m] \in \mathbb{C}^{|\Omega^{\text{NRIS}}|_c \times N}$ and ${\mathbf{\Phi}}^{\text{RIS}}[m] \in \mathbb{C}^{|\Omega^{\text{RIS}}|_c \times N_{\text{RIS}}}$} can be calculated as
\begin{subequations}
	\label{eq_PseudoInverse}	
	\begin{align}
		{\mathbf{\Phi}}^{\text{NRIS}}[m] = &({{\mathbf{\tilde W}}_{:,\Omega^{\text{NRIS}} }^{\text{NRIS}}}[m])^\dag {{\mathbf{Y}}^{\text{NRIS}}}[m],\\
		{\mathbf{\Phi}}^{\text{RIS}}[m] = &({{\mathbf{\tilde W}}_{:,\Omega^{\text{RIS}} }^{\text{RIS}}}[m])^\dag {{\mathbf{Y}}^{\text{RIS}}}[m].
	\end{align}	
\end{subequations}
At the end of each iteration, the residual should be updated as follows
\begin{subequations}
	\label{eq_UpdateResidual}	
	\begin{align}
		{\mathbf{R}}^{{\text{NRIS}}}[m] =& {\mathbf{Y}}^{{\text{NRIS}}}[m] - {\mathbf{\tilde W}}_{:,{\Omega ^{{\text{NRIS}}}}}^{{\text{NRIS}}}[m]  {\mathbf{\Phi }}^{{\text{NRIS}}}[m],\\
		{\mathbf{R}}^{{\text{RIS}}}[m] = &{\mathbf{Y}}^{{\text{RIS}}}[m] - {\mathbf{\tilde W}}_{:,{\Omega ^{{\text{RIS}}}}}^{{\text{RIS}}}[m]  {\mathbf{\Phi }}^{{\text{RIS}}}[m].
	\end{align}	
\end{subequations}
After (\ref{eq_cor_A_R}) to (\ref{eq_UpdateResidual}) are iterated multiple times and the stop criterion is reached, we can obtain the final channel as follows
\begin{subequations}
	\label{eq_EstCha}	
	\begin{align}
		{\mathbf{\hat h}}^{{\text{BU}}}[m] =& {\mathbf{D}}_{:,\Omega^{\text{NRIS}} }^{{\text{NRIS}}}[m]{\mathbf{\Phi }}^{{\text{NRIS}}}[m],\\
		{\mathbf{\hat h}}^{{\text{RU}}}[m] =& {\mathbf{D}}_{:,\Omega^{\text{RIS}} }^{{\text{RIS}}}[m]{\mathbf{\Phi }}^{{\text{RIS}}}[m],
	\end{align}	
\end{subequations}

The LA-GMMV-OMP algorithm is summarized in Algorithm \ref{alg_PgOMP-MMV_new}, where the LA-GMMV-OMP algorithm degenerates into the GMMV-OMP algorithm if steps \ref{step_OMP_2}-\ref{step_OMP_5} are not performed.
For brevity, some subscripts and superscripts for partial variables are omitted.

\subsection{Proposed Complete Dictionary based Localization (CDL) Scheme}\label{subsec_loc}
The CDL scheme is summarized in Algorithm \ref{CDL}, where the overall idea is to use PGD algorithm to refine $\hat \theta^{\text{BU}}_0$ as shown in steps \ref{alg_PGD_begin}-\ref{alg_PGD_end}, and PHD to refine $\hat \theta^{\text{RU}}_0$ as shown in steps \ref{alg_PHD_begin}-\ref{alg_PHD_end}.
Finally, the UE can be located by line intersection.
\begin{algorithm}[htbp]
	\begin{small}
		\label{alg_PgOMP-MMV_new}
		\caption{Proposed LA-GMMV-OMP Algorithm}
		\LinesNumbered 
		\KwIn{received pilot ${\mathbf{Y}}$, equivalent combining matrix ${\mathbf{\bar W}}$, threshold to terminate $\varpi_{\text{OMP}}$, the maximum number of iterations in the LA-GMMV-OMP algorithm $L_\text{max}$}
		\KwOut{estimated channel $\mathbf{\hat h}$}
		Initialization ${{\mathbf{R}}} = {{\mathbf{R}}}_0 = {\mathbf{Y}}$, $\Omega  = \{ \emptyset \} $\; \label{step_OMP_1}
		Generate the FSPRD $\mathbf{W}$ as Algorithm \ref{alg_FreqSelec}\;
		Calculate $\mathbf{\tilde W}$ using ${\mathbf{\bar W}}$ and $\mathbf{D}$ as (\ref{eq_A_tilde_BS})\;
		\For{$i = \{ 1, 2, \cdots, L_\text{max}\} $}{
			\For{$m = \{ 1, 2, \cdots, M\} $}{
				Calculate the correlation matrix ${\mathbf{\Gamma}}[m]$ as (\ref{eq_cor_A_R})\;
			}
			\If{$i=1$}{\label{step_OMP_2}
				Obtain coarse AoAs $\hat \theta_0^{\text{BU}}$, $\hat \theta_0^{\text{RU}}$ as (\ref{eq_OMP_co_theta})\;\label{step_OMP_3}
				Obtain fine estimations of AoA and distance ($\hat \theta_0^{\text{BU}}$,$\hat r_0^{\text{BU}}$), ($\hat \theta_0^{\text{RU}}$,$\hat r_0^{\text{RU}}$) by the CDL scheme\;\label{step_OMP_4}
				Update the FSPRD used in the step 3 as (\ref{eq_appendDic}) and calculate the new ${\mathbf{\Gamma}}[m]$ as (\ref{eq_cor_A_R})\;
			}\label{step_OMP_5}			
			Find out new support set, $\gamma$, as (\ref{eq_FindSupSet}) and (\ref{eq_UpdateSupSet})\;
			Update the support set $\Omega = \Omega \cup \gamma$ \;
			
			\For{$m = \{ 1, 2, \cdots, M\} $}{
				Calculate the orthogonal projection as (\ref{eq_PseudoInverse})\;
				Update the residual $\textbf{R}[m]$ as (\ref{eq_UpdateResidual})\;
			}
			\textbf{if} $||\bf{R}||_{\text{F}}$/$||\bf{R}_0||_{\text{F}}$ $>$ $\varpi_{\text{OMP}}$, \textbf{break}\;
			$\bf{R}_0$ = $\bf{R}$\;
		}
		Acquire the estimated channel $\mathbf{\hat h}$ as (\ref{eq_EstCha})\;
	\end{small}
\end{algorithm}

From the first iteration of CE process in the LA-GMMV-OMP algorithm, the coarse location of the UE can be acquired.
Since the AoA and distance corresponding to each element in FSPRD is known to be generated by Algorithm \ref{alg_FreqSelec} (especially (\ref{theta_sample})), we can deduce the corresponding coarse AoA estimations of the LoS paths from the UE to the BS and the RIS as
\begin{subequations}
	\label{eq_OMP_co_theta}	
	\begin{align}
		\hat \theta_0^{{\text{BU}}} = & ({{2\left\lceil {{i^{{\text{NRIS}}}}/S} \right\rceil  - 1}})/({{\varsigma N}}) - 1,\\
		\hat \theta_0^{{\text{RU}}} = &({2\left\lceil {{i^{{\text{RIS}}}}/S} \right\rceil  - 1})/({{\varsigma N_{\text{RIS}}}}) - 1,\label{eq_theta_RU_coarse}
	\end{align}	
\end{subequations}
respectively, where $i^{\text{NRIS}} = \mathop {\arg \max\limits_{i}}{{\mathbf{\Upsilon }}_{i}^{{\text{NRIS}}}}$ and $i^{\text{RIS}} = \mathop {\arg \max\limits_{i}}{{\mathbf{\Upsilon }}_{i}^{{\text{RIS}}}}$.
Although ${{\mathbf{\Upsilon }}^{{\text{NRIS}}}}$ is not only related to the AoA from the UE to the BS, but also related to the distance from the UE to the BS, the distance is too imprecise to locate the UE since the distance from the UE to the BS is sampled in the manner of inverse proportional function.
It is the same with the ${{\mathbf{\Upsilon }}^{{\text{RIS}}}}$.
{\color{black}
Additionally, since the UE is assumed to be equipped with one omni-directional antenna, there is no AoD information in the uplink stage and the UE's location cannot be estimated by NLoS paths.
Meanwhile, in THz systems, the energy of NLoS paths is too small compared to that of the LoS path to extract the UE's location, and the area of the scatterer can not be negligible, which further increases the difficulty of high-precision localization.
Further, although the UE can be equipped with multiple antennas, the estimation of the UE's location through the AoDs of NLoS paths is also limited by the number of antennas at the UE side, where the AoD resolution is poor if the number of antennas is small.}
Therefore, only $\hat \theta_0^{{\text{BU}}}$ and $\hat \theta_0^{{\text{RU}}}$ can be sought to locate the UE relatively accurately by line intersection, where the anchors are the BS and RIS.
By combining the $\hat \theta_0^{\text{BU}}$ and $\hat \theta_0^{\text{RU}}$ with the locations of the BS and the RIS, denoted as (${x^{{\text{BS}}}}$, ${y^{{\text{BS}}}}$) and (${x^{{\text{RIS}}}}$, ${y^{{\text{RIS}}}}$), the location of the UE can be calculated as
\begin{align}
	\label{eq_OMP_coarse_UE_x_y}
		\hat x^{{\text{UE}}} & = ({y^{{\text{RIS}}}} - {y^{{\text{BS}}}} + {k^{{\text{BU}}}}{x^{{\text{BS}}}} - {k^{{\text{RU}}}}{x^{{\text{RIS}}}})/({{{k^{{\text{BU}}}} - {k^{{\text{RU}}}}}}) \notag \\ 
		\hat y^{{\text{UE}}} & = ({{k^{{\text{BU}}}}{y^{{\text{RIS}}}} - {k^{{\text{RU}}}}{y^{{\text{BS}}}} + {k^{{\text{BU}}}}{k^{{\text{RU}}}}({x^{{\text{BS}}}} - {x^{{\text{RIS}}}})})/({{{k^{{\text{BU}}}} - {k^{{\text{RU}}}}}}), \notag \\
\end{align}
where ${k^{\text{BU}}=-\tan (\frac{\pi}{2}-\vartheta^{\text{B}}-\arcsin (\hat \theta_0^{{\text{BU}}}))}\text{ and }{k^{\text{RU}}=\tan(}$
${\frac{\pi}{2}-\vartheta^{\text{R}}-\arcsin (\hat \theta_0^{{\text{RU}}}))}$ are the slope of the line from the BS to the UE and the slope of the line from the RIS to the UE, respectively, as Fig. \ref{fig_hyperbola} shows.
The BS and the RIS are set on the $x$-axis, so $y_{\text{BS}}=y_{\text{RIS}}=0$.
Then, we can get the distance through
\begin{subequations}
	\label{eq_OMP_coarse_UE_r}	
	\begin{align}
		\hat r_0^{{\text{BU}}} =& \sqrt {{{(\hat x^{{\text{UE}}} - x^{{\text{BS}}})}^2} + {{(\hat y^{{\text{UE}}} - y^{{\text{BS}}})}^2}},\\
		\hat r_0^{{\text{RU}}} =& \sqrt {{{(\hat x^{{\text{UE}}} - x^{{\text{RIS}}})}^2} + {{(\hat y^{{\text{UE}}} - y^{{\text{RIS}}})}^2}}.
	\end{align}	
\end{subequations}

However, the estimated  ($\hat \theta_0^{{\text{BU}}}$, $\hat \theta_0^{{\text{RU}}}$) and ($\hat r_0^{{\text{BU}}}$, $\hat r_0^{{\text{RU}}}$) are limited by the number of FSPRD grids, leading to limited localization accuracy.
Previous off-grid methods are in far-field region without the BSE as \cite{offgrid_2018_TVT}\cite{offgrid_2021_TWC} or in near-field region without the BSE as \cite{2022_TCOM_LinglongDai_PolarDomain}, which cannot be directly applied to the HFNF with severe BSE in this paper.
Therefore, an off-grid method is proposed to solve this challenging localization problem.
\subsubsection{\textbf{Using the polar-domain gradient descent algorithm to refine the AoA from the UE to the BS}}\label{subsubsec_PGD}
$\quad$

\begin{figure}[!t]
	\centering
	\subfigure[The loss function is obtained from (\ref{eq_OMP_GD_absolute_phase}).]{
		\begin{minipage}[t]{0.45\linewidth}
			\centering
			\includegraphics[width=1.6in]{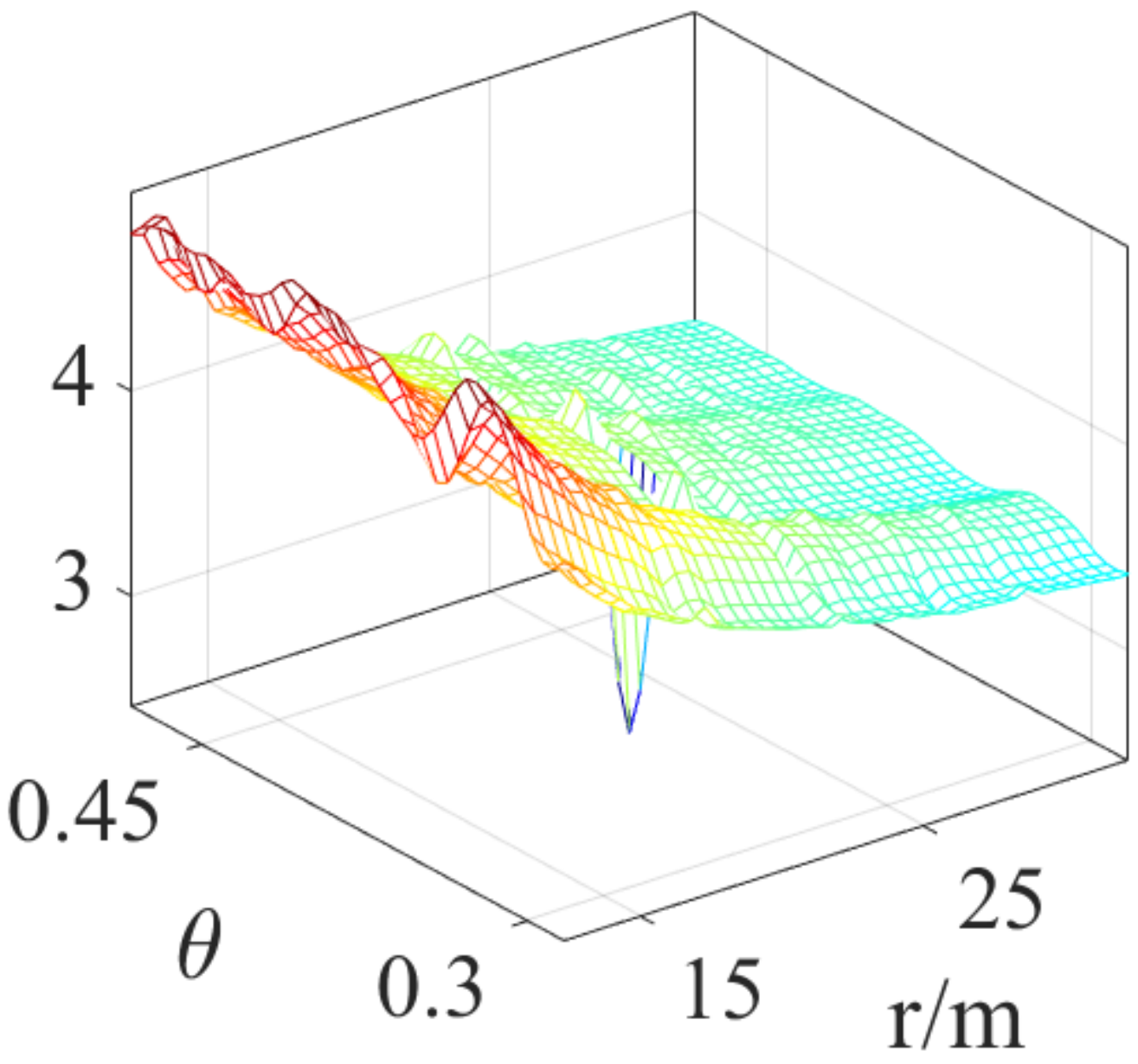}
			\label{abs_phase}
		\end{minipage}%
	}%
	\hspace{4mm}
	\subfigure[The loss function is obtained from (\ref{eq_OMP_GD_relative_phase}).]{
		\begin{minipage}[t]{0.45\linewidth}
			\centering
			\includegraphics[width=1.7in]{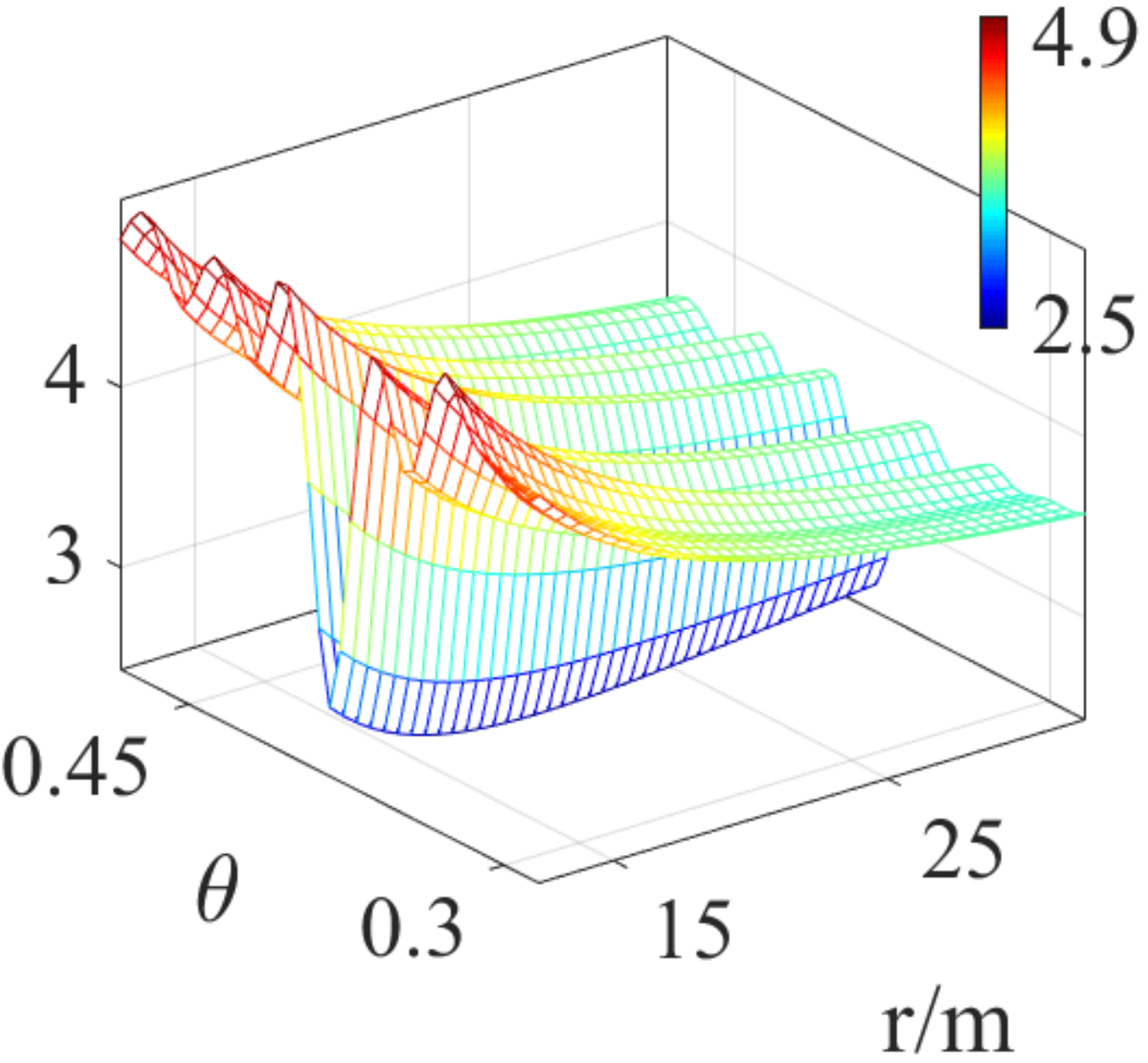}
			\label{rel_phase}
		\end{minipage}%
	}%
	\centering
	\caption{The view of the absolute value of the loss function when signal to noise ratio (SNR) is $20$ dB, where AoA and distance are independent values. The parameters are $N=256$, $f_c=0.1$THz, $B=10$GHz, $P^{\text{NRIS}}=4$, $N_{\text{RF}}=4$. The AoA can be obtained more simply and accurately from (b).}
	\label{abs_rel_phase_loss}
\end{figure}

Particularly, we only use PGD to estimate the AoA from the UE to the BS to sense the UE's location.
As mentioned earlier, if the AoA is estimated with the following loss function as in conventional off-grid methods \cite{2022_TCOM_LinglongDai_PolarDomain},\cite{offgrid_2018_TVT}
\begin{equation}
	\label{eq_OMP_GD_absolute_phase}
	{{v^{{\text{NRIS}}}}} = {\sum\nolimits_{m = 1}^M {\left\| {{\mathbf{Y}}^{{\text{NRIS}}}[m] - {{{\mathbf{\bar W}}}^{{\text{NRIS}}}}  {\hat{\mathbf{h}}}^{{\text{BU}}}[m]} \right\|_{\text{F}}^2} },
\end{equation}
the results will be affected by the accuracy of the estimated distance, as shown in Fig. \ref{abs_phase}.
The reason is that the $e^{ - j{k_m}\bar{r}_{l,g}^{{\text{BU}}}}$ of (\ref{near_field_channel}) plays a key role in (\ref{eq_OMP_GD_absolute_phase}),
so the estimated inaccurate distance results in an inaccurate AoA estimation.
Note that $e^{ - j{k_m}\bar{r}_{l,g}^{{\text{BU}}}}$ in (\ref{near_field_channel}) is determined by the total distance from the UE to the center of the BS antenna array at the subcarrier $f_m$.
On account of the fully connected antenna architecture of hybrid beamforming, we let the first RF-chain of the combiner in one of all time slots, e.g. ${{\mathbf{\bar W}}_{1,:}^{{\text{NRIS}}}}$, to be 
\begin{equation}
	\label{eq_OMP_FirstLine_PhaseShifter}
	\left\{ {\begin{array}{*{20}{c}}
			{\begin{array}{*{20}{c}}
					{\underbrace {\begin{array}{*{20}{c}}
								0\ ...\ 0 
						\end{array}}_{\frac{{N - 1}}{2}}}&\frac{1}{{\sqrt{N_{\text{RF}}} }}&{\underbrace {\begin{array}{*{20}{c}}
								0\ ...\ 0 
						\end{array}}_{\frac{{N - 1}}{2}}} 
				\end{array}{\text{, if }}N{\text{ is odd}}} \\ 
			{\begin{array}{*{10}{c}}
					{\underbrace {\begin{array}{*{10}{c}}
								0\ ...\ 0 
						\end{array}}_{\frac{{N - 2}}{2}}}&{\begin{array}{*{10}{c}}
							\frac{1}{{\sqrt{N_{\text{RF}}} }}&\frac{1}{{\sqrt{N_{\text{RF}}} }} 
					\end{array}}&{\underbrace {\begin{array}{*{10}{c}}
								0\ ...\ 0 
						\end{array}}_{\frac{{N - 2}}{2}}{\text{, if }}N{\text{ is even}}} 
			\end{array}} 
	\end{array}} \right.,
\end{equation}
while each element in ${{\mathbf{\bar W}}_{2:\text{end},:}^{{\text{NRIS}}}}$ is set as (\ref{eq_W_NRIS}).
The settings of (\ref{eq_OMP_FirstLine_PhaseShifter}) can be used to remove $e^{ - j{k_m}\bar{r}_{l,g}^{{\text{BU}}}}$ in (\ref{near_field_channel}) so that the phase of the center antenna will be zero and the phases of other antennas are the relative values of the center antenna.
The concrete operating step is as follows
\begin{equation}
	\label{eq_Y_bar}
	\left\{	
	\begin{aligned}
		{\mathbf{\bar Y}}_{i}^{{\text{NRIS}}}[m] & = {\mathbf{Y}}_{i}^{{\text{NRIS}}}[m],{\text{ for }}i=1
		\\
		{\mathbf{\bar Y}}_{i}^{{\text{NRIS}}}[m] & = {{\mathbf{S}}_{i}}[m]{\sqrt {\sum\limits_{m = 1}^M {|{\mathbf{Y}}_{i}^{{\text{NRIS}}}[m]{|^2}} } }/{\sqrt {\sum\limits_{m = 1}^M {|{{\mathbf{S}}_{i}}[m]{|^2}} } }\\
		{\text{ for }}i&=2,\cdots,N_{\text{RF}}P^{\text{NRIS}}
	\end{aligned}	
	\right.,
\end{equation}
where ${{\mathbf{S}}_{i}}[m] = {{\mathbf{Y}}_{i}^{{\text{NRIS}}}[m]}/{{\mathbf{Y}}_{1}^{{\text{NRIS}}}[m]}$.
We then obtain the new loss function as
	\begin{equation}
		\label{eq_OMP_GD_relative_phase}
		\begin{aligned}
			{v^{{\text{NRIS}}}} 
			& = \sum\nolimits_{m = 1}^M {\left\| {{\mathbf{\bar Y}}^{{\text{NRIS}}}[m] - {{{\mathbf{\bar W}}}^{{\text{NRIS}}}}  {\mathbf{\bar h}}^{{\text{BU}}}[m]} \right\|_{\text{F}}^2}\\
			& = \left\| {{\mathbf{\bar Y}}_{}^{{\text{NRIS}}} - {{{\mathbf{\bar W}}}^{{\text{NRIS}}}}  {\mathbf{\bar h}}_{}^{{\text{BU}}}} \right\|_{\text{F}}^2,
		\end{aligned}
	\end{equation}
where ${\mathbf{\bar{h}}}_{}^{{\text{BU}}}$ denotes the equivalent LoS channel and can be expressed as 
\begin{equation}
	\label{rel_LoS_channel}
	{\mathbf{\bar h}}^{{\text{BU}}}[m] = \hat {\boldsymbol{\alpha}}_{0}^{{\text{BU}}}[m] \odot{\mathbf{b}}_{0}^{{\text{BU}}}[m]({f_m},\hat{\theta} _0^{{\text{BU}}},\hat{r}_0^{{\text{BU}}}),
\end{equation}
where $\hat{\boldsymbol{\alpha}} _{0}^{{\text{BU}}}[m]$ denotes the estimated channel gain of the LoS path on the $m$-th subcarrier,
$\hat{r}_0^{{\text{BU}}}$ denotes the estimated distance of the LoS path between the center of the BS antenna and the UE,
and $\hat{\theta} _0^{{\text{BU}}}$ denotes the estimated AoA of the LoS path from the UE to the center of the BS antenna.
Since ${\mathbf{\bar h}}^{{\text{BU}}}[m]$ is the LoS channel , $\hat{\boldsymbol{\alpha}} _{0}^{{\text{BU}}}[m]$ can be obtained approximately by the Friis formula and the known attenuation due to absorption of atmospheric gases in the given frequency band.
Note that there are two differences between (\ref{rel_LoS_channel}) and (\ref{near_field_channel}). The first difference is that we only consider the LoS path of the channel since the influence of NLoS paths in THz is the secondary factor and the information of the UE's location is only included in the LoS path if the UE is equipped with one antenna.
The second difference is that we eliminate the term $e^{ - j{k_m}\bar{r}_{l,g}^{{\text{BU}}}}$ in order to obtain a better property of the loss function.
It can be seen from Fig. \ref{rel_phase} that the AoA localization result will not be affected by the accuracy of the estimated distance if the loss function (\ref{eq_OMP_GD_relative_phase}) is adopted.
Therefore the PGD algorithm can be adopted and the gradient of ${v^{{\text{NRIS}}}}$ with respect to ${{ \hat{\theta} _0^{{\text{BU}}}}}$ is
\color{black}
\begin{align}
	\label{derivative}
	& {{\partial {v^{{\text{NRIS}}}}}}/{{\partial \hat \theta _0^{{\text{BU}}}}}   \notag\\
	& = {{\partial {\text{tr}}[{{({\mathbf{\bar Y}}_{}^{{\text{NRIS}}} - {{{\mathbf{\bar W}}}^{{\text{NRIS}}}}{\mathbf{\bar h}}_{}^{{\text{BU}}})}^H}({\mathbf{\bar Y}}_{}^{{\text{NRIS}}} - {{{\mathbf{\bar W}}}^{{\text{NRIS}}}}{\mathbf{\bar h}}_{}^{{\text{BU}}})]}}/{{\partial \hat \theta _0^{{\text{BU}}}}} \notag\\
	&=  - {\text{tr}}[{({\mathbf{\bar Y}}_{}^{{\text{NRIS}}})^H}{{{\mathbf{\bar W}}}^{{\text{NRIS}}}}{{\partial {\mathbf{\bar h}}_{}^{{\text{BU}}}}}/{{\partial \hat \theta _0^{{\text{BU}}}}}] \notag\\
	& - {\text{tr}}[{({{{\mathbf{\bar W}}}^{{\text{NRIS}}}}{{\partial {\mathbf{\bar h}}_{}^{{\text{BU}}}}}/{{\partial \hat \theta _0^{{\text{BU}}}}})^H}{\mathbf{\bar Y}}_{}^{{\text{NRIS}}}]  \notag\\
	&+ {\text{tr}}[{({{{\mathbf{\bar W}}}^{{\text{NRIS}}}}{{\partial {\mathbf{\bar h}}_{}^{{\text{BU}}}}}/{{\partial \hat \theta _0^{{\text{BU}}}}})^H}{{{\mathbf{\bar W}}}^{{\text{NRIS}}}}{\mathbf{\bar h}}_{}^{{\text{BU}}}] \notag\\
	& + {\text{tr}}[{({{{\mathbf{\bar W}}}^{{\text{NRIS}}}}{\mathbf{\bar h}}_{}^{{\text{BU}}})^H}{{{\mathbf{\bar W}}}^{{\text{NRIS}}}}{{\partial {\mathbf{\bar h}}_{}^{{\text{BU}}}}}/{{\partial \hat \theta _0^{{\text{BU}}}}}]  \notag\\
	&=-2{\mathscr{R}{\{\text{tr}[}}{({{{\mathbf{\bar W}}}^{{\text{NRIS}}}}{{\partial {\mathbf{\bar h}}_{}^{{\text{BU}}}}}/{{\partial \hat \theta _0^{{\text{BU}}}}})^H}{\mathbf{\bar Y}}_{}^{{\text{NRIS}}}{\text{]\} }}  \notag\\
	&+2{\mathscr{R}{\{\text{tr}[ }}{({{{\mathbf{\bar W}}}^{{\text{NRIS}}}}{\mathbf{\bar h}}_{}^{{\text{BU}}})^H}{{{\mathbf{\bar W}}}^{{\text{NRIS}}}}{{\partial {\mathbf{\bar h}}_{}^{{\text{BU}}}}}/{{\partial \hat \theta _0^{{\text{BU}}}}}{\text{]\} }} \notag\\
	&=  - 2  \mathscr{R}\{ {\text{tr}}[{({{{\mathbf{\bar W}}}^{{\text{NRIS}}}}{\boldsymbol{\hat \alpha}}_{0}^{{\text{BU}}}[m] \odot {\partial {\mathbf{ b}}_{0}^{{\text{BU}}}}/{\partial \hat \theta _0^{{\text{BU}}}})^H}{\mathbf{\bar Y}}_{}^{{\text{NRIS}}}]\}  \notag\\ 
	&+ {\text{2}}  \mathscr{R}\{ {\text{tr}}[{({{{\mathbf{\bar W}}}^{{\text{NRIS}}}}{\mathbf{\bar h}}_{}^{{\text{BU}}})^H}{{{\mathbf{\bar W}}}^{{\text{NRIS}}}}{\boldsymbol{\hat \alpha}}_{0}^{{\text{BU}}}[m] \odot {\partial {\mathbf{ b}}_{0}^{{\text{BU}}}}/{\partial \hat \theta _0^{{\text{BU}}}}]\}.  \notag\\
\end{align}
\color{black}
The gradient of the $n$-th element of ${\mathbf{b}_0^{{\text{BU}}}[m]}$ with respect to ${{ \hat{\theta} _0^{{\text{BU}}}}}$ is derived as
\begin{small}
	\begin{equation}
		\label{derivative_h}
		\begin{aligned}
			\left. {{\partial {\mathbf{b}}_0^{{\text{BU}}}[m]}/{\partial \hat{\theta} _0^{{\text{BU}}}}} \right|_{n}  
			& = \beta   {\hat{r}_0^{{\text{BU}}}{{\mathbf{\delta }}_n}d}/{\sqrt {{{(\hat{r}_0^{{\text{BU}}})}^2} + {\mathbf{\delta }}_n^2{d^2} - 2\hat{r}_0^{{\text{BU}}}\hat{\theta} _0^{{\text{BU}}}{{\mathbf{\delta }}_n}d} }, \\ 
		\end{aligned}
	\end{equation}
\end{small}
where $\beta  = {\operatorname{e} ^{ - j\frac{{2\pi }}{{{\lambda _m}}}(\sqrt {{{(\hat{r}_0^{{\text{BU}}})}^2} + {\mathbf{\delta }}_n^2{d^2} - 2\hat{r}_0^{{\text{BU}}}\hat{\theta} _0^{{\text{BU}}}{{\mathbf{\delta }}_n}d}  - \hat{r}_0^{{\text{BU}}})}} ( j\frac{{2\pi }}{{{\lambda _m}}})$.
The specific process of the PGD can be seen in steps \ref{alg_PGD_begin}-\ref{alg_PGD_end} of the Algorithm \ref{CDL}, where the step size $\Delta$ in each iteration is determined by the Armijo-Goldstein condition\cite{ref_ArmijoWolfe}\footnote{
{\color{black}
We have tried the Wolfe condition in simulations to find the appropriate step size, but the estimation accuracy of AoAs is similar to using the Armijo-Goldstein condition.
The Armijo-Goldstein condition is to ensure that the step size is not too large, whereas the Wolfe condition is to ensure that the step size is neither too small nor too large.
Considering the computational complexity of Armijo-Goldstein condition is lower than the Wolfe condition, we adopted the Armijo-Goldstein condition here.}}.
\subsubsection{\textbf{Using the polar-domain hierarchical dictionary to refine the AoA from the UE to the RIS}}
$\quad$

Compared to ${{\mathbf{\bar W}}^{{\text{NRIS}}}}$, ${{\mathbf{\bar W}}^{{\text{RIS}}}}$ is related to the frequency.
Therefore, the PGD cannot be used to refine the AoA from the UE to the RIS since ${\mathbf{Y}}^{\text{RIS}}$ cannot be converted to ${\mathbf{\bar Y}}^{\text{RIS}}$ as ${\mathbf{Y}}^{\text{NRIS}}$ to ${\mathbf{\bar Y}}^{\text{NRIS}}$.
Instead, polar-domain hierarchical dictionary (PHD) is used.

Since the coarse $\hat \theta^{\text{RU}}_0$ is passed by the LA-GMMV-OMP, where the FSPRD is used, the AoA spacing is $\frac{1}{\varsigma N_{\text{RIS}}}$ in angle domain (the sine of the real angle), as can be seen in (\ref{theta_sample}).
Therefore, the first search of PHD ranges from $\hat \theta^{\text{RU}}_0 - \frac{1}{\varsigma N_{\text{RIS}}}$ to $\hat \theta^{\text{RU}}_0 + \frac{1}{\varsigma N_{\text{RIS}}}$.
If the number of grids in each search is $N_{\text{PHD}}$, the general formula for the AoA of the first search can be expressed as
\begin{equation}
	\label{eq_first_PHD}
	\hat \theta^{\text{RU}}_0 - \frac{1}{\varsigma N_{\text{RIS}}} + \frac{2(i-1)}{(N_{\text{PHD}}-1)\varsigma N_{\text{RIS}}}, \forall i = 1,2,\cdots,N_{\text{PHD}}.
\end{equation}
Similarly, the general formula for the AoA of the $k$-th ($k\geq 2$) search can be expressed as
\begin{small}
\begin{equation}
	\label{eq_kth_PHD}
	\begin{small}
		\begin{gathered}
			\hat \theta^{\text{RU}}_0[k-1] - \frac{1}{\varsigma N_{\text{RIS}} (\frac{N_{\text{PHD}}-1}{2})^{k-1}} + \frac{2(i-1)}{(N_{\text{PHD}}-1)\varsigma N_{\text{RIS}}(\frac{N_{\text{PHD}}-1}{2})^{k-1}}, \hfill \\
			\qquad \qquad \qquad \qquad \quad \forall i = 1,2,\cdots,N_{\text{PHD}},
		\end{gathered}
	\end{small}	
\end{equation}
\end{small}
where $\hat \theta^{\text{RU}}_0[k-1]$ is the AoA estimation in the $(k-1)$-th search and $\hat \theta^{\text{RU}}_0[0] \triangleq \hat \theta^{\text{RU}}_0$.
In the $k$-th search, we denote the set of the search range as $\mathcal{R}_k$, whose $i$-th element is given by (\ref{eq_kth_PHD}).
Then, we generate $N_{\text{PHD}}$ girds of dictionary ${\mathbf{D}}^{\text{R}}[m,k] \in \mathbb{C}^{{N_{{\text{RIS}}}} \times N_{\text{PHD}}}$ with the same distance $\hat r_0^{\text{RU}}$ but different angles, which are given by $\mathcal{R}_k$.
Next, ${\mathbf{\Gamma }}^{{\text{R}}}[m,k] \in \mathbb{C}^{N_{\text{PHD}}}$, the correlation matrix on the $m$-th subcarrier in the $k$-th search, can be calculated as
\begin{equation}
	\label{eq_HieDic_cor}
	{\mathbf{\Gamma }}^{{\text{R}}}[m,k] = |{({\mathbf{\bar W}}^{{\text{RIS}}}[m]{\mathbf{D}}^{{\text{R}}}[m,k])^H}  {{\mathbf{Y}}^{{\text{RIS}}}[m]}|.
\end{equation}
Benefiting from the MMV property, the estimation of the AoA index from the UE to the RIS in the $k$-th search can be obtained as
\begin{equation}
	\label{eq_Fine_AoA_RU}
	\hat i _k^{{\text{R}}} = \mathop {\arg \max\limits_i }  \sum\nolimits_{m = 1}^M {|{\mathbf{\Gamma }}^{{\text{R}}}_i[m,k]{|^2}}.
\end{equation}
Therefore, we can obtain the estimation of the AoA from the UE to the RIS in the $k$-th search as
\begin{small}
\begin{equation}
	\begin{gathered}
	\label{eq_kth_AoA_UERIS}
		\hat \theta^{\text{RU}}_0[k] = \hfill \\ \hat \theta^{\text{RU}}_0[k-1] - \frac{1}{\varsigma N_{\text{RIS}} (\frac{N_{\text{PHD}}-1}{2})^{k-1}} + \frac{2(\hat i _k^{{\text{R}}}-1)}{(N_{\text{PHD}}-1)\varsigma N_{\text{RIS}}(\frac{N_{\text{PHD}}-1}{2})^{k-1}}.	
	\end{gathered}
\end{equation}
\end{small}
Once the search range $\frac{2}{\varsigma N_{\text{RIS}} (\frac{N_{\text{PHD}}-1}{2})^{k}}$ of $(k+1)$-th iteration is less than the threshold $\varpi_{\text{PHD}}$, which is given in advance, $\hat \theta^{\text{RU}}_0[k]$ will output as the final result.
The CDL scheme is summarized in Algorithm \ref{CDL}.

\begin{algorithm}[!t]
	\begin{small}
		\label{CDL}
		\caption{Proposed CDL Scheme}
		\LinesNumbered 
		\KwIn{received pilot ${\mathbf{Y}}$, equivalent combining matrix ${\mathbf{\bar W}}$, coarse AoA estimation $\hat \theta_0^{{\text{BU}}}$ and $\hat \theta_0^{{\text{RU}}}$, the maximum number of iteration in PGD $I_{\text{max}}$, threshold to terminate the PGD $\varpi_{\text{PGD}}$, threshold to terminate the PHD $\varpi_{\text{PHD}}$, number of points per search in PHD $N_{\text{PHD}}$}
		\KwOut{fine AoA estimation $\hat \theta_0^{{\text{BU}}}$ and $\hat \theta_0^{{\text{RU}}}$, fine distance estimation $\hat r_0^{{\text{BU}}}$ and $\hat r_0^{{\text{RU}}}$, fine estimation of UE's location ($\hat x^{\text{UE}},\hat  y^{\text{UE}}$)}
		Obtain ($\hat x^{\text{UE}}, \hat y^{\text{UE}}$), $r_0^{\text{BU}}$, $r_0^{\text{RU}}$ from (\ref{eq_OMP_coarse_UE_x_y}) and (\ref{eq_OMP_coarse_UE_r})\;
		\textbf{/* Refine $\hat \theta_0^{{\text{BU}}}$* /}\;\label{alg_PGD_begin}
		Obtain $\mathbf{\bar Y^{\text{NRIS}}}$ and ${v^{{\text{NRIS}}}}$ as (\ref{eq_Y_bar}) and (\ref{eq_OMP_GD_relative_phase}), respectively\;
		
		\While{iteration $<$ $I_{\text{max}}$}{
			Obtain the $\nabla_{}=\frac{{\partial {v^{{\text{NRIS}}}}}}{{\partial \hat{\theta} _0^{{\text{BU}}}}}$ as (\ref{derivative})\;		
			
			Use Armijo-Goldstein condition to obtain the step $\Delta$ based on the $\nabla_{}$ and $\hat \theta _0^{{\text{BU}}}$\;
			
			Update the $\hat \theta _0^{{\text{BU}}}$ as $\hat \theta_0^{{\text{BU}}} = \hat \theta_0^{{\text{BU}}} - \Delta  \nabla$ and ${v^{{\text{NRIS}}}}$ as (\ref{eq_OMP_GD_relative_phase})\;	
			
			\textbf{if} $|\Delta  \nabla|<\varpi_{\text{PGD}}$, \textbf{break}\;		
		}\label{alg_PGD_end}	
		
		\textbf{/* Refine $\hat \theta_0^{{\text{RU}}}$ */}\;\label{alg_PHD_begin}
		\For{k=1,2, $\cdots$}{
		Generate the search range $\mathcal{R}_k$ as (\ref{eq_kth_PHD})\;
		
		Generate ${\mathbf{\Gamma }}^{{\text{R}}}[m,k] \in \mathbb{C}^{N_{\text{PHD}}}$, the correlation matrix on the $m$-th subcarrier in the $k$-th search, as (\ref{eq_HieDic_cor})\;
		
		Obtain $\hat i _k^{{\text{R}}}$, the index estimation of the AoA from the UE to the RIS in the $k$-th search, as (\ref{eq_Fine_AoA_RU})\;
		
		Obtain  $\hat \theta^{\text{RU}}_0[k]$, the estimation of the AoA from the UE to the RIS in the $k$-th search, as (\ref{eq_kth_AoA_UERIS})\;
		
		\textbf{if} $\frac{2}{\varsigma N_{\text{RIS}} (\frac{N_{\text{PHD}}-1}{2})^{k}}<\varpi_{\text{PHD}}$, $\hat \theta^{\text{RU}}_0 = \hat \theta^{\text{RU}}_0[k]$ and \textbf{break}\;
		}\label{alg_PHD_end}
		\textbf{/* Obtain $\hat r_0^{{\text{BU}}}$, $\hat r_0^{{\text{RU}}}$ and ($\hat x^{\text{UE}}, \hat y^{\text{UE}}$)*/}\;
		Use $\hat \theta _0^{{\text{RU}}}$ and $\hat \theta _0^{{\text{BU}}}$ to obtain the fine UE's location ($\hat x^{\text{UE}}, \hat y^{\text{UE}}$) as (\ref{eq_OMP_coarse_UE_x_y}) and $\hat r_0^{\text{BU}}$,$\hat r_0^{\text{RU}}$ as (\ref{eq_OMP_coarse_UE_r})\;
	\end{small}
\end{algorithm}

After estimating the AoA from the UE to the BS and the RIS using the PGD algorithm and the PHD, respectively, the precise location of the UE, ($\hat x^{\text{UE}}$, $\hat y^{\text{UE}}$), can be obtained through (\ref{eq_OMP_coarse_UE_x_y}).
Then, combined with the coordinate of the BS and the RIS, the distance from the UE to the BS, $\hat r_0^{\text{BU}}$, and the distance from the UE to the RIS, $\hat r_0^{\text{RU}}$, can be obtained as (\ref{eq_OMP_coarse_UE_r}).
Finally, a new FSPRD can be updated by adding the new element generated with ($\hat \theta_0^{\text{BU}}, \hat r_0^{\text{BU}}$)/($\hat \theta_0^{\text{RU}}, \hat r_0^{\text{RU}}$) to the old one.
For every subcarrier $f_m$, the HFNF steering vector $\mathbf{b}^{\text{NRIS}}[m]$ and $\mathbf{b}^{\text{RIS}}[m]$ are generated as (\ref{near_field_steervector}) and the new FSPRD can be generated as follows
\begin{subequations}
	\label{eq_appendDic}	
	\begin{align}
		\label{eq_appendDic_NRIS}
		{\mathbf{D}}^{{\text{NRIS}}}[m] =& [{\mathbf{D}}^{{\text{NRIS}}}[m]\textbf{}\;{{\mathbf{b}}^{\text{NRIS}}[m]}],\\
		\label{eq_appendDic_RIS}
		{\mathbf{D}}^{{\text{RIS}}}[m] =& [{\mathbf{D}}^{{\text{RIS}}}[m]\textbf{}\;{{\mathbf{b}}^{\text{RIS}}[m]}].
	\end{align}	
\end{subequations}
Then, the new FSPRD can be used in (\ref{eq_A_tilde_BS})-(\ref{eq_EstCha}) to improve the CE performance.
Moreover, in the subsequent downlink transmission, the estimated channel can be used for precoding and beamforming\cite{ref_RevisionPrecode1, ref_RevisionPrecode2}.

\subsection{Complexity Analysis}
\color{black}
The computational complexity associated with localization is listed as follows.

\leftline{\textbf{1) PSOMP\cite{2022_TCOM_LinglongDai_PolarDomain}:}}\rightline{{$\mathcal{O}$($\varsigma({P^{{\text{NRIS}}}} + {P^{{\text{RIS}}}}){N_{{\text{RF}}}}N^2SM$)}}
	
\leftline{\textbf{2) GMMV-OMP:}}\rightline{{$\mathcal{O}$($\varsigma({P^{{\text{NRIS}}}} + {P^{{\text{RIS}}}}){N_{{\text{RF}}}}N^2SM$)}}
	
\leftline{\textbf{3) LA-GMMV-OMP (CDL):}}
\rightline{{$\mathcal{O}$($\varsigma({P^{{\text{NRIS}}}} + {P^{{\text{RIS}}}}){N_{{\text{RF}}}}N^2SM$)}}
\rightline{+$\mathcal{O}$(${P^{{\text{NRIS}}}}{N_{{\text{RF}}}}NM{I_{\text{PGD} }}$) + $\mathcal{O}$(${P^{{\text{RIS}}}}{N_{{\text{RF}}}}NMI_{\text{PHD}}$)}
$\mathcal{O}$($\varsigma({P^{{\text{NRIS}}}} + {P^{{\text{RIS}}}}){N_{{\text{RF}}}}N^2SM$) is the complexity in PSOMP and (LA-)GMMV-OMP to calculate the correlation between the FSPRD and the received pilot ${{\mathbf{Y}}^{\text{NRIS}}}$, ${{\mathbf{Y}}^{\text{RIS}}}$.
Once the index of FSPRD, which has the largest correlation between the FSPRD and received pilot, is acquired, the UE's location can be determined immediately from (\ref{eq_OMP_coarse_UE_x_y}).
The extra complexity $\mathcal{O}$(${P^{{\text{NRIS}}}}{N_{{\text{RF}}}}NM{I_{\text{PGD}}}$) in LA-GMMV-OMP (CDL) is to utilize the PGD to refine the AoA from the UE to the BS, where $I_{\text{PGD}}$ denotes the number of total iterations in PGD algorithm, and it is less than 10 in iterations.
The extra complexity $\mathcal{O}$(${P^{{\text{RIS}}}}{N_{{\text{RF}}}}NMI_{\text{PHD}}$) in LA-GMMV-OMP (CDL) is to utilize the PHD to refine the AoA from the UE to the RIS, where $I_{\text{PHD}}$ denotes the total number of search grids.

\textcolor{black}{
As for the complexity of CE, the computational complexity of PSOMP and GMMV-OMP, $\mathcal{O}$($\varsigma \hat L P {N_{{\text{RF}}}}N^2SM$), are of the same order of magnitude with the same number of iterations $\hat L$, where $P$ can be replaced by $P^{\text{NRIS}}$ and $P^{\text{RIS}}$ if $\bf{h}^{\text{BU}}$ and $\bf{h}^{\text{RU}}$ are estimated, respectively.
The complexity of LA-GMMV-OMP is based on the complexity of GMMV-OMP plus its extra localization complexity.}
\color{black}

\section{Proposed Location Sensing Scheme not Relying on Channel Estimation}\label{sec:PDL}
\subsection{Problem Formulation}
The aim of the proposed location sensing scheme not relying on channel estimation is to sense the UE's location from the received signal with low overhead and without the need for CE, and the schematic diagram of this scheme is shown in Fig. \ref{fig_hyperbola}.
Compared to the proposed joint channel and location sensing scheme, the prior information of the ToA can be utilized and there is no need to guarantee the performance of the CE.
Therefore, the pilot overhead and the size of the FSPRD in this scheme can be greatly reduced.
This scheme shares the same received signal model, which is described detailedly from (\ref{NRIS_rcv_pilot_unit}) to (\ref{RIS_rcv_pilot}), as the joint channel and location sensing scheme.
Then, we can formulate the following optimization problem of this scheme,
\textcolor{black}{
\begin{align}
	\footnotesize
	\label{eq_OptPro2}
		&\mathop {\min }\limits_{\hat x^{{\text{UE}}},\hat y^{{\text{UE}}}} \;\;\;\;\;\Sigma_{m = 1}^M {\Big|\!\Big| {{{\mathbf{Y}}^{{\text{NRIS}}}}[m] - {{{\mathbf{\bar W}}}^{{\text{NRIS}}}}{\hat {{\mathbf{ h}}}^{{\text{BU}}}_{\text{LoS}}}[m](\hat x^{{\text{UE}}},\hat y^{{\text{UE}}})} \Big|\!\Big|_{\text{F}}^2}  \hfill \notag\\
		&\qquad\qquad+\Sigma_{m = 1}^M {\Big|\!\Big| {{{\mathbf{Y}}^{{\text{RIS}}}}[m] - {{{\mathbf{\bar W}}}^{{\text{RIS}}}}[m]{{{\mathbf{\hat h}}}^{{\text{RU}}}_{\text{LoS}}}[m](\hat x^{{\text{UE}}},\hat y^{{\text{UE}}}) }}\hfill \notag\\
		&{{\qquad\qquad\qquad\qquad\qquad\qquad- {{{\mathbf{\check W}}}^{{\text{RIS}}}}{{{\mathbf{\hat h}}}^{{\text{BU}}}_{\text{LoS}}}[m](\hat x^{{\text{UE}}},\hat y^{{\text{UE}}})} \Big|\!\Big|_{\text{F}}^2}  \hfill \notag\\
&	\;\;\;{\text{s}}.{\text{t}}.\qquad\qquad{\text{C1 -- C6}}, \hfill \notag\\
\end{align}}
where constraints C1 to C6 are the same as the optimization problem (\ref{eq_OptPro1}) and ${{{\mathbf{\hat h}}}^{{\text{BU}}}_{\text{LoS}}}$ ( or ${{{\mathbf{\hat h}}}^{{\text{RU}}}_{\text{LoS}}}$) denotes the estimated LoS path from the UE to the BS (or RIS).
Different from the joint channel and location sensing scheme, the location sensing scheme not relying on channel estimation only estimates the UE's location ($\hat x^{{\text{UE}}},\hat y^{{\text{UE}}}$), whose information is completely included in the LoS path.
Therefore, only LoS paths are utilized for localization, and this treatment is justified since the loss of NLoS paths in the THz band is too severe\cite{ref_2022_TCOM_NanYang_ChongHan}.
\subsection{Proposed Partial Dictionary based Localization (PDL) Scheme}
First, $\mathbf{W}^{\text{NRIS}}[p]$ is set randomly and the RIS is turned off when AoA and delay from the UE to the BS are estimated from $\mathbf{Y}^{\text{NRIS}}$, which is assumed not affected by the RIS.
Second, the RIS is turned on and we can obtain $\mathbf{Y}^{\text{RIS}}$ by setting $\mathbf{W}^{\text{RIS}}[p]$ as (\ref{eq_W_allFreq}) to estimate the delay from the UE to the RIS.  
The delay can be estimated from different subcarriers based on an algorithm modified from MUSIC. 
We utilize the ${{\mathbf{Y}}^{{\text{NRIS}}}}$ and ${{\mathbf{Y}}^{{\text{RIS}}}}$ to do eigenvector decomposition (EVD) as
\begin{subequations}	
	\label{spectral_decomposition}	
	\begin{align}
		{{({\mathbf{Y}}^{{\text{NRIS}}})}^H}{\mathbf{Y}}^{{\text{NRIS}}} =& {\mathbf{E}}_{}^{{\text{NRIS}}}{\Lambda ^{{\text{NRIS}}}}{({\mathbf{E}}_{}^{{\text{NRIS}}})^{H}},\\
		{{({\mathbf{Y}}^{{\text{RIS}}})}^H}{\mathbf{Y}}^{{\text{RIS}}} =& {\mathbf{E}}_{}^{{\text{RIS}}}{\Lambda ^{{\text{RIS}}}}{({\mathbf{E}}_{}^{{\text{RIS}}})^{H}},
	\end{align}	
\end{subequations}
where ${\mathbf{E}}_{}^{{\text{NRIS}}}$ and ${\mathbf{E}}_{}^{{\text{RIS}}}$ are the matrices composed of the eigenvectors of ${{({\mathbf{Y}}^{{\text{NRIS}}})}^H}{\mathbf{Y}}^{{\text{NRIS}}}$ and ${{({\mathbf{Y}}^{{\text{RIS}}})}^H}{\mathbf{Y}}^{{\text{RIS}}}$, respectively.
We assume that the eigenvalues are arranged from the largest to the smallest in ${\Lambda ^{{\text{NRIS}}}}$ and ${\Lambda ^{{\text{RIS}}}}$ without loss of generality,
and the eigenvectors corresponding to all but the largest eigenvalues are treated as the noise subspace as
\begin{subequations}
	\label{noise_subspaces}	
	\begin{align}
		{\bar {\mathbf{E}}}_{}^{\text{NRIS}} =& {\mathbf{E}}_{:,2:\text{end}}^{{\text{NRIS}}},\\
		{\bar {\mathbf{E}}}_{}^{{\text{RIS}}} =& {\mathbf{E}}_{:,2:\text{end}}^{{\text{RIS}}}.
	\end{align}	
\end{subequations}
By defining ${\mathbf{f}} \in {\mathbb{C}^{1 \times M}}$ as the frequency vector in $M$ subcarriers, we generate the vector $\mathbf{a}(\tau)=e^{j2\pi \tau \mathbf{f}}\in {\mathbb{C}^{1 \times M}}$ at $M$ subcarriers and estimate $\hat \tau^{{\text{NRIS}}}$, the delay from the UE to the BS directly, and $\hat \tau^{{\text{RIS}}}$, the delay from the UE to the BS via the RIS, as 
\begin{subequations}
	\label{tau_est}	
	\begin{align}
		\label{tau_est_1}	
		\hat \tau^{{\text{NRIS}}}{\text{ }} =& \mathop {\arg \max\limits_\tau}  (1/({{\mathbf{a}}(\tau )  {\bar{\mathbf{E}}}_{}^{{\text{NRIS}}}  {{({\mathbf{a}}(\tau )  {\bar{\mathbf{E}}}_{}^{{\text{NRIS}}})}^H}})),\\
		\label{tau_est_2}	
		\hat \tau^{{\text{RIS}}} =& \mathop {\arg \max\limits_\tau } (1/({{\mathbf{a}}(\tau )  {\bar{\mathbf{E}}}_{}^{{\text{RIS}}}  {{({\mathbf{a}}(\tau )  {\bar{\mathbf{E}}}_{}^{{\text{RIS}}})}^H}})),
	\end{align}	
\end{subequations}
respectively.

However, the above modified MUSIC algorithm is computationally complex in two aspects, one is EVD and one is spectral peak search.
The computational complexity of spectral peak search can be greatly reduced by the idea of the hierarchical dictionary, but the complexity of EVD is not easy to reduce.
Since only the LoS path is required to be estimated, the dimensionality of the noise subspace ($M-1$) is known.
Therefore, by subspace analysis\cite{ref_MUSICnoEVD}, it is possible to obtain the noise subspaces, which are equivalent to the spaces spanned by the eigenvectors corresponding to the eigenvalues other than the largest one.
We take the process of estimating $\hat \tau^{{\text{NRIS}}}$, which is the same as $\hat \tau^{{\text{RIS}}}$, as an example.
We calculate the autocorrelation matrix of $({\mathbf{Y}}^{{\text{NRIS}}})^H = ({{\mathbf{\bar W}}^{{\text{NRIS}}}}  {\mathbf{h}}^{{\text{BU}}})^H + ({\mathbf{N}}^{{\text{NRIS}}})^H \in \mathbb{C}^{M\times P^{\text{NRIS}}N_{\text{RF}}}$ as follows,
\begin{align}
	\label{eq_autocorr_Y}
	E(({\bf{Y}}^{\text{NRIS}})^H{\bf{Y}}^{\text{NRIS}}) & = 
	({\bf{h}}^{\text{BU}})^H ({\bf{W}}^{\text{NRIS}})^H {\bf{W}}^{\text{NRIS}} {\bf{h}}^{\text{BU}} \notag \\
	&+
	E(({\bf{N}}^{\text{NRIS}})^H {\bf{N}}^{\text{NRIS}}),
\end{align}
where $E(\cdot)$ is the expected operator.
By approximating the (\ref{eq_autocorr_Y}) and utilizing $	E(({\bf{N}}^{\text{NRIS}})^H {\bf{N}}^{\text{NRIS}}) = \sigma^2 P^{\text{NRIS}} N_{\text{RF}} \mathbf{I}$, where $\sigma^2$ is the assumed known noise power, the denoising autocorrelation matrix $\tilde{\mathbf{Y}}^{\text{NRIS}} \in \mathbb{C}^{M \times M}$ can be denoted as
\begin{align}
	\label{eq_approAutocorrY}	
	\tilde{\mathbf{Y}}^{\text{NRIS}} & \triangleq ({\bf{Y}}^{\text{NRIS}})^H{\bf{Y}}^{\text{NRIS}} - \sigma^2 P^{\text{NRIS}} N_{\text{RF}} \mathbf{I} \notag \\
	& \approx ({\bf{h}}^{\text{BU}})^H ({\bf{W}}^{\text{NRIS}})^H {\bf{W}}^{\text{NRIS}} {\bf{h}}^{\text{BU}},
\end{align}
where $\tilde{\mathbf{Y}}^{\text{NRIS}}$ is rank-deficient and its maximum rank is $P^{\text{NRIS}} N_{\text{RF}}$.
Then, we can obtain the orthogonal complement of $\tilde{\mathbf{Y}}^{\text{NRIS}}$ and hence the noise subspace of $E(({\bf{Y}}^{\text{NRIS}})^H{\bf{Y}}^{\text{NRIS}})$.
The matrix $\tilde{\mathbf{Y}}^{\text{NRIS}}$ can be partitioned as the block matrix as follows
\begin{equation}
	\label{eq_BlockMatrix}
	\tilde{\mathbf{Y}}^{\text{NRIS}} = [\tilde{\mathbf{Y}}^{\text{NRIS}}_1\  \tilde{\mathbf{Y}}^{\text{NRIS}}_2],
\end{equation}
where $\tilde{\mathbf{Y}}^{\text{NRIS}}_1 \in \mathbb{C}^{M \times P^{\text{NRIS}} N_{\text{RF}}}$ and $\tilde{\mathbf{Y}}^{\text{NRIS}}_2 \in \mathbb{C}^{M \times (M - P^{\text{NRIS}} N_{\text{RF}})}$.
Since $\tilde{\mathbf{Y}}^{\text{NRIS}}$ is rank-deficient, we can assume that there exist ${\bf{G}} = [{\bf{G}}_1^H \  -{\bf{G}}_2^H]^H
\in \mathbb{C}^{M\times (M-P^{\text{NRIS}} N_{\text{RF}})}$ that makes 
\begin{equation}
	\label{eq_YG0}
	\tilde{\mathbf{Y}}^{\text{NRIS}} {\bf{G}} = \bf{0},
\end{equation}
where ${\bf{G}}_1 \in \mathbb{C}^{P^{\text{NRIS}} N_{\text{RF}} \times (M-P^{\text{NRIS}} N_{\text{RF}})}$ and ${\bf{G}}_2 \in \mathbb{C}^{(M-P^{\text{NRIS}} N_{\text{RF}}) \times (M-P^{\text{NRIS}} N_{\text{RF}})}$.
Therefore, ${\bf{G}}_1$ can be denoted as
\begin{equation}
	\label{eq_G_1}
	{\bf{G}}_1 = 	 ({\tilde{{\bf{Y}}}_1^{\text{NRIS}}})^\dag {\tilde{\bf{Y}}}_2^{\text{NRIS}} {\bf{G}}_2
\end{equation}
By taking the conjugate transpose of (\ref{eq_YG0}), we can obtain ${\bf{G}}^H \tilde{\mathbf{Y}}^{\text{NRIS}} = \bf{0}$ and the projection matrix to the subspace spanned by G can be calculated as 
	\begin{align}
		\label{eq_Proj_G}
		{\bf{P}}_{{\bf{G}}} & = {{\bf{G}}} ({{\bf{G}}}^H {{\bf{G}}})^{-1}{{\bf{G}}}^{H} \notag \\
		&= \left[ {\begin{array}{*{20}{c}}
				{\tilde {{\bf{Y}}}}_{1}^{\dag}{\tilde {{\bf{Y}}}}_{2} \\
				-\bf{I}
		\end{array}} \right]
		(\left[ {\begin{array}{*{20}{c}}
				{\tilde {{\bf{Y}}}}_{1}^{\dag}{\tilde {{\bf{Y}}}}_{2} \\
				-\bf{I}
		\end{array}} \right]^H
		\left[ {\begin{array}{*{20}{c}}
				{\tilde {{\bf{Y}}}}_{1}^{\dag}{\tilde {{\bf{Y}}}}_{2} \\
				-\bf{I}
		\end{array}} \right])^{-1}
		\left[ {\begin{array}{*{20}{c}}
				{\tilde {{\bf{Y}}}}_{1}^{\dag}{\tilde {{\bf{Y}}}}_{2} \\
				-\bf{I}
		\end{array}} \right]^H,
	\end{align}
where ${\bf{P}}_{{\bf{G}}}$ is only associated with $	\tilde{\mathbf{Y}}^{\text{NRIS}}$.
Therefore, (\ref{tau_est_1}) can be equivalently converted to the problem as follows
\begin{equation}
	\label{eq_searchProj_G}
	\hat \tau^{\text{NRIS}}=
	\text{arg}\ \mathop{\max}\limits_\tau
	1/
		[{\bf{a}}(\tau)
		{\bf{P}}_{{\bf{G}}}
		{\bf{a}}^H(\tau)].
\end{equation}
The process of solving for $\hat \tau^{\text{RIS}}$ is the same as for $\hat \tau^{\text{NRIS}}$ and we omit it for simplicity.
Since $\hat \tau^{{\text{NRIS}}}$ and $\hat \tau^{{\text{RIS}}}$ include the same common clock offset, we can obtain the time difference between the UE arriving at the BS and the UE arriving at the RIS as
\begin{equation}
	\label{tau_TDoA}
	{\hat \tau ^{{\text{TDoA}}}}{= \hat \tau^{{\text{NRIS}}} - (}\hat \tau^{{\text{RIS}}} - {r^{{\text{B}}2{\text{R}}}}/c).
\end{equation}
In reality, $\hat \tau ^{{\text{TDoA}}}$ may be positive or negative, depending on the distance between the BS and the UE and the distance between the RIS and the UE.
Without loss of generality, we assume $\hat \tau ^{{\text{TDoA}}}$ is positive here.
Then we can obtain one side of the hyperbola and the UE must be around, which can be seen in Fig. \ref{fig_hyperbola}.
For convenience, we put the BS and the RIS at the focal points of the hyperbola, and let them lie on the $x$-axis.
Therefore, we can acquire the standard equation of the hyperbola as 
\begin{equation}
	\label{hyperbola_equation}
	{{x^2}}/{(a^{\text{h}})^2} - {{y^2}}/{(b^{\text{h}})^2} = 1,
\end{equation}
where ${a^{\text{h}}} = \frac{{c{\hat \tau ^{{\text{TDoA}}}}}}{2}$ and $({b^{\text{h}}})^2 = (c^{\text{h}})^2 - (a^{\text{h}})^2 = {(\frac{{{r_{{\text{B2R}}}}}}{2})^2} - (a^{\text{h}})^2$.
After that, we can lock the UE on the hyperbola, where the FSPRD ${{\mathbf{D}}^{\text{h}}}[m] \in \mathbb{C}^{N \times T}$ can be generated to sense the UE's location on grids.
$N$ is the number of the BS antenna and $T$ is the number of grids on the hyperbola.
There are many ways to generate the FSPRD, and the method we adopt is to generate a group of lines from the BS and intersect the hyperbola.
The group of angles can be denoted as 
\begin{equation}
	\label{eq_BS_hyper_angle}
	[{\hat \theta ^{\text{h}}},\;{\hat \theta ^{\text{h}}} + \Delta \theta ,\;{\hat \theta ^{\text{h}}} + 2\Delta \theta \;,\; \cdots ,\;{\hat \theta ^{\text{h}}} + (T - 1)\Delta \theta ],
\end{equation}
where ${\hat \theta ^{\text{h}}}$ is the initial angle value between the line and the $x$-axis, and $\Delta \theta$ is the angle spacing.
\begin{algorithm}[htbp]
	\begin{small}
		\label{PDL}
		\caption{Proposed PDL Scheme}
		\LinesNumbered 
		\KwIn{received pilot ${{\mathbf{Y}}}$, equivalent combining matrix ${{\mathbf{\bar W}}}$, the maximum number of iteration $I_{\text{max}}$, threshold to terminate the PGD $\varpi_{\text{PGD}}$}
		\KwOut{fine AoA estimation $\hat \theta _0^{{\text{BU}}}$, fine distance estimation $\hat r_0^{{\text{BU}}}$, fine estimation of UE's location ($\hat x^{\text{UE}}, \hat y^{\text{UE}}$)}

		Obtain the denoising autocorrelation matrix $\tilde{\mathbf{Y}}^{\text{NRIS}}$ as (\ref{eq_approAutocorrY})\;\label{PDL_step_begin}
		Construct ${\bf{G}} = [{\bf{G}}_1^H \  -{\bf{G}}_2^H]^H$ to make ${\bf{G}}$ satisfy (\ref{eq_YG0}), and obtain the relationship between the ${\bf{G}}_1$ and ${\bf{G}}_2$ as shown in (\ref{eq_G_1})\;
		Obtain the projection matrix, $\mathbf{P_G}$, onto the subspace spanned by $\mathbf{G}$ as (\ref{eq_Proj_G})\;
		Obtain ${\hat \tau} ^{\text{NRIS}}$ by searching spectral peak according to (\ref{eq_searchProj_G}) with the idea of hierarchical dictionary\;\label{PDL_step_end}
		The procedure for obtaining ${\hat \tau} ^{\text{RIS}}$ is the same as for obtaining ${\hat \tau} ^{\text{NRIS}}$, as shown from step \ref{PDL_step_begin} to \ref{PDL_step_end}.
		
		Obtain the $\hat \tau^{{\text{TDoA}}}$ as (\ref{tau_TDoA}) and the hyperbola equation as (\ref{hyperbola_equation})\;
		Generate FSPRD ${{\mathbf{D}}^{\text{h}}}$ as Algorithm \ref{alg_FreqSelec} on the hyperbola and obtain the coarse estimation $\hat \theta _0^{{\text{BU}}}$ as (\ref{RISAL_coarse_angle})\;
		
		Combine the hyperbola and $\hat \theta _0^{{\text{BU}}}$ with the location of the BS, (${x^{{\text{BS}}}}$, ${y^{{\text{BS}}}}$), to obtain $\hat r _0^{{\text{BU}}}$ using (\ref{coarse_UE_x_y}) and (\ref{coarse_UE_r})\;
		
		Obtain $\mathbf{\bar Y^{\text{NRIS}}}$ and ${v^{{\text{NRIS}}}}$ as (\ref{eq_Y_bar}) and (\ref{eq_OMP_GD_relative_phase}), respectively\;	       
		
		\While{iteration $<$ $I_{\text{max}}$}{
			Obtain the $\nabla_{}=\frac{{\partial {v^{{\text{NRIS}}}}}}{{\partial \hat{\theta} _0^{{\text{BU}}}}}$ as (\ref{derivative})\;		
			
			Use Armijo-Goldstein condition to update the step $\Delta$ based on the $\nabla_{}$ and  $\hat \theta _0^{{\text{BU}}}$\;
					
			Update the $\hat \theta _0^{{\text{BU}}}$ as $\hat \theta_0^{{\text{BU}}} = \hat \theta_0^{{\text{BU}}} - \Delta  \nabla$ and ${v^{{\text{NRIS}}}}$ as (\ref{eq_OMP_GD_relative_phase})\;
			
			\textbf{if} $|\Delta  \nabla|<\varpi_{\text{PGD}}$, \textbf{break}\;
		}
		
		Obtain ($\hat x^{\text{UE}}, \hat y^{\text{UE}}$) and $\hat r _0^{{\text{BU}}}$ using $\hat \theta _0^{{\text{BU}}}$, (\ref{coarse_UE_x_y}) and (\ref{coarse_UE_r})\;
	\end{small}
\end{algorithm} 	
Then we can obtain the coarse AoA from the UE to the BS by the correlation as
\begin{equation}
	\label{RISAL_coarse_angle}
	{\hat \theta}_0^{\text{BU}} = {\hat \theta}^{\text{h}} + (\hat t -1) \Delta \theta,
\end{equation}
where ${\hat t = {\text{arg max}}_t  \sum\nolimits_{m = 1}^M {{{({{{\mathbf{\bar W}}}^{{\text{NRIS}}}}{\mathbf{D}}_{:,t}^{\text{h}}[m])}^H}{\mathbf{Y}}^{{\text{NRIS}}}[m]} }$.
By combining the $\hat \theta _0^{\text{BU}}$ with the location of the BS, denoted as (${x^{{\text{BS}}}}$, ${y^{{\text{BS}}}}$), the location of the UE can be calculated as the intersection of the line and the hyperbola
\begin{small}
	\begin{align}
		\label{coarse_UE_x_y}
		\begin{aligned}
			\vspace{-2mm}
			\hat x^{{\text{UE}}} & = \frac{{\sqrt {4{k^4}{a^4}(x^{{\text{BS}}})^2 + 4{a^2}({b^2} - {a^2}{k^2})({k^2}(x^{{\text{BS}}})^2 + {b^2})} }}{{2({b^2} - {a^2}{k^2})}} \\ 
			& + \frac{{ - 2{a^2}{k^2}{x^{{\text{BS}}}}}}{{2({b^2} - {a^2}{k^2})}} \\ 
			\hat y^{{\text{UE}}} & = k({\hat x}^{{\text{UE}}} - {x^{{\text{BS}}}})
		\end{aligned}
		,
	\end{align}
\end{small}
where ${k=-\tan (\frac{\pi}{2}-\vartheta^{\text{B}}-\arcsin (\hat \theta _0^{{\text{BU}}}))}$ is the slope of the line from the BS to the UE, as Fig. \ref{fig_hyperbola} shows.
In (\ref{coarse_UE_x_y}), ${y^{{\text{BS}}}}$ is set to 0, which is the same setting as that in the simulation.
Then, we can get the distance $\hat r_0^{\text{BU}}$ as
\begin{equation}
	\label{coarse_UE_r}
	\hat r_0^{{\text{BU}}} = \sqrt {{(\hat x^{{\text{UE}}} - x^{{\text{BS}}})}^2 + {{(\hat y^{{\text{UE}}} - y^{{\text{BS}}})}^2}}.
\end{equation}
However, the estimated  $\hat \theta_0^{{\text{BU}}}$ is limited to the number of grids, leading to limited localization accuracy.
Therefore, the PGD algorithm is also employed to refine the AoA from the UE to the BS.
The step of PGD in PDL is similar to the one in CDL and Algorithm \ref{PDL} contains complete steps of the proposed PDL scheme.

\subsection{Complexity Analysis}
\textcolor{black}{
As a comparison of PDL, full-digital estimating signal parameter via rotational invariance techniques (ESPRIT) \cite{2021_JSAC_AnwenLiao}, full-digital MUSIC \cite{2022_JSAC_RISaidedSensing}, and hybrid beamforming MUSIC \cite{ref_HybridMUSIC} are used to estimate the AoA from the UE to the BS, which is combined with the TDoA to sense the UE's location.
Their steps of obtaining the TDoA are the same as the PDL.}

\color{black}
The computational complexity of the proposed algorithm and the baseline algorithms is provided as follows.

\begin{small}
		\leftline{1) \textbf{ESPRIT (full-digital)\cite{2021_JSAC_AnwenLiao}:}}
		\rightline{$\mathcal{O}$($M^3+(M^2+M)R$) + $\mathcal{O}$($N^3+N^2M$)}
	
		\leftline{2) 	\textbf{MUSIC (full-digital)\cite{2022_JSAC_RISaidedSensing}:}}
		\rightline{$\mathcal{O}$($M^3+(M^2+M)R$) +	$\mathcal{O}$($N^3+N^2M+QN^2$)}
	
		\leftline{3) 	\textbf{MUSIC (hybrid)\cite{ref_HybridMUSIC}:}}
		\rightline{$\mathcal{O}$($M^3+(M^2+M)R$) +	$\mathcal{O}$($(P^{\text{NRIS}}N_{\text{RF}})^3+(P^{\text{NRIS}}N_{\text{RF}})^2Q$)}
		\rightline{ + $\mathcal{O}$($P^{\text{NRIS}}N_{\text{RF}}NM$)}
	
		\leftline{4) 	\textbf{PDL (hybrid):}}
		\rightline{$\mathcal{O}$($M^3+(M^2+M)R$) +	$\mathcal{O}$($(P^{\text{NRIS}}N_{\text{RF}}NM)(I_{\text{PGD}}+T)$)}
\end{small}

$R$ denotes the number of distance grids to get the $\hat \tau^{\text{RIS}}$ and $\hat \tau^{\text{NRIS}}$,
$Q$ denotes the number of angle grids searching in the whole space,
$T$ denotes the number of grids generated on the hyperbola, and
$I_{\text{PGD}}$ denotes the number of total iterations in PGD of the PDL scheme.
$I_{\text{PGD}}$ is less than 10 in iterations.
\color{black}
\section{Simulation Results}\label{sec:Simulation}
\subsection{Simulation Setup}\label{sec_Setup}
We consider the system model as shown in Fig. \ref{fig_hyperbola} and Fig. \ref{WDBAL_fig}.
The BS and the RIS receive the uplink signals from the active UE within a sector of radius $R_{\text{s}}=100$m and central angle $90^{\circ}$.
The UE is set within the effect Rayleigh distance as (\ref{effective_Rayleigh_distance}) if the UE is in the near-field region.
Unless otherwise specified, the simulation setup is detailed as follows:
$N=256$, 
$N_{\text{RF}}=4$, 
$N_{\text{RIS}}=256$, 
$f_c=0.1$ THz, 
$B=10$ GHz,
$M=2048$,
$P=30$dBm,
$\vartheta^{\text{B}}=\frac{\pi}{4}$,
$\vartheta^{\text{R}}=\frac{\pi}{4}$,
$S=10$,
$L_\text{max}=20$,
$\varpi_{\text{OMP}}=0.85$ ($N_s=6$),
$\varpi_{\text{OMP}}=0.95$ ($N_s=1$),
$\varpi_{\text{PGD}}=1\times 10^{-7}$,
$\varpi_{\text{PHD}}=2\times 10^{-5}$,
$N_{\text{PHD}}=41$,
$\hbar=0.1$,
$I_\text{max}=20$,
$\varsigma = 2$.
\textcolor{black}{
According to the ITU-R P.676-12, the molecular absorption coefficient $\boldsymbol{\alpha}^{\text{A}}[m]$ can be set to $-0.45$ dB/km from $0.095$ THz to $0.105$ THz\cite{ref_2022_WC_ChongHan}.
In the CDL scheme, $P^{\text{NRIS}}=16$, $P^{\text{RIS}}=32$ and it uses only 64 of the 2048 subcarriers to take advantage of the MMV property and reduce computational complexity.
These 64 subcarriers are composed of the first subcarrier extracted from each of the 32 consecutive subcarriers in the 2048 subcarriers.}
In the PDL scheme, $P^{\text{NRIS}}=8$ and $P^{\text{RIS}}=16$.
Therefore, in the CDL scheme, the compression ratio of estimation of $\bf{h}^{\text{BU}}$ or extraction of the UE's location from $\bf{h}^{\text{BU}}$ is $P^{\text{NRIS}}N_{\text{RF}}/N=1/4$.
The compression ratio of estimation of $\bf{h}^{\text{RU}}$ or extraction of the UE's location from $\bf{h}^{\text{RU}}$ is $P^{\text{RIS}}/N=1/8$, since the rank of the channel matrix between the RIS and the BS is almost 1.
In the PDL scheme, the compression ratio of extraction of the UE's location from $\bf{h}^{\text{BU}}$ is $P^{\text{NRIS}}N_{\text{RF}}/N=1/8$ and the one of extraction of the UE's location from $\bf{h}^{\text{RU}}$ is $P^{\text{RIS}}/N=1/16$.
The time slots in the CDL scheme are twice as much as that in the PDL scheme since the number of observations cannot be too few in order to guarantee the performance of the CE.
The noise power spectrum density at the receiver is set as $\sigma_{\text{NSD}}^{2}=-174$ dBm/Hz and the transmit power of the UE $P$ is set from 15 dBm to 45 dBm in evaluating the CE performance.
The channel is modeled as the cluster-sparse multi-path channel that is widely used in mmWave/THz systems.
There are 3 scatterers between the UE and the BS, and 3 scatterers between the UE and the RIS.
Therefore, the clusters number of the channel between the UE and the BS and between the UE and the RIS are both 3.
Moreover, there are 6 paths in each cluster.
The BS is set at (-10$\sqrt{2}$ m, 0 m) and the RIS is set at (10$\sqrt{2}$ m, 0 m) in the near-field region as shown in Fig. \ref{fig_hyperbola}.
In the far-field condition, the BS is set at (-20$\sqrt{2}$ m, 0 m) and the RIS is set at (20$\sqrt{2}$ m, 0 m).
The UE is set at (5.96 m, -10.1 m) and (11.83 m, -20.2 m) in the near-field and far-field region, respectively.
The scatterers location are set randomly in the whole region.
Each LoS complex channel coefficient in $\boldsymbol{\alpha}^{\text{F}}[m]$ can be calculated from the Friis transmission formula as $\alpha^{{\text{NRIS}}}[m] = {e^{j{\varepsilon ^{{\text{NRIS}}}}}}\sqrt {{G_{\text{T}}}{G_{\text{R}}}\lambda _m^2} /4\pi R$ and $\alpha^{{\text{RIS}}}[m] = {e^{j{\varepsilon ^{{\text{RIS}}}}}}\sqrt {{G_{\text{T}}}{G_{\text{R}}}{S_{{\text{eff}}}}\lambda _m^2} /\sqrt {{{(4\pi )}^3}{{({R_1})}^2}{{({R_2})}^2}}$ \cite{2022_JSAC_RISaidedSensing},
where ${\alpha^{\text{RIS}}[m]}$ and ${\alpha^{\text{NRIS}}[m]}$ are denoted as the LoS channel gain from the UE to the BS directly and from the UE to the BS via the RIS, respectively.
$R$ denotes the distance from the UE to the certain BS antenna element.
$R_1$, $R_2$ denote the distance from the UE to the certain element of the RIS, the distance from the certain element of the RIS to the certain BS antenna element, respectively.
$G_\text{T}$ and $G_\text{R}$ denote antenna gain of the UE and the BS, respectively.
$\varepsilon ^{\text{NRIS}},\varepsilon ^{\text{RIS}} \in \mathcal{U}[0, 2\pi)$ are the phase shift by the channel.
${\alpha}^{\text{S}}[m] \sim\mathcal{CN}(0, 1)$ denotes the small-scale complex channel gain of the NLoS path.
$S_{\text{eff}}$, which is set to $(N_{\text{RIS}}d)^2\ \text{m}^2$ in the simulation, is the effective reflection area of the RIS.
As for the NLoS complex channel gain, defined similarly as $\alpha^{{\text{RIS}}}[m]$, multi-hop paths and the scattering area of the scatterers are taken into account.
In this paper, we only consider $2$-hop paths and assume that the scattering area equal to $3\text{m}^2$.
The root mean square error (RMSE) is considered as the accuracy evaluation of the UE's location, which can be defined as ${\text{RMS}}{{\text{E}}_\vartheta }  = [{\sum\nolimits_{n = 1}^{N_{\text{it}}} {{{({{\hat \vartheta }_n} - {\vartheta _{{\text{real}}}})}^2}} }/{N_{\text{it}}}]^{0.5}$ and ${\text{RMS}}{{\text{E}}_r}  = [{\sum\nolimits_{n = 1}^{N_{\text{it}}} {{{({{\hat r}_n} - {r_{{\text{real}}}})}^2}} }/{N_{\text{it}}}]^{0.5}$,
where ${{\hat \vartheta }_n}$ and ${{\hat r }_n}$ are the estimation results of the AoA and distance from the UE to the BS every iteration $n$, respectively, ${r_{{\text{real}}}}$ and ${\vartheta_{{\text{real}}}}$ are the real values, $N_{\text{it}}$ is the number of the total iterations.
The normalized mean square errors (NMSE) is considered as the accuracy evaluation of CE, which can be defined as ${\text{NMSE  =  }}||{\mathbf{h}} - {\mathbf{\hat h}}||_{\text{F}}^2/||{\mathbf{h}}||_{\text{F}}^2$, where $\mathbf{h}$ is the real channel matrix while $\mathbf{\hat h}$ is the estimated channel matrix.

\begin{figure*}[!t]
	\color{black}
	\subfigure[$\mathbf{h}^{\text{BU}}$is the near-field channel.]{
		\begin{minipage}[t]{0.25\linewidth}
			\includegraphics[width=1.6in]{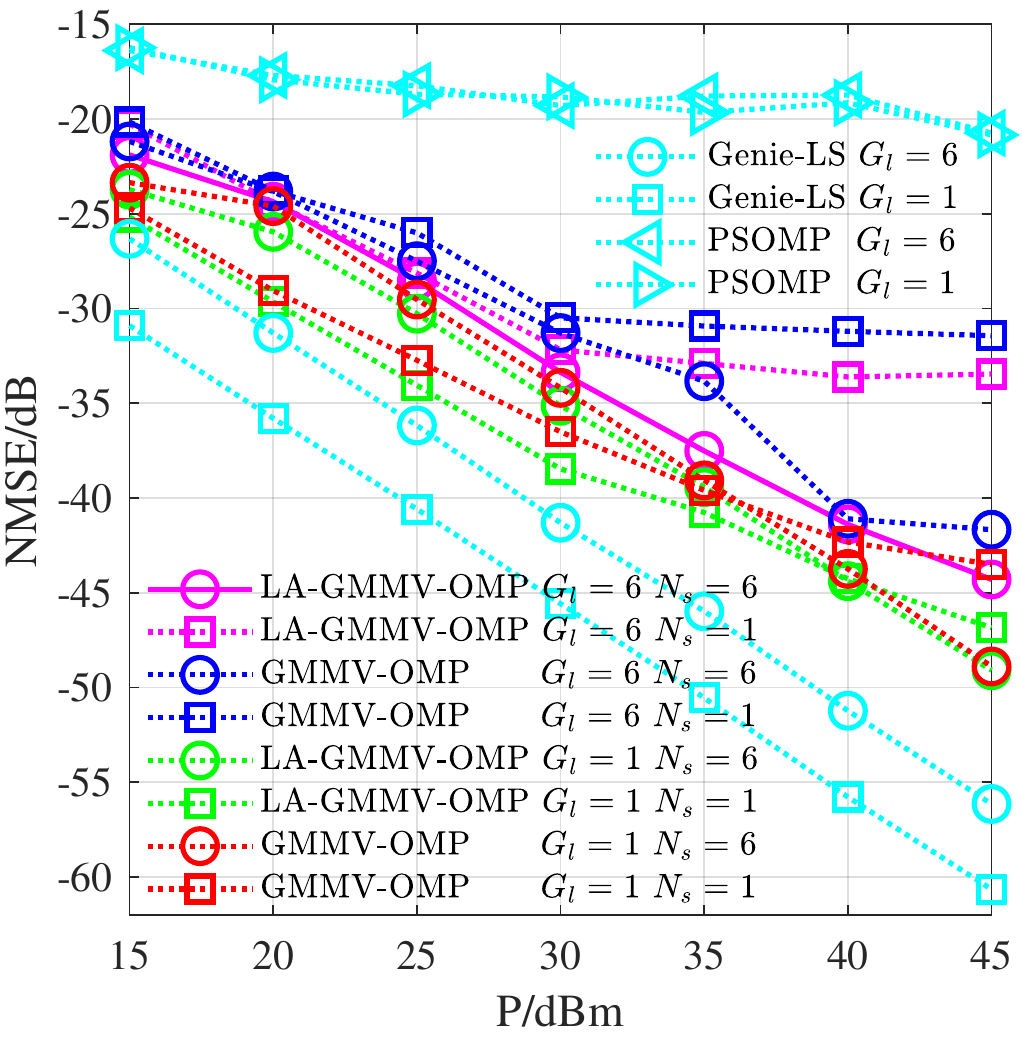}
			\label{fig_CE_BSUE_near}
		\end{minipage}%
	}%
	\subfigure[$\mathbf{h}^{\text{BU}}$is the far-field channel.]{
		\begin{minipage}[t]{0.25\linewidth}
			\includegraphics[width=1.6in]{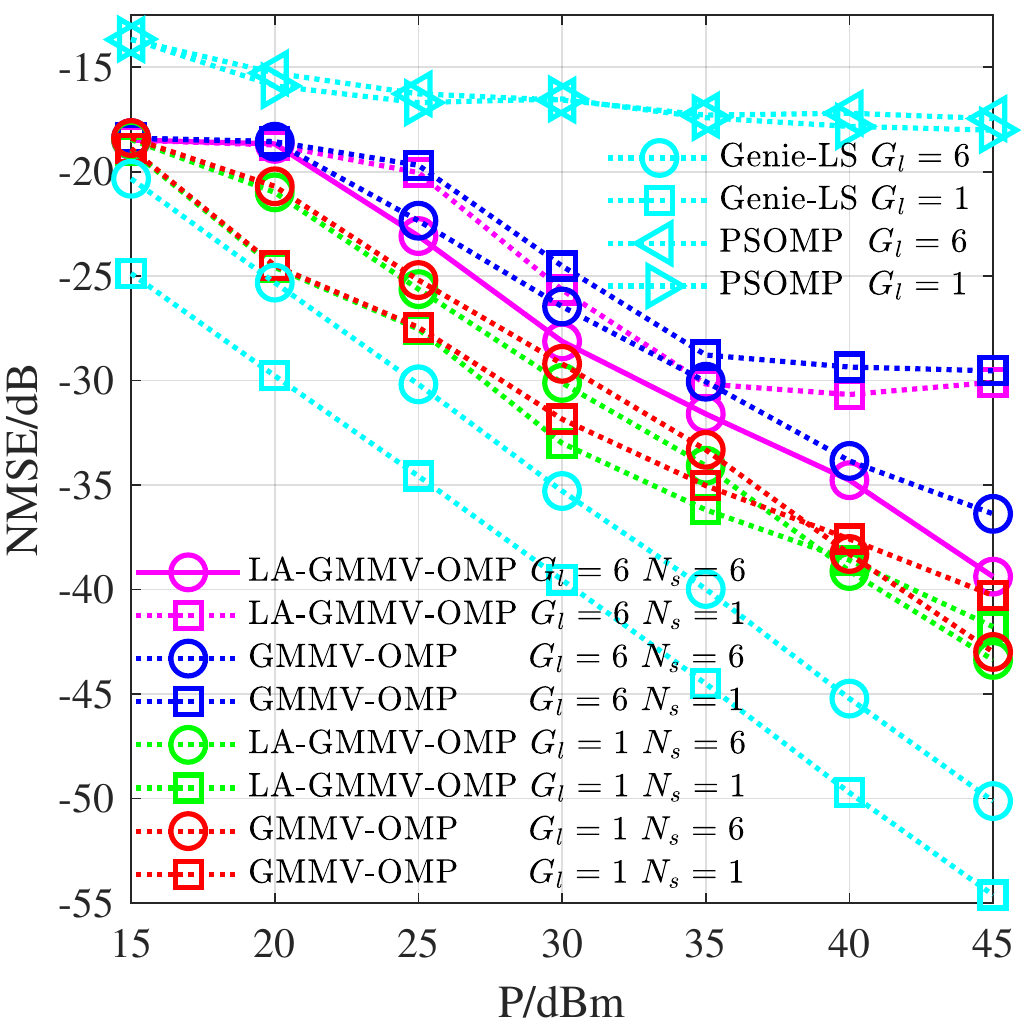}
			\label{fig_CE_BSUE_far}
		\end{minipage}%
	}%
	\subfigure[$\mathbf{h}^{\text{RU}}$ is the near-field channel.]{
		\begin{minipage}[t]{0.25\linewidth}
			\includegraphics[width=1.6in]{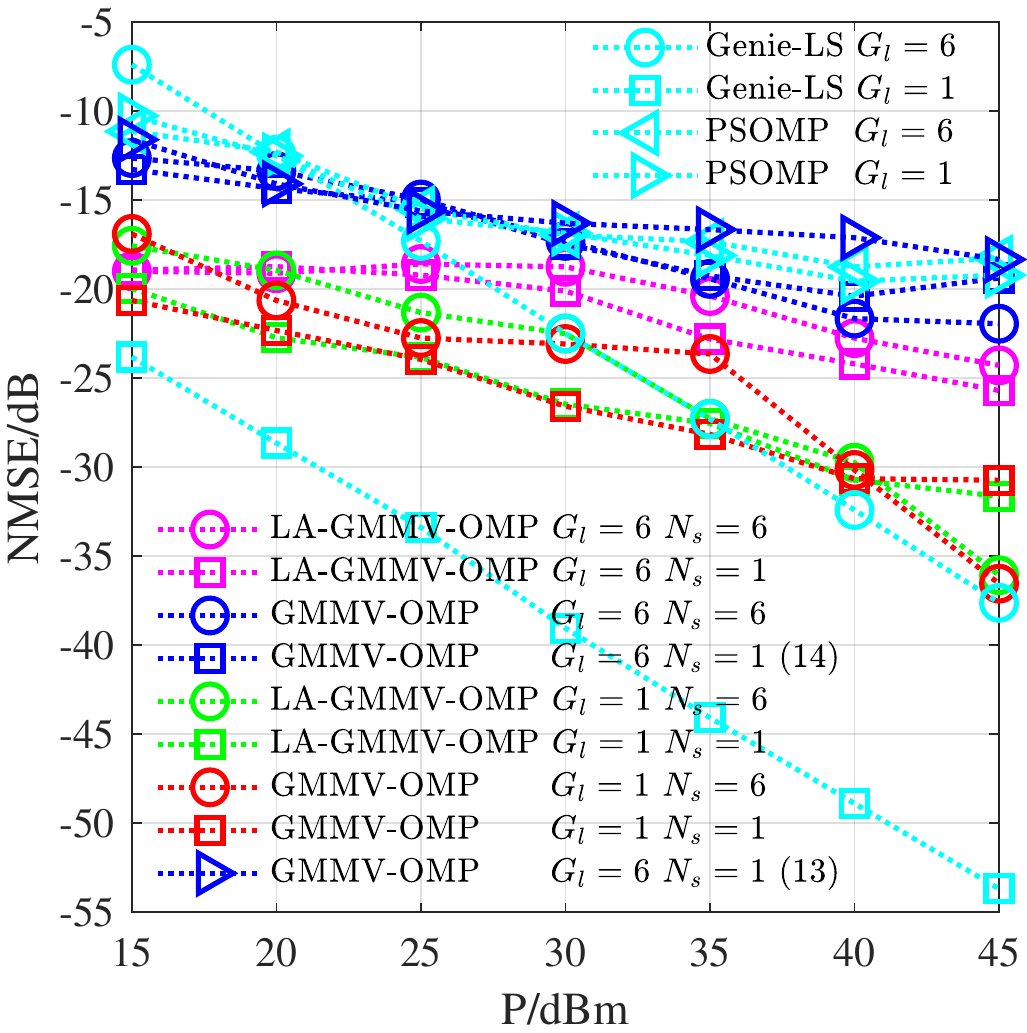}
			\label{fig_CE_RISUE_near}
		\end{minipage}%
	}%
	\subfigure[$\mathbf{h}^{\text{RU}}$ is the far-field channel.]{
		\begin{minipage}[t]{0.25\linewidth}
			\includegraphics[width=1.6in]{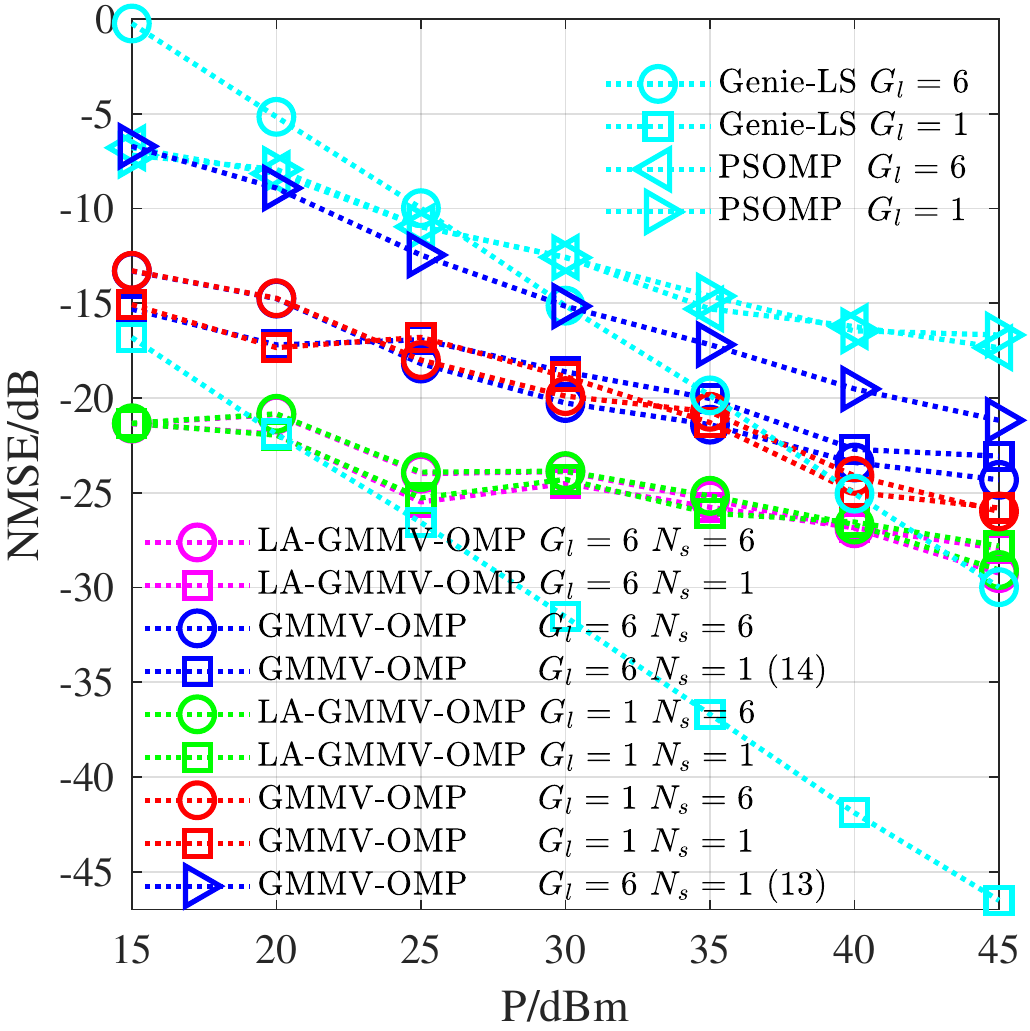}
			\label{fig_CE_RISUE_far}
		\end{minipage}%
	}%
	\centering
	\vspace{-3mm}	
	\caption{The CE performance of $\mathbf{h}^{\text{BU}}$is shown in (a) and (b) while that of $\mathbf{h}^{\text{RU}}$is shown in (c) and (d).}
	\label{fig_CE}
	\vspace{-3mm}
\end{figure*}

In the following simulation results, Genie-least square (LS), PSOMP \cite{2022_TCOM_LinglongDai_PolarDomain}, GMMV-OMP, LA-GMMV-OMP are simulated to compare the performance of CE in the near-field (Fig. \ref{fig_CE_BSUE_near}, Fig. \ref{fig_CE_RISUE_near}) and in the far-field (Fig. \ref{fig_CE_BSUE_far}, Fig. \ref{fig_CE_RISUE_far}) condition.
PSOMP and Genie-LS are the baseline algorithms and they are used to prove the effectiveness of GMMV-OMP under the HFNF BSE.
The comparison between the GMMV-OMP and LA-GMMV-OMP is to observe that whether the UE's location is beneficial to CE under the condition of this paper.

\begin{figure*}[h]
	\color{black}
	\flushleft
	\subfigure[RMSE$_\vartheta$ vs transmit power.]{
		\begin{minipage}[t]{0.25\linewidth}
			\vspace{-3mm}
			\flushleft
			\includegraphics[width=1.6in]{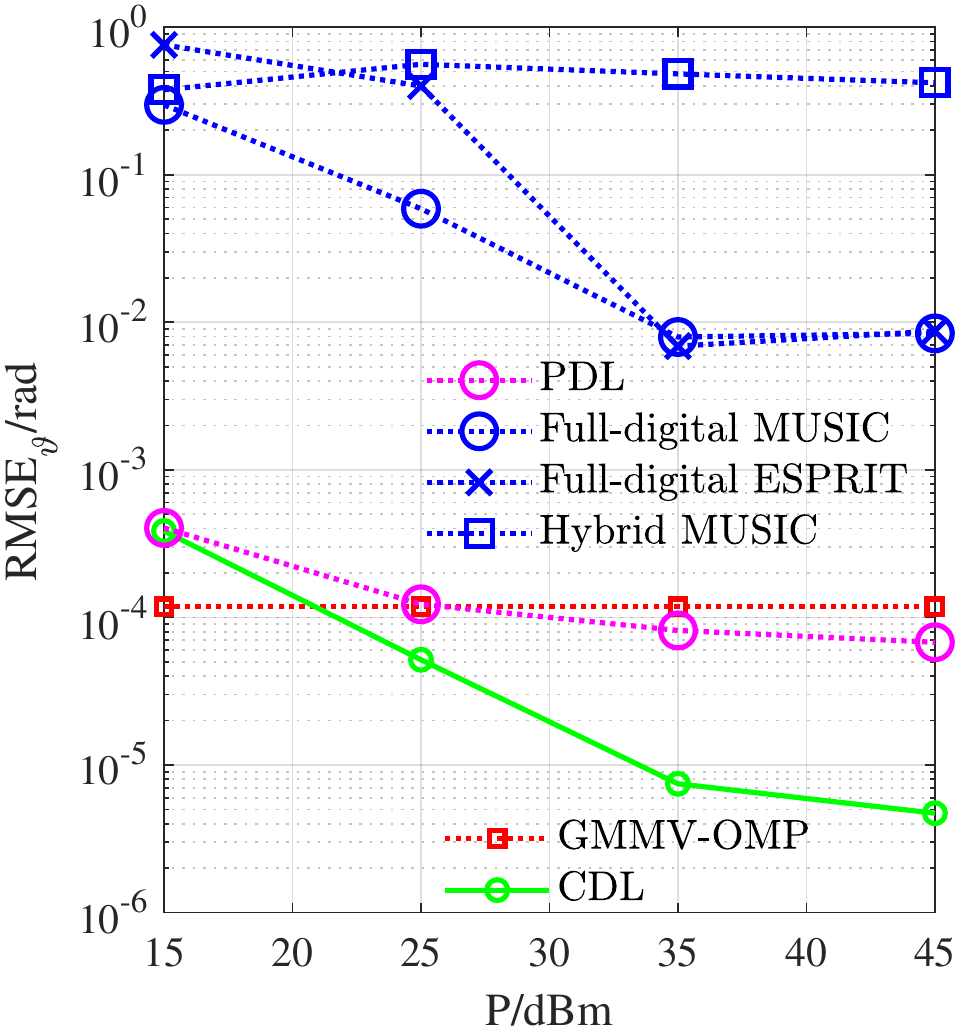}
			\label{fig_loc_RMSE_Pt_near_theta}
		\end{minipage}%
	}%
	\subfigure[RMSE$_r$ vs transmit power.]{
		\begin{minipage}[t]{0.25\linewidth}
			\vspace{-3mm}
			\flushleft
			\includegraphics[width=1.6in]{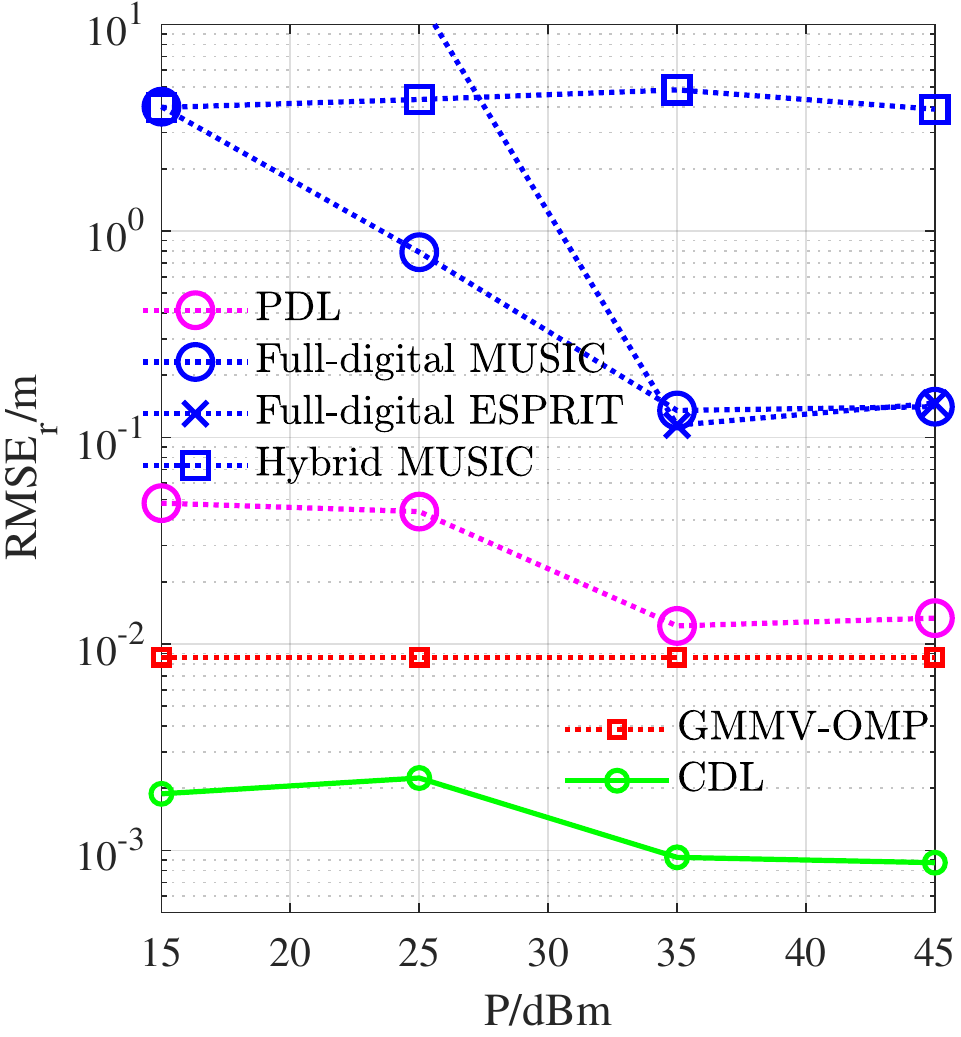}
			\label{fig_loc_RMSE_Pt_near_r}
		\end{minipage}%
	}%
	\subfigure[RMSE$_\vartheta$ vs transmit power.]{
		\begin{minipage}[t]{0.25\linewidth}
			\vspace{-3mm}
			\flushleft
			\includegraphics[width=1.6in]{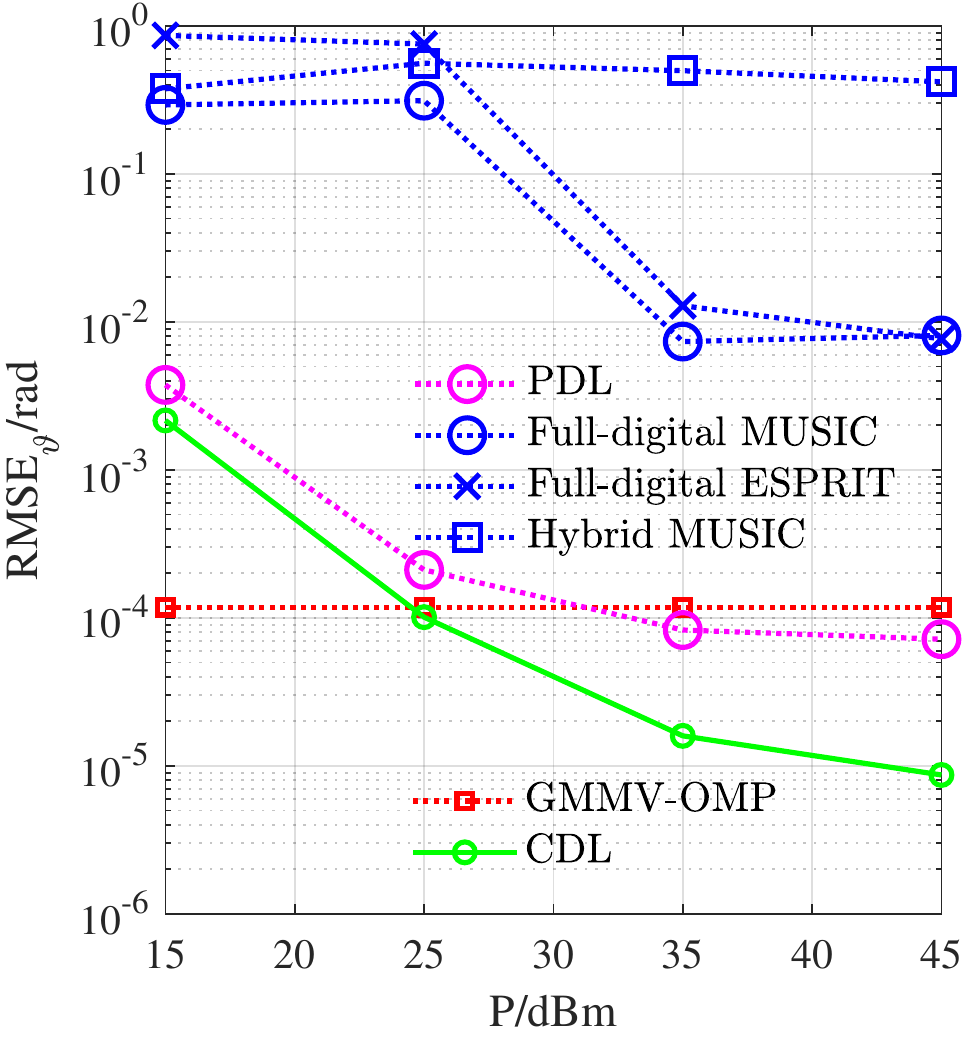}
			\label{fig_loc_RMSE_Pt_far_theta}
		\end{minipage}%
	}%
	\subfigure[RMSE$_r$ vs transmit power.]{
		\begin{minipage}[t]{0.25\linewidth}
			\vspace{-3mm}
			\flushleft
			\includegraphics[width=1.6in]{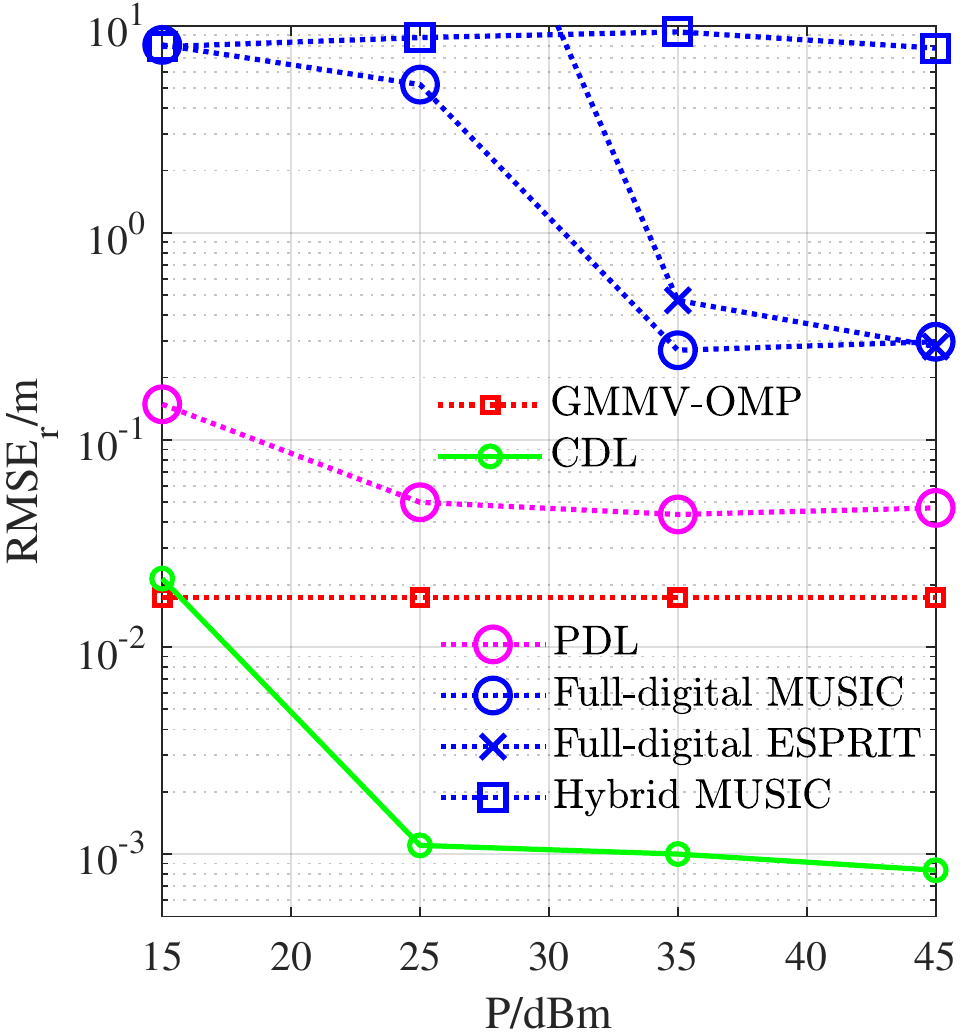}
			\label{fig_loc_RMSE_Pt_far_r}
		\end{minipage}%
	}%
	\centering	
	\vspace{-3mm}
	\caption{RMSE performance of $\vartheta$ and $r$ versus the transmit power with the BSE: the channel of (a) and (b) is in the near-field region, while that of (c) and (d) is in the far-field region.}
	\label{fig_loc_RMSE_Pt}
		\vspace{-4mm}
\end{figure*}

\begin{figure}[!h]
	\centering
	\color{black}
	\subfigure[$\text{RMSE}_{\vartheta}$ vs transmit power.]{
		\begin{minipage}[t]{0.5\linewidth}
			\vspace{-3mm}
			\flushleft			\includegraphics[width=1.6in]{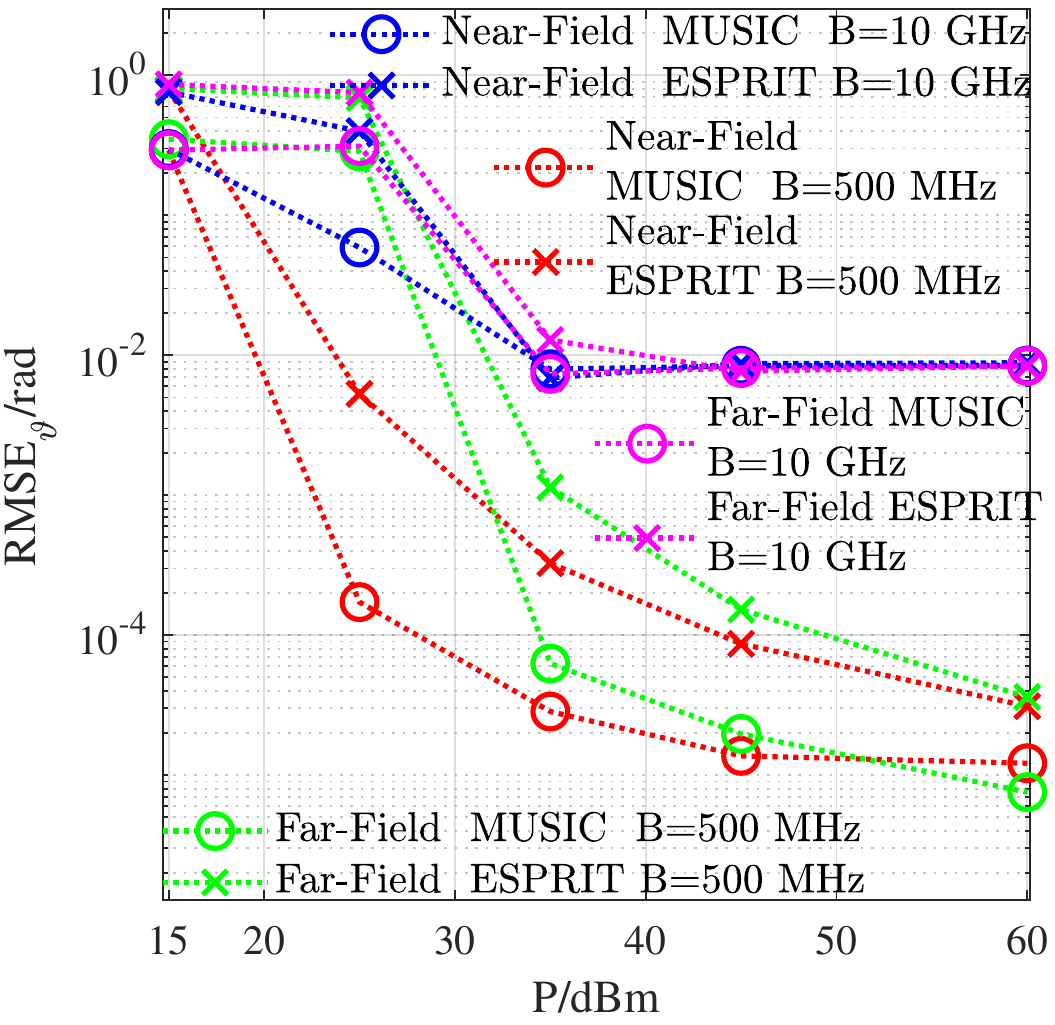}
			\label{fig_loc_Pt_NoBSE_Theta}
		\end{minipage}%
	}%
	\subfigure[ $\text{RMSE}_{r}$ vs transmit power.]{
		\begin{minipage}[t]{0.5\linewidth}
			\vspace{-3mm}
			\flushleft			\includegraphics[width=1.6in]{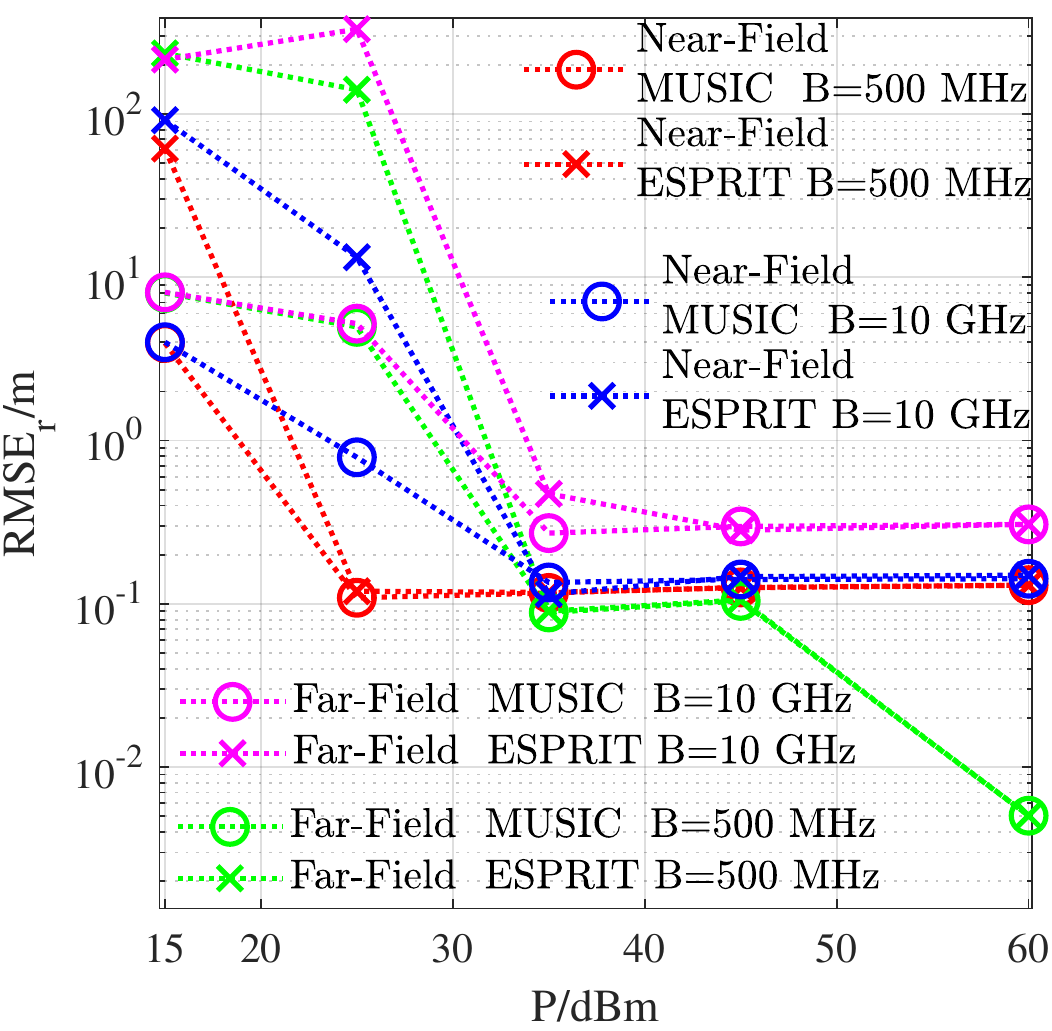}
			\label{fig_loc_Pt_NoBSE_Range}
		\end{minipage}%
	}%
	\centering	
	\vspace{-3mm}
	\caption{The localization performance of the MUSIC and ESPRIT algorithms with or without BSE in the HFNF versus the transmit power.}
	\label{fig_loc_Pt_NoBSE}
	\vspace{-4mm}
\end{figure}

MUSIC \cite{2022_JSAC_RISaidedSensing} based on full-digital beamforming, MUSIC based on hybrid beamforming \cite{ref_HybridMUSIC}(to illustrate that the conventional MUSIC algorithm is difficult to work stably in the hybrid-beamforming structure), ESPRIT \cite{2021_JSAC_AnwenLiao} based on full-digital beamforming are the baseline algorithms to prove the effectiveness of the PDL and CDL scheme under the HFNF BSE in Fig. \ref{fig_loc_RMSE_Pt} \ref{fig_loc_Pt_NoBSE} and \ref{fig_loc_Base_NumP}.
It is worth noting that although ESPRIT and MUSIC-based algorithms are used to estimate the $\theta^{\text{BU}}$, their methods to estimate the TDoA before estimating the $\theta^{\text{BU}}$ are the same as PDL.
Moreover, the on-grid localization result of GMMV-OMP in step \ref{step_OMP_3} of the Algorithm \ref{alg_PgOMP-MMV_new} is also shown.
\textcolor{black}{
The effects of several parameters on the localization performance of CDL schemes and PDL schemes are also investigated in Fig. \ref{fig_loc_Base_NumP}, \ref{fig_loc_UEDistance_NumRIS}, and \ref{fig_loc_NumBS_Bandwidth}, from which we derive some interesting insights.}

\subsection{Sensing Results of UE's Channel}
In Fig. \ref{fig_CE_BSUE_near} and Fig. \ref{fig_CE_BSUE_far}, different algorithms are implemented to compare their CE performance of $\bf{h}^{\text{BU}}$ under the HFNF BSE.
\textcolor{black}{
The performance of Genie-LS is the best since the support set in it is fully correct, so it can be a lower bound for the proposed algorithms.}
The performance of PSOMP is the worst, because the dictionary used in PSOMP is PTM.
Although the near-field is taken into consideration in PTM, the BSE is not yet settled.
The use of FSPRD in GMMV-OMP not only takes the near-field into consideration, but also solves the beam squint problem.
Therefore, the performance of GMMV-OMP is better than that of PSOMP.
The performance of LA-GMMV-OMP is better than that of GMMV-OMP since the former is assisted by the relatively accurate UE location estimation from the CDL scheme, which illustrates that the information of the UE's location can improve the performance of the CE.
Moreover, the CE performance of the proposed algorithms show almost no performance difference between the near-field and the far-field except for the difference brought by the received SNR, which indicates that the proposed algorithm can work stably in the HFNF.

\textcolor{black}{
As for the value of $G_l$, if $G_l\ne 1$, i.e., the channel has a cluster structure, all CE algorithms will degrade.
The performance of algorithms other than the Genie-LS algorithm degrades because accurate support set is more difficult to obtain when the channel has a cluster structure.
For Genie-LS, the angles and distances in the support set are too close to each other, so the pseudo-inverse operation performs poorly.
As a result, the CE performance of Genie-LS deteriorates.
As for the value of $N_s$, if $N_s=6$ and $G_l=6$ (we choose $N_s=6$ since every cluster we generated has 6 paths\footnote{How to determine the specific value of $N_s$ in the implementation of the algorithm is not a trivial task, and how to determine it in the absence of a prior on the channel clusters needs further study in the future.}), we can relatively correctly select the atoms corresponding to the scatterer with the largest energy in the current iteration.
However, if only one atom with the highest energy is selected each time, i.e., $N_s=1$, the energy of the other atoms with the same angle and distance will be weakened in future iterations, since this path has already been weakened from the residual after each iteration of the OMP-based algorithm.
These atoms with reduced energy are difficult to pick out correctly in the future iterations and hence the CE performance degrades.
However, if $N_s$ is 6, its CE performance is worse than that of 1 at low transmit power but better than that of 1 at high transmit power under the condition that the channel does not have a cluster structure ($G_l=1$).
This is because selecting multiple atoms simultaneously is more likely to select the wrong atoms at low transmit power, while more likely to select the right atoms at the high transmit power.
Moreover, when $N_s > 1$, the algorithm has fewer adaptive iterations and faster running speed.}

In Fig. \ref{fig_CE_RISUE_near} and Fig. \ref{fig_CE_RISUE_far}, different algorithms are implemented to compare their CE performance of $\bf{h}^{\text{RU}}$ under the HFNF BSE.
The estimation results of $\bf{h}^{\text{RU}}$ are generally consistent with those of $\bf{h}^{\text{BU}}$, except three main differences.
\textcolor{black}{
Firstly, if $\bf{h}^{\text{RU}}$ has a cluster structure, the performance of the Genie-LS will become so poor that proposed algorithms can outperform this so-called lower bound.
On the one hand, it suffers the same drawback as that of $\bf{h}^{\text{BU}}$ since the poor performance of the pseudo-inverse in LS.
On the other hand, the rank of $\mathbf{W}^{\text{RIS}}$ and $\mathbf{H}^{\text{BR}}[m]$ is almost one so that the condition number of ${\mathbf{\bar W}}^{{\text{RIS}}}[m]$ is very large.
The property of sensing matrix ${\mathbf{\bar W}}^{{\text{RIS}}}[m]$ used to estimate $\bf{h}^{\text{RU}}$ is much poorer than that of sensing matrix ${\mathbf{\bar W}}^{{\text{NRIS}}}$ used to estimate $\bf{h}^{\text{BU}}$.
Thus, even if the algorithms other than Genie-LS adaptively stop the iteration without finding all the correct indices, their CE performance is better than that of Genie-LS with all the correct indices, demonstrating the superiority of the proposed algorithms and their adaptive iteration stopping criterion.
}
Secondly, due to the poor properties of the sensing matrix ${\mathbf{\bar W}}^{{\text{RIS}}}[m]$ in the estimation of $\bf{h}^{\text{RU}}$, and the small energy of the NLoS paths, even the channel cluster structure is overwhelmed.
Therefore, there is no obvious advantage or disadvantage to choosing multiple atoms simultaneously as opposed to choosing only one atom.
Thirdly, if the near-field channel $\bf{h}^{\text{RU}}$ does not have cluster structure, the estimation of the UE location does not give a clear benefit to the CE performance, since the atom selected by the GMMV-OMP is very close to the actual location of the UE under our simulation settings.
Moreover, it is better to design the combiner using (\ref{eq_W_allFreq}) than (\ref{eq_W_centralFreq}), since the energy in different frequency is more balanced when (\ref{eq_W_allFreq}) is adopted.
If (\ref{eq_W_centralFreq}) is adopted, the energy in the center frequency is higher while the energy of the marginal frequency is lower.

In conclusion, GMMV-OMP performs much better than PSOMP, whether in the near-field region or in the far-field region, whether through the RIS or not.
Compared with GMMV-OMP, LA-GMMV-OMP has about 2 dB NMSE performance gain, which demonstrates that the information of the UE's location has some benefits to the CE.
When estimating channels with cluster structure, the value of $N_s$ can be larger to obtain better CE performance and lower iteration numbers.
Moreover, the CE performance of $\bf{h}^{\text{RU}}$ is worse than that of $\bf{h}^{\text{BU}}$, since the property of the sensing matrix of the former is much poorer than that of the latter and the received SNR of the former is much lower than that of the latter.
\subsection{Sensing Results of UE's Location}\label{sec_SimRltLoc}
Localization accuracy of different algorithms are evaluated in terms of AoA and distance, namely $\theta^{\text{BU}}_0$ and $r^{\text{BU}}_0$, under HFNF BSE.
\subsubsection{\textbf{Localization Performance of Different Algorithms}}
$\quad$

Fig. \ref{fig_loc_RMSE_Pt} shows the variation of the localization performance of different algorithms with the transmit power.
As for Fig. \ref{fig_loc_RMSE_Pt_near_theta} and Fig. \ref{fig_loc_RMSE_Pt_far_theta}, the CDL scheme has the highest AoA localization accuracy followed by the PDL scheme.
The use of redundant dictionaries allows the on-grid GMMV-OMP algorithm to achieve good AoA estimation accuracy, but once the index of the FSPRD at the location of the UE is accurately found, the AoA localization accuracy reaches the highest.
Even if full-digital beamforming is adopted, the localization accuracy of MUSIC and ESPRIT algorithms is still inferior to the proposed PDL and CDL scheme with hybrid beamforming.
This is because MUSIC and ESPRIT algorithms are suitable for the far-field region without the BSE.
However, in near-field \textcolor{black}{XL-array} systems, the steering vector is not only dependent on the AoA from the UE to the BS, but also dependent on the distance from the UE to the BS.
Due to the energy spread effect in \cite{2022_TCOM_LinglongDai_PolarDomain}, one near-field steering vector corresponds to multiple far-field steering vectors, breaking the far-field premise of MUSIC and ESPRIT algorithms.
What's more, the HFNF BSE leads to the shift of the angles and distances on different subcarriers.
Hence, the severe HFNF BSE will reduce the localization accuracy.
When the hybrid beamforming is adopted, the MUSIC algorithm has poor performance and the ESPRIT algorithm cannot work at all due to the incomplete rotation invariance.
As for Fig. \ref{fig_loc_RMSE_Pt_near_r} and Fig. \ref{fig_loc_RMSE_Pt_far_r}, the CDL scheme has the highest distance localization accuracy, but the second highest distance localization accuracy is achieved by the GMMV-OMP algorithm, not the PDL scheme.
This illustrates that localization through two AoAs is more accurate than localization through one AoA and one TDoA, where the BS and the RIS are the two anchors.
Comparing the AoA localization accuracy and the distance localization accuracy at high transmit power, we can find that distance localization accuracy of the PDL scheme is limited by TDoA estimation, because the AoA localization accuracy of the PDL scheme is higher than that of the GMMV-OMP algorithm, but the distance localization accuracy of the former is lower than that of the latter.

Fig. \ref{fig_loc_Pt_NoBSE} shows the localization performance of the baselines with and without BSE in the HFNF\footnote{\label{footnote_bandwidth10GHz}In the simulations, for the sake of fairness, the noise power at different bandwidths is set as the noise power at 10 GHz, thus, the performance of the proposed schemes is evaluated solely in terms of the bandwidth impact (other than SNR).}.
It can be seen from Fig. \ref{fig_loc_RMSE_Pt} that the localization performance of the conventional MUSIC and ESPRIT algorithms, which only work well in the far-field narrowband scenarios without the BSE, is much poorer than the proposed schemes and the reason is shown in Fig. \ref{fig_loc_Pt_NoBSE}.
The BSE is the main factor, as can be seen from the clear performance improvement of the two conventional algorithms at a bandwidth of 500 MHz, where the BSE is no obvious.
The near-field spherical wave is the secondary factor since the received SNR in the far-field is lower than that in the near-field at the same transmit power.
However, when the transmit power is high, it can be seen that the performance of the far-field reaches or even exceeds that of the near-field one at 500 MHz, which proves that the near-field spherical wave has a negative effect on the performance of these conventional algorithms.

\begin{figure*}[!t]
	\centering
	\color{black}
	\subfigure[CDF of the $\text{RMSE}_{\vartheta}$.]{
	\begin{minipage}[t]{0.25\linewidth}
			\vspace{-3mm}
			\flushleft			\includegraphics[width=1.6in]{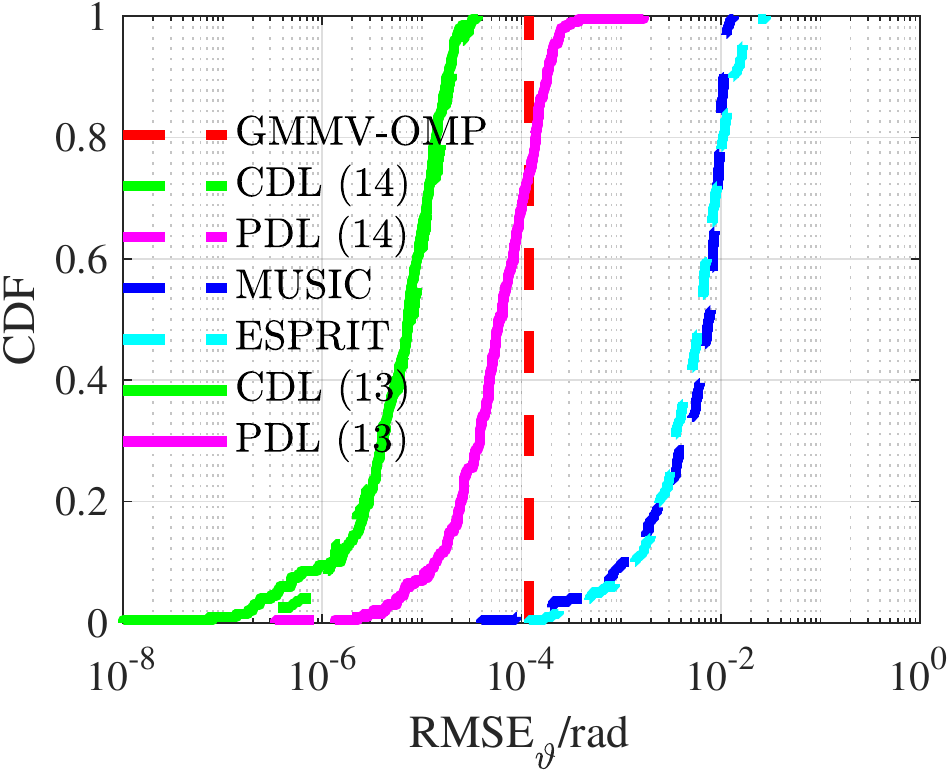}
			\label{fig_loc_Base_theta}
		\end{minipage}%
	}%
	\subfigure[CDF of the $\text{RMSE}_{r}$.]{
		\begin{minipage}[t]{0.25\linewidth}
			\vspace{-3mm}
			\flushleft			\includegraphics[width=1.6in]{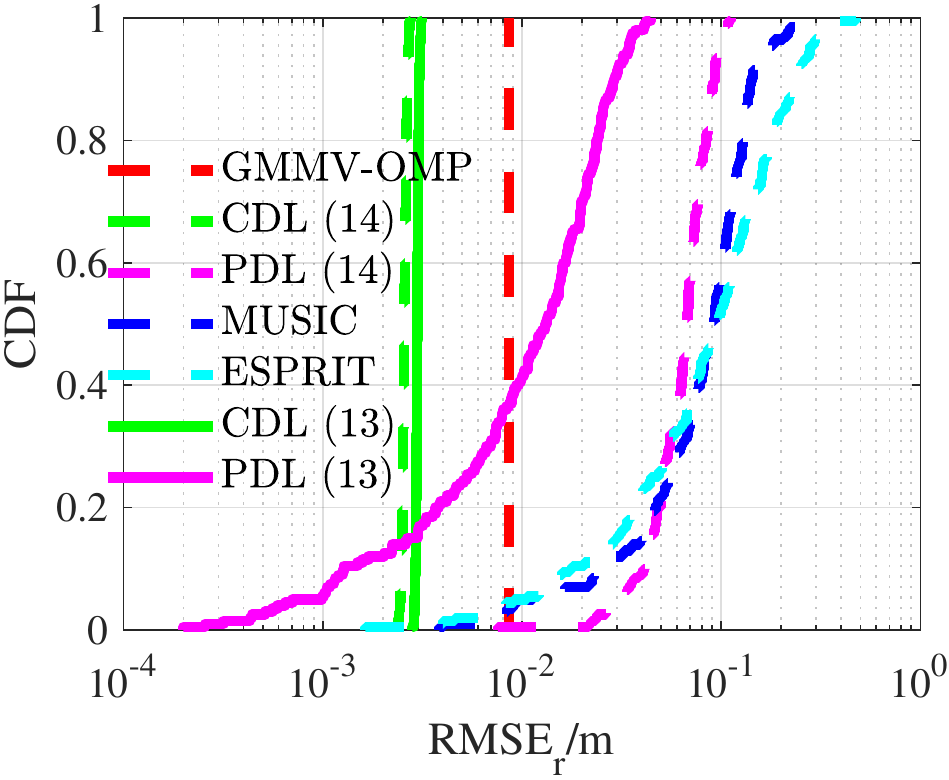}
			\label{fig_loc_Base_r}
		\end{minipage}%
	}%
	\subfigure[CDF of the $\text{RMSE}_{\vartheta}$.]{
		\begin{minipage}[t]{0.25\linewidth}
			\vspace{-3mm}
			\flushleft
			\includegraphics[width=1.6in]{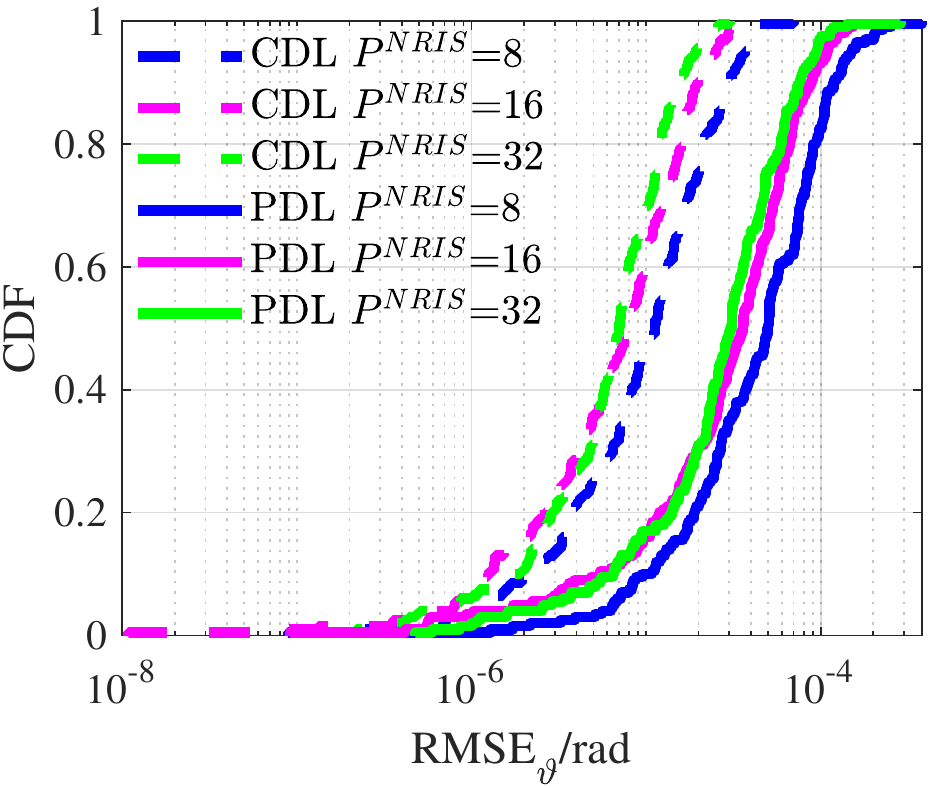}
			\label{fig_loc_NumP_theta}
		\end{minipage}%
	}%
	\subfigure[CDF of the $\text{RMSE}_{r}$.]{
		\begin{minipage}[t]{0.25\linewidth}
			\vspace{-3mm}
			\flushleft
			\includegraphics[width=1.6in]{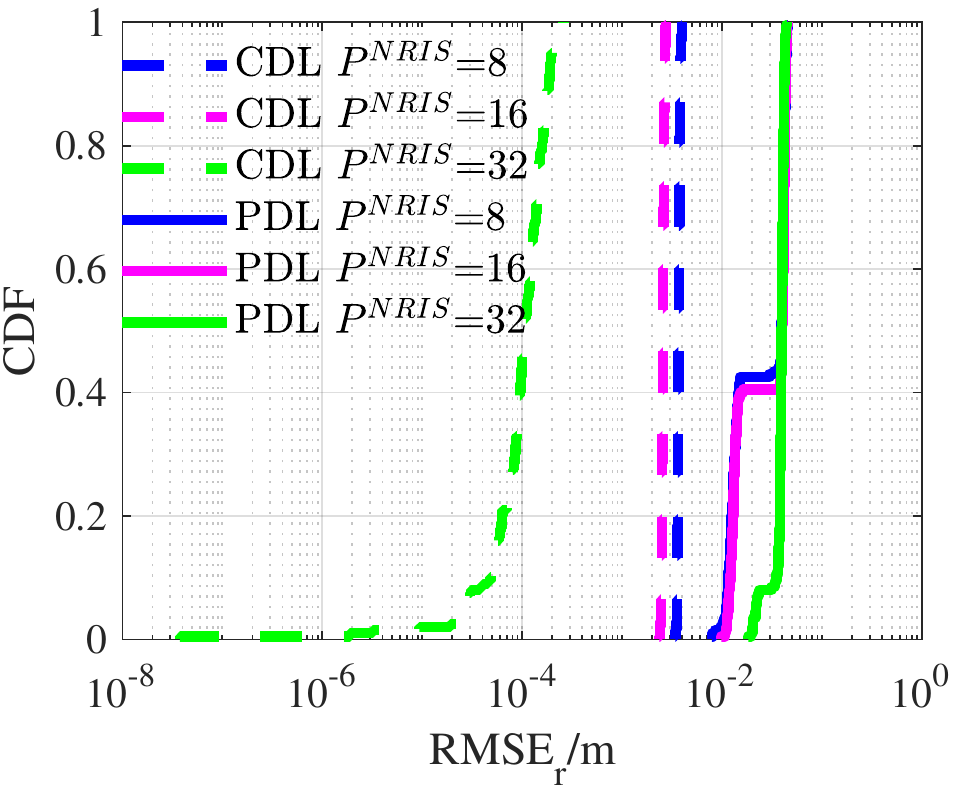}
			\label{fig_loc_NumP_r}
		\end{minipage}%
	}%
	\centering	
	\vspace{-3mm}
	\caption{(a) and (b) show the localization accuracy of different algorithms. (c) and (d)  reflect the effect of the number of the uplink time slots on the localization accuracy.
	Except for the parameters in the legend, the values of other parameters follow the default values in Section \ref{sec_Setup}.
	}
	\label{fig_loc_Base_NumP}
\end{figure*}

Fig. \ref{fig_loc_Base_theta} and \ref{fig_loc_Base_r} show the cumulative distribution function (CDF) of the AoA localization RMSE and the distance localization RMSE for different algorithms.
The results coincide with those in Fig. \ref{fig_loc_RMSE_Pt}.
It is worth noting that no matter whether the combiner is designed according to (\ref{eq_W_centralFreq}) or (\ref{eq_W_allFreq}), it will not affect the localization performance of CDL.
However, for the PDL scheme, using (\ref{eq_W_centralFreq}) can achieve better distance localization accuracy.
The reason is that only the phase difference among different subcarriers of the channel is considered in the PDL scheme, but the phase difference influenced by the sensing matrix is not considered. 

\begin{figure*}[!t]
	\centering
	\color{black}
	\subfigure[CDF of the $\text{RMSE}_{\vartheta}$.]{
		\begin{minipage}[t]{0.25\linewidth}
			\vspace{-3mm}
			\flushleft
			\includegraphics[width=1.6in]{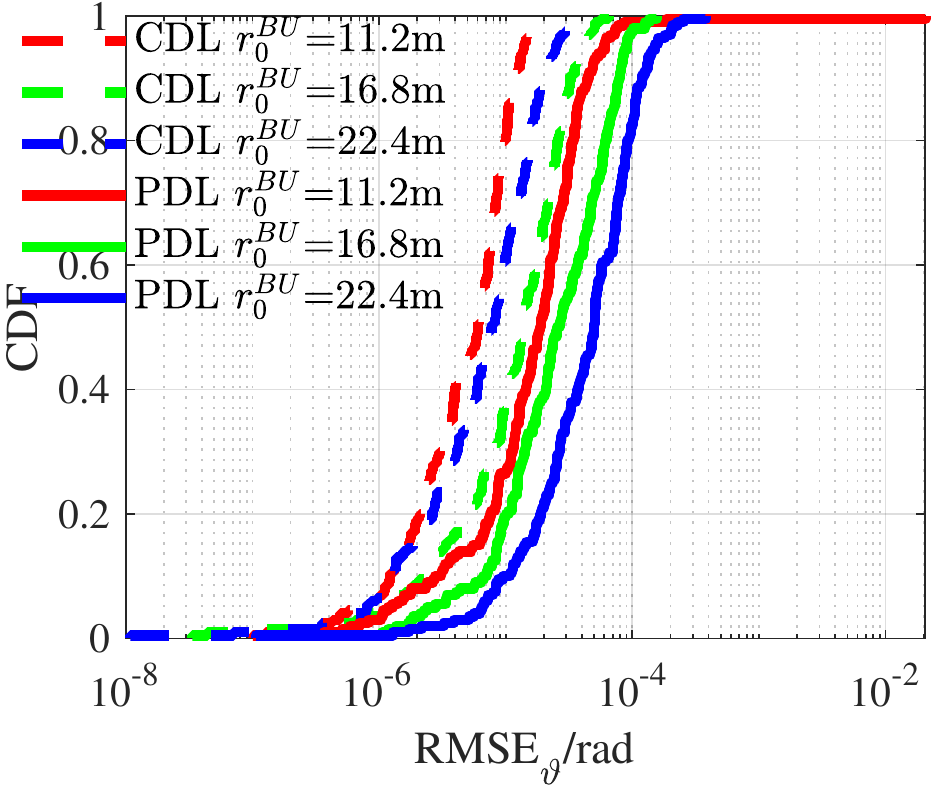}
			\label{fig_loc_UEDistance_theta}
		\end{minipage}%
	}%
	\subfigure[CDF of the $\text{RMSE}_{r}$.]{
		\begin{minipage}[t]{0.25\linewidth}
			\vspace{-3mm}
			\flushleft
			\includegraphics[width=1.6in]{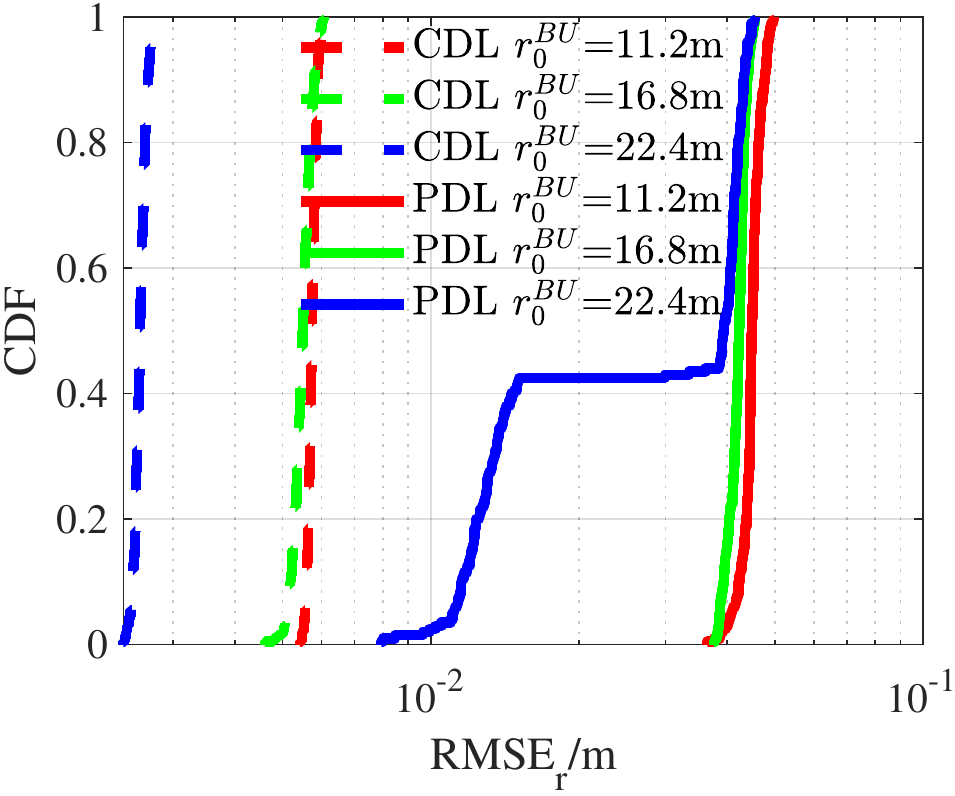}
			\label{fig_loc_UEDistance_r}
		\end{minipage}%
	}%
	\subfigure[CDF of the $\text{RMSE}_{\vartheta}$.]{
		\begin{minipage}[t]{0.25\linewidth}
			\vspace{-3mm}
			\flushleft
			\includegraphics[width=1.6in]{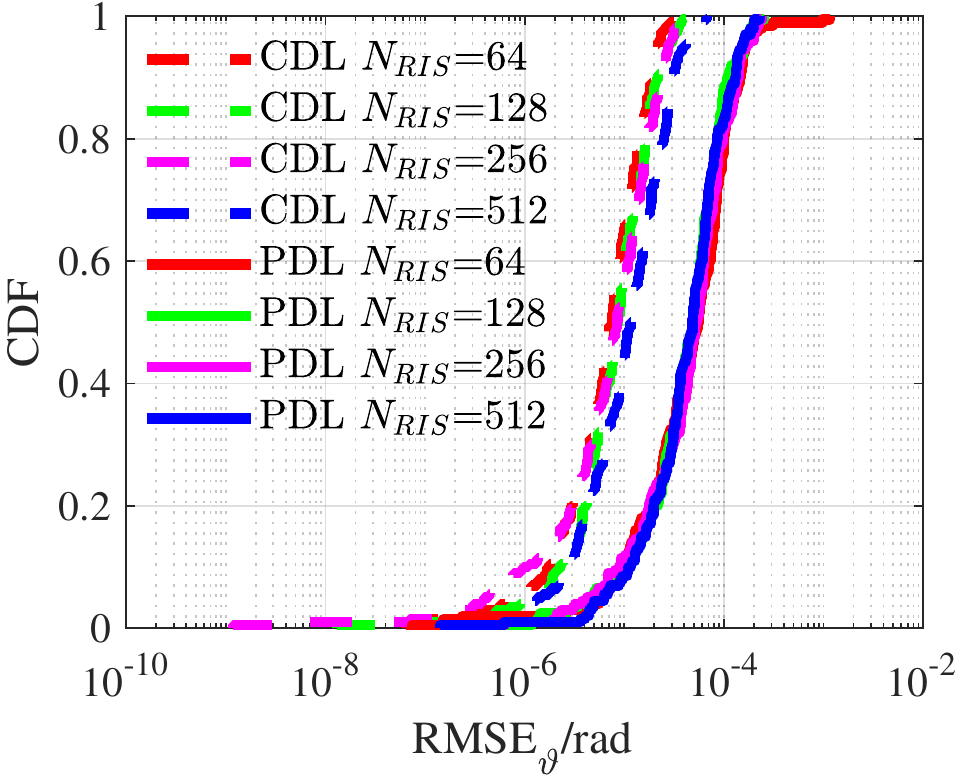}
			\label{fig_loc_NumRIS_theta}
		\end{minipage}%
	}%
	\subfigure[CDF of the $\text{RMSE}_{r}$.]{
		\begin{minipage}[t]{0.25\linewidth}
			\vspace{-3mm}
			\flushleft
			\includegraphics[width=1.6in]{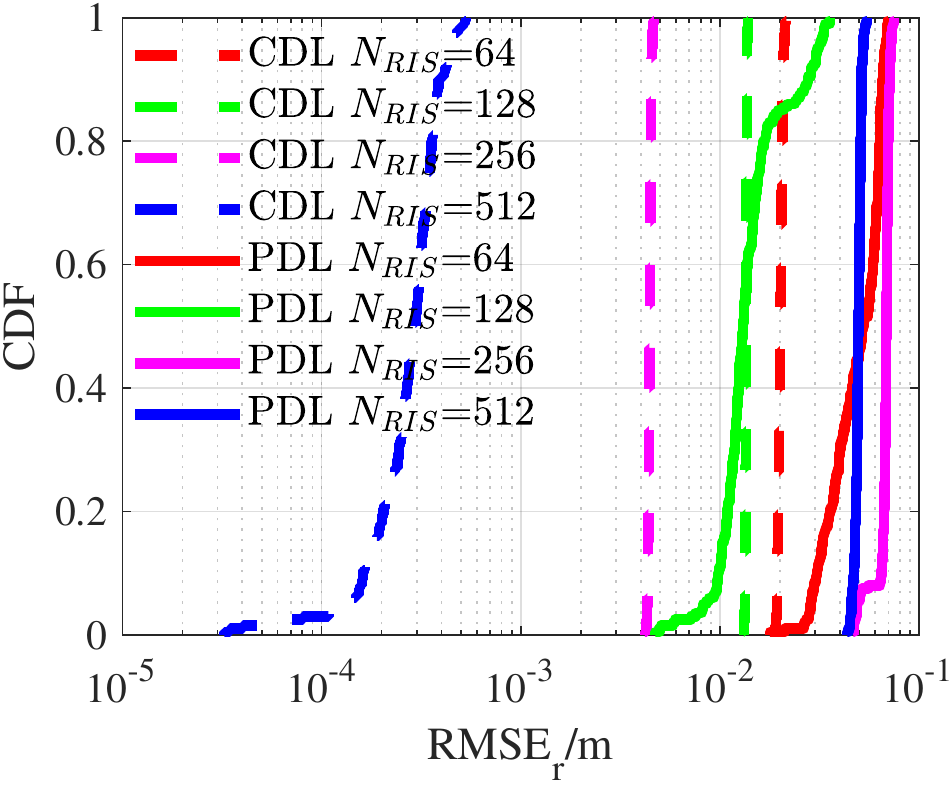}
			\label{fig_loc_NumRIS_r}
		\end{minipage}%
	}%
	\centering	
	\vspace{-3mm}
	\caption{(a) and (b) reflect the effect of the distance from the UE to the BS on the localization accuracy. (c) and (d)  reflect the effect of the number of the RIS elements on the localization accuracy.
	Except for the parameters in the legend, the values of other parameters follow the default values in Section \ref{sec_Setup}.}
	\label{fig_loc_UEDistance_NumRIS}
\end{figure*}
\begin{figure*}[!t]
	\centering
	\color{black}
	\subfigure[CDF of the $\text{RMSE}_{\vartheta}$.]{
		\begin{minipage}[t]{0.25\linewidth}
			\vspace{-3mm}
			\flushleft
			\includegraphics[width=1.6in]{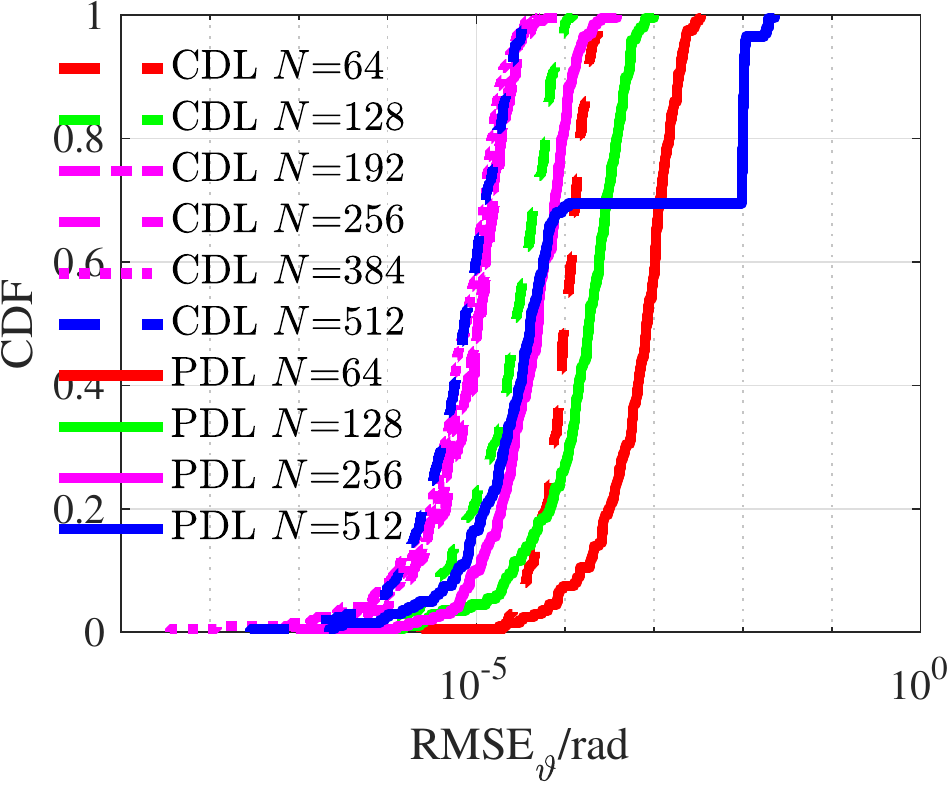}
			\label{fig_loc_NumBS_theta}
		\end{minipage}%
	}%
	\subfigure[CDF of the $\text{RMSE}_{r}$.]{
		\begin{minipage}[t]{0.25\linewidth}
			\vspace{-3mm}
			\flushleft
			\includegraphics[width=1.6in]{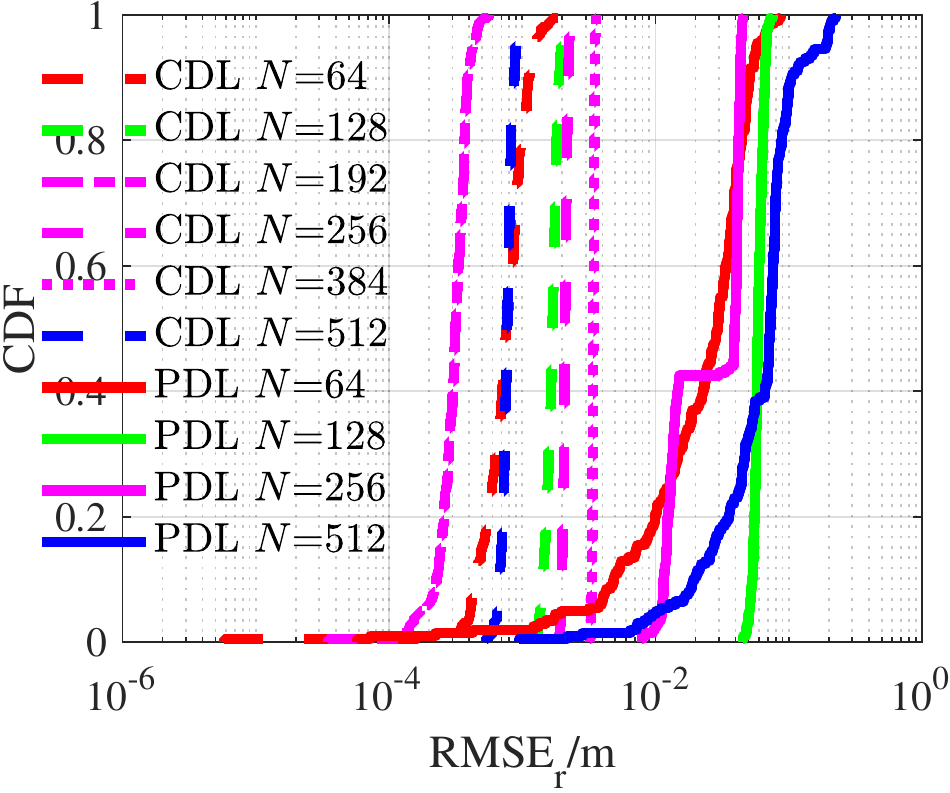}
			\label{fig_loc_NumBS_r}
		\end{minipage}%
	}%
	\subfigure[CDF of the $\text{RMSE}_{\vartheta}$.]{
		\begin{minipage}[t]{0.25\linewidth}
			\vspace{-3mm}
			\flushleft
			\includegraphics[width=1.6in]{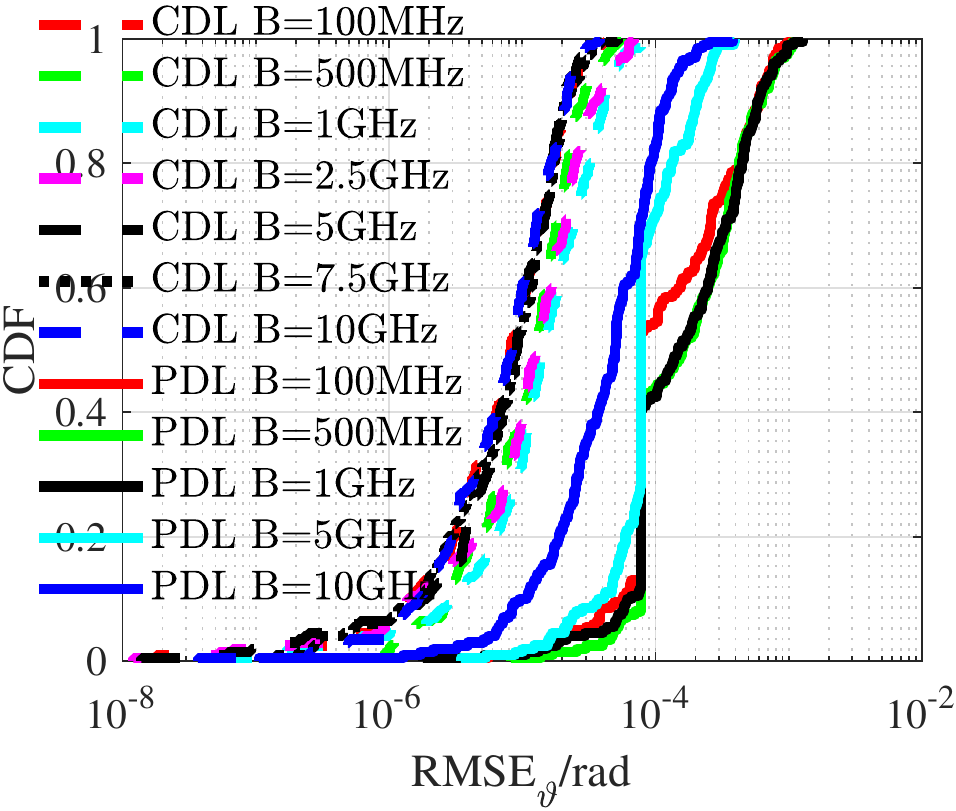}
			\label{fig_loc_Bandwidth_theta}
		\end{minipage}%
	}%
	\subfigure[CDF of the $\text{RMSE}_{r}$.]{
		\begin{minipage}[t]{0.25\linewidth}
			\vspace{-3mm}
			\flushleft
			\includegraphics[width=1.6in]{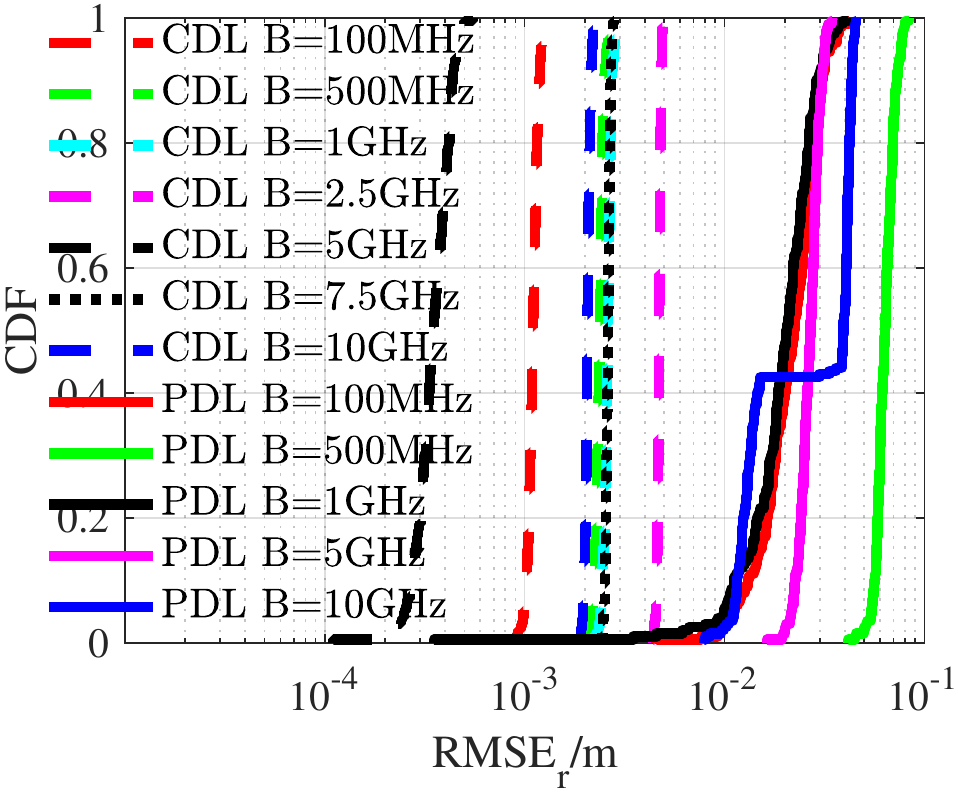}
			\label{fig_loc_Bandwidth_r}
		\end{minipage}%
	}%
	\centering	
	\vspace{-3mm}
	\caption{(a) and (b) reflect the effect of the number of BS antennas on the localization accuracy. (c) and (d)  reflect the effect of the bandwidth on the localization accuracy.
	Except for the parameters in the legend, the values of other parameters follow the default values in Section \ref{sec_Setup}.}
		\label{fig_loc_NumBS_Bandwidth}
		\vspace{-3mm}
\end{figure*}

\subsubsection{\textbf{Effects of Different Parameters on the Localization Performance of CDL and PDL Schemes}}
$\quad$

{\color{black}
Fig. \ref{fig_loc_NumP_theta} and \ref{fig_loc_NumP_r} show the CDF of RMSE$_\vartheta$ and RMSE$_r$ for different number of time slots.
The value of $P^{\text{NRIS}}$ are \{8, 16, 32\} and the corresponding value of $P^{\text{RIS}}$ are \{16, 32, 64\}.
The localization accuracy of the CDL scheme and the AoA localization accuracy of the PDL scheme increase with the number of observed time slots, but the distance localization performance of the PDL scheme does not change significantly with the increase of the number of observed time slots, which means that the delay estimation performance of the PDL scheme has reached saturation when the number of observed time slots is small.
This is because the size of the RIS elements and the size of the BS antenna array limit the further improvement of the PDL scheme's TDoA accuracy under underdetermined observation in the hybrid-beamforming structure.

Fig. \ref{fig_loc_UEDistance_theta} and \ref{fig_loc_UEDistance_r} show the CDF of RMSE$_\vartheta$ and RMSE$_r$ for different distances from the UE to the BS, i.e., $r_0^{\text{BU}}$, where $\theta_0^{\text{BU}}$ remains constant.
The difference in localization performance caused by different distances mainly comes from the variation of the received power, since the HFNF BSE has been overcome in the proposed localization schemes.
At the same transmit power, the farther away the UE is from the BS, the smaller SNR of ${\mathbf{Y}}^{{\text{NRIS}}}[m]$.
However, under our simulation settings, the farther away the UE from the BS means the closer the UE to the RIS, i.e., the larger SNR of ${\mathbf{Y}}^{{\text{RIS}}}[m]$.
Under the influence of these two opposing factors, the AoA localization accuracy of the CDL scheme decreases first and then increases, while the distance localization accuracy keeps increasing.
For the PDL scheme, the AoA localization accuracy keeps decreasing while the distance localization accuracy keeps increasing.


Fig. \ref{fig_loc_NumRIS_theta} and \ref{fig_loc_NumRIS_r} show the CDF of RMSE$_\vartheta$ and RMSE$_r$ for different number of the RIS elements $N_{\text{RIS}}$.
As $N_{\text{RIS}}$ increases, we can observe the following changes that affect the final localization performance.
1) The effective reflection area of the RIS will increase and hence the received SNR increases;
2) The AoA resolution from the UE to the RIS will increase;
3) The BS combiner is only aligned with the center of the RIS, so the phase difference between the rest of the elements at the RIS and each antenna at the BS becomes larger;
4) Increasing the RIS area leads to the larger delay estimation errors of the PDL scheme\footnote{\label{footnote_DistanceErr}The estimation error of ${\hat \tau}^{\text{TDoA}}$ acquired from the PDL scheme is related to the size of RIS and BS array since the signals of the UE arrive at different elements of these arrays have different delays.};
5) When the number of observed time slots remains constant, the reduction of the compression ratio leads to a deterioration of localization performance.
Based on the aforementioned analysis, as $N_{\text{RIS}}$ increases, the AoA localization accuracy of the CDL scheme remains constant, while the distance localization accuracy of the CDL scheme keeps increasing.
For the PDL scheme, the AoA localization accuracy remains constant, while the distance localization accuracy is non-monotonic.


Fig. \ref{fig_loc_NumBS_theta} and \ref{fig_loc_NumBS_r} show the CDF of RMSE$_\vartheta$ and RMSE$_r$ for different number of the BS array elements $N$.
As $N$ increases, we can observe the following changes that affect the final localization performance.
1) The AoA resolution from the UE to the BS will increase;
2) The BS combiner is only aligned with the center of the RIS, so the phase difference between the rest of the elements at the RIS and each antenna at the BS becomes larger;
3) Increasing the aperture of the BS antenna array leads to larger distance estimation errors of the PDL scheme\textsuperscript{\ref{footnote_DistanceErr}}.
Based on the aforementioned analysis, as $N$ increases, the AoA localization accuracy of the CDL scheme keeps increasing, while the distance localization accuracy of the CDL scheme is non-monotonic.
For the PDL scheme, both the AoA and distance localization accuracy are non-monotonic.

Fig. \ref{fig_loc_Bandwidth_theta} and \ref{fig_loc_Bandwidth_r} show the CDF of RMSE$_\vartheta$ and RMSE$_r$ for different size of bandwidth $B$\textsuperscript{\ref{footnote_bandwidth10GHz}}.
As $B$ increases, we can observe the following factors that affect the final localization performance.
1) The design of the BS combiner $\mathbf{W}^{\text{RIS}}[p]$ shown in (\ref{eq_W_allFreq}) is heuristic and may not be optimal;
2) The proposed PDL scheme cannot make the delay estimation free from the unknown HFNF steering vector under the large array and underdetermined observations.
Based on the aforementioned analysis, as $B$ increases, the AoA localization accuracy of the CDL scheme remains constant, while the distance localization accuracy of the CDL scheme is non-monotonic.
For the PDL scheme, both the AoA and distance localization accuracy are non-monotonic.
}


In conclusion, the CDL and PDL schemes are all superior to the baseline algorithms, both in the near-field and far-field.
At the same time, we tested the localization performance of the two schemes under different simulation conditions and found some interesting compromises.
Since the CDL scheme is closely linked with the (LA)-GMMV-OMP algorithm, thus, if there is a need for both CE and the UE's location, the CDL scheme can be adopted.
If there is only the need for the UE's location, the PDL scheme can be adopted.

\section{Conclusion}\label{sec:Conclusion}
\vspace{-1mm}
In this paper, we have investigated how to sense the UE’s channel and location under the hybrid BSE for THz \textcolor{black}{XL-array} systems.
First, a procedure for generating the FSPRD was proposed to model the dictionary of the HFNF steering vectors under the BSE on grids.
Then, we proposed two RIS-assisted localization paradigms depending on whether the UE's channel was required to be estimated when sensing the UE's location.
Finally, the two proposed schemes were tested under the hybrid BSE, and the simulation results showed that the proposed schemes outperformed the baseline algorithms in terms of the UE's location sensing performance and the UE's CE performance.
{\color{black}
Possible future research directions based on this work include more efficient design of the following three elements in the HFNF BSE scenario: the beam training procedure, the combiner at the BS and the reflection at the RIS.}

\bibliographystyle{IEEEtran}
\bibliography{ref}

\begin{thebibliography}{10}
\providecommand{\url}[1]{#1}
\csname url@samestyle\endcsname
\providecommand{\newblock}{\relax}
\providecommand{\bibinfo}[2]{#2}
\providecommand{\BIBentrySTDinterwordspacing}{\spaceskip=0pt\relax}
\providecommand{\BIBentryALTinterwordstretchfactor}{4}
\providecommand{\BIBentryALTinterwordspacing}{\spaceskip=\fontdimen2\font plus
\BIBentryALTinterwordstretchfactor\fontdimen3\font minus
  \fontdimen4\font\relax}
\providecommand{\BIBforeignlanguage}[2]{{%
\expandafter\ifx\csname l@#1\endcsname\relax
\typeout{** WARNING: IEEEtran.bst: No hyphenation pattern has been}%
\typeout{** loaded for the language `#1'. Using the pattern for}%
\typeout{** the default language instead.}%
\else
\language=\csname l@#1\endcsname
\fi
#2}}
\providecommand{\BIBdecl}{\relax}
\BIBdecl

\bibitem{2018_CST_localization}
J.~A. del Peral-Rosado \emph{et~al.}, ``{Survey of cellular mobile radio
  localization methods: From 1G to 5G},'' \emph{IEEE Commun. Surv. Tutor.},
  vol.~20, no.~2, pp. 1124--1148, 2018.

\bibitem{2018_TCOM_RSSindoor}
Z.~Lin \emph{et~al.}, ``{3-D indoor positioning for millimeter-wave massive
  MIMO systems},'' \emph{IEEE Trans. Commun.}, vol.~66, no.~6, pp. 2472--2486,
  Jun. 2018.

\bibitem{2017_TSP_DirectLoc_ToA}
N.~Garcia \emph{et~al.}, ``{Direct localization for massive MIMO},'' \emph{IEEE
  Trans. Signal Process.}, vol.~65, no.~10, pp. 2475--2487, May 2017.

\bibitem{2015_CC_TDoA}
H.~Xiong \emph{et~al.}, ``{TDOA localization algorithm with compensation of
  clock offset for wireless sensor networks},'' \emph{China Commun.}, vol.~12,
  no.~10, pp. 193--201, Oct. 2015.

\bibitem{2ToA1TDoA}
M.~Van~Eeckhaute \emph{et~al.}, ``{Low complexity iterative localization of
  time-misaligned terminals in cellular networks},'' \emph{IEEE Trans. Veh.
  Technol.}, vol.~67, no.~11, pp. 10\,730--10\,739, Nov. 2018.

\bibitem{ref_PositioningIn5GNetworks}
S.~Dwivedi \emph{et~al.}, ``Positioning in {5G} networks,'' \emph{IEEE Commun.
  Mag.}, vol.~59, no.~11, pp. 38--44, 2021.

\bibitem{ref_3GPP_38215}
3GPP, ``{Tech. Spec. (TS) 38.215: physical layer measurements},'' ,v. 17.2.0,
  2022.

\bibitem{2018_TWC_AoA}
A.~Shahmansoori \emph{et~al.}, ``{Position and orientation estimation through
  millimeter-wave MIMO in 5G systems},'' \emph{IEEE Trans. Wireless Commun.},
  vol.~17, no.~3, pp. 1822--1835, Mar. 2018.

\bibitem{2022_JSAC_RISaidedSensing}
X.~Shao \emph{et~al.}, ``{Target sensing with intelligent reflecting surface:
  Architecture and performance},'' \emph{IEEE J. Sel. Areas Commun.}, vol.~40,
  no.~7, pp. 2070--2084, Jul. 2022.

\bibitem{ref_TWC_wangwei}
W.~Wang and W.~Zhang, ``Joint beam training and positioning for intelligent
  reflecting surfaces assisted millimeter wave communications,'' \emph{IEEE
  Trans. Wireless Commun.}, vol.~20, no.~10, pp. 6282--6297, 2021.

\bibitem{ref:RIS_assisted_localization_bound}
Z.~Wang \emph{et~al.}, ``Location awareness in beyond {5G} networks via
  reconfigurable intelligent surfaces,'' \emph{IEEE J. Sel. Areas Commun.},
  vol.~40, no.~7, pp. 2011--2025, 2022.

\bibitem{ref_ShicongLiu}
S.~Liu \emph{et~al.}, ``Deep denoising neural network assisted compressive
  channel estimation for {mmWave} intelligent reflecting surfaces,'' \emph{IEEE
  Trans. Veh. Technol.}, vol.~69, no.~8, pp. 9223--9228, 2020.

\bibitem{ref_ZiweiWan}
Z.~Wan \emph{et~al.}, ``Terahertz massive {MIMO} with holographic
  reconfigurable intelligent surfaces,'' \emph{IEEE Trans. on Commun.},
  vol.~69, no.~7, pp. 4732--4750, 2021.

\bibitem{ref:OMP}
J.~Lee, G.-T. Gil, and Y.~H. Lee, ``Channel estimation via orthogonal matching
  pursuit for hybrid {MIMO} systems in millimeter wave communications,''
  \emph{IEEE Trans. on Commun.}, vol.~64, no.~6, pp. 2370--2386, 2016.

\bibitem{ref:ZhenGao_Letter}
Z.~Gao \emph{et~al.}, ``Channel estimation for millimeter-wave massive {MIMO}
  with hybrid precoding over frequency-selective fading channels,'' \emph{IEEE
  Commun. Lett.}, vol.~20, no.~6, pp. 1259--1262, 2016.

\bibitem{2022_TCOM_LinglongDai_PolarDomain}
M.~Cui and L.~Dai, ``{Channel estimation for extremely large-scale MIMO:
  Far-field or near-field?}'' \emph{IEEE Trans. Commun.}, vol.~70, no.~4, pp.
  2663--2677, Apr. 2022.

\bibitem{gOMP}
J.~Wang, S.~Kwon, and B.~Shim, ``{Generalized orthogonal matching pursuit},''
  \emph{IEEE Trans. Signal Process.}, vol.~60, no.~12, pp. 6202--6216, Dec.
  2012.

\bibitem{offgrid_2018_TVT}
C.~Hu \emph{et~al.}, ``{Super-resolution channel estimation for mmWave massive
  MIMO with hybrid precoding},'' \emph{IEEE Trans. Veh. Technol.}, vol.~67,
  no.~9, pp. 8954--8958, Sep. 2018.

\bibitem{offgrid_2021_TWC}
N.~González-Prelcic \emph{et~al.}, ``{Wideband channel tracking and hybrid
  precoding for mmWave MIMO systems},'' \emph{IEEE Trans. Wireless Commun.},
  vol.~20, no.~4, pp. 2161--2174, Apr. 2021.

\bibitem{ref_ICC_loc_RIS}
Abu-Shaban \emph{et~al.}, ``Near-field localization with a reconfigurable
  intelligent surface acting as lens,'' in \emph{ICC 2021 - IEEE Int. Conf.
  Commun.}, 2021, pp. 1--6.

\bibitem{ref_GFF_loc}
H.~Luo and F.~Gao, ``Beam squint assisted user localization in near-field
  communications systems,'' \emph{arXiv preprint arXiv:2205.11392}, 2022.

\bibitem{2019_TSP_AoA}
B.~Zhou, A.~Liu, and V.~Lau, ``{Successive localization and beamforming in 5G
  mmwave MIMO communication systems},'' \emph{IEEE Trans. Signal Process.},
  vol.~67, no.~6, pp. 1620--1635, Mar. 2019.

\bibitem{2021_JSAC_AnwenLiao}
A.~Liao \emph{et~al.}, ``{Terahertz ultra-massive MIMO-based aeronautical
  communications in space-air-ground integrated networks},'' \emph{IEEE J. Sel.
  Areas Commun.}, vol.~39, no.~6, pp. 1741--1767, Jun. 2021.

\bibitem{ref_hybrid_GaoFeifei}
F.~Gao \emph{et~al.}, ``Wideband beamforming for hybrid massive {MIMO}
  terahertz communications,'' \emph{IEEE J. Sel. Areas Commun.}, vol.~39,
  no.~6, pp. 1725--1740, 2021.

\bibitem{ref_Tse}
D.~Tse and P.~Viswanath, \emph{Fundamentals of wireless communication}.\hskip
  1em plus 0.5em minus 0.4em\relax Cambridge university press, 2005.

\bibitem{ref_2022_TCOM_NanYang_ChongHan}
A.~Shafie \emph{et~al.}, ``Spectrum allocation with adaptive sub-band bandwidth
  for terahertz communication systems,'' \emph{IEEE Trans. on Commun.},
  vol.~70, no.~2, pp. 1407--1422, 2022.

\bibitem{ref_2011_TWC_663}
J.~M. Jornet and I.~F. Akyildiz, ``Channel modeling and capacity analysis for
  electromagnetic wireless nanonetworks in the terahertz band,'' \emph{IEEE
  Trans. Wireless Commun.}, vol.~10, no.~10, pp. 3211--3221, 2011.

\bibitem{ref_2022_WC_ChongHan}
C.~Han \emph{et~al.}, ``Molecular absorption effect: A double-edged sword of
  terahertz communications,'' \emph{IEEE Wireless Commun.}, pp. 1--8, 2022.

\bibitem{2023_IEEENetwork_NanYang_ChongHan}
A.~Shafie \emph{et~al.}, ``Terahertz communications for {6G} and beyond
  wireless networks: Challenges, key advancements, and opportunities,''
  \emph{IEEE Netw.}, pp. 1--8, 2022.

\bibitem{ref_RevisionPrecode1}
Q.~Yuan \emph{et~al.}, ``Deep learning-based hybrid precoding for terahertz
  massive {MIMO} communication with beam squint,'' \emph{IEEE Commun. Lett.},
  vol.~27, no.~1, pp. 175--179, 2023.

\bibitem{ref_RevisionPrecode2}
Y.~Wu \emph{et~al.}, ``{3-D} hybrid beamforming for terahertz broadband
  communication system with beam squint,'' \emph{IEEE Trans. Broadcast.},
  vol.~69, no.~1, pp. 264--275, 2023.

\bibitem{2021_arXiv_LinglongDai_effectiveRD}
M.~Cui \emph{et~al.}, ``{Near-field wideband beamforming for extremely large
  antenna array},'' \emph{arXiv preprint arXiv:2109.10054}, 2021.

\bibitem{ref_ArmijoWolfe}
L.~Armijo, ``Minimization of functions having {Lipschitz} continuous first
  partial derivatives,'' \emph{Pacific Journal of mathematics}, vol.~16, no.~1,
  pp. 1--3, 1966.

\bibitem{ref_MUSICnoEVD}
S.~Marcos \emph{et~al.}, ``The propagator method for source bearing
  estimation,'' \emph{Signal processing}, vol.~42, no.~2, pp. 121--138, 1995.

\bibitem{ref_HybridMUSIC}
Y.~Chen \emph{et~al.}, ``Millidegree-level {Direction-of-Arrival} estimation
  and tracking for terahertz ultra-massive {MIMO} systems,'' \emph{IEEE Trans.
  Wireless Commun.}, vol.~21, no.~2, pp. 869--883, 2022.

\end{thebibliography}


\vfill

\end{spacing}
\end{document}